\documentclass[aps,prd,preprint,superscriptaddress,showpacs,amsfonts]{revtex4}

\begin{document}


\title{Conservative formulations of general relativistic kinetic theory}



\author{Christian Y. Cardall}
\email[]{ccardall@mail.phy.ornl.gov}
\affiliation{Physics Division, Oak Ridge National Laboratory, Oak Ridge,
 	TN 37831-6354}
\affiliation{Department of Physics and Astronomy, University of Tennessee,
	Knoxville, TN 37996-1200} 
\affiliation{Joint Institute for Heavy Ion Research, Oak Ridge National
	Laboratory, Oak Ridge, TN 37831-6374}

\author{Anthony Mezzacappa}
\email[]{mezz@mail.phy.ornl.gov}
\affiliation{Physics Division, Oak Ridge National Laboratory, Oak Ridge,
 	TN 37831-6354}


\date{\today}

\begin{abstract}
Experience with core-collapse supernova simulations shows that 
accurate accounting of total particle number and 4-momentum 
can be
a challenge for computational radiative transfer. 
This 
accurate accounting
would be facilitated by the use of particle number and 4-momentum
transport equations that allow transparent conversion between volume and surface integrals in both configuration and momentum space. 
Such conservative formulations of general relativistic kinetic theory in multiple spatial dimensions are presented in this paper, and their relevance to core-collapse supernova simulations is described.
\end{abstract}


\maketitle

\def\InvisibleSpace{}
\def\InvisibleComma{}
\def\Mvariable{}
\def\Mfunction{}
\def\dispSFNumberedEquationmath{}
\def\inlineSFinmath{}
\def\multsp{}
\def\IndentingNewLine{}

\def\Tilde{\sim}
\def\ScriptCapitalL{{\cal L}}
\def\ScriptCapitalN{{\cal N}}
\def\ScriptCapitalT{{\cal T}}
\def\ScriptCapitalO{{\cal O}}
\def\ScriptCapitalE{{\cal E}}
\def\ScriptN{{\mathfrak n}}
\def\ScriptE{{\mathfrak e}}
\def\ScriptP{{\mathfrak p}}

\def\RawWedge{"705E}
\def\_{"7016}
\def\overvar#1#2{\mathaccent #2 #1}


\section{Introduction}
\label{sec:introduction}

The state of the art in core-collapse supernova simulations now
includes energy- and angle-dependent neutrino transport 
\cite{rampp00,mezzacappa01,liebendoerfer01b,rampp02,liebendoerfer02,thompson02,janka02,janka02b}. 
However, experience 
shows that simultaneous conservation of both energy and lepton 
number is difficult numerically \cite{mezzacappa93b,liebendoerfer02}.  
This challenge motivates us to develop
new conservative formulations of relativistic kinetic theory that 
are specifically 
attuned to
the need for accurate accounting of particle number and energy in
numerical simulations of radiative transfer problems.

Before describing the conservative formulations of kinetic theory
we seek, we explain why energy- and angle-dependent neutrino transport
is necessary in supernova simulations and detail the magnitude of the 
challenge of energy conservation.

Sophisticated
treatments of neutrino transport are necessary because the ultimate
energy source of the supernova explosion---the gravitational potential
energy of the stellar progenitor's core---is eventually converted almost
completely into neutrinos. 
Some of the gravitational potential energy is lost to escaping
neutrinos during core collapse, but most of it is converted into a thermal
bath of dense nuclear matter, photons, electron/positron pairs, and
trapped neutrinos deep inside
the nascent neutron 
star.  Neutrinos, having the weakest interactions, are the most efficient
means of cooling; they diffuse outward on a time scale of seconds towards 
a semi-transparent region near the surface of the newly forming neutron star, 
and eventually
escape with about 99\% of the released gravitational energy. 
In modeling the conversion of gravitational potential energy into 
neutrino fluxes, energy- and
angle-dependent neutrino transport is necessary to accurately follow the
transition from quasi-isotropic diffusion to forward-peaked free streaming.
In this transition region energy is transferred from the neutrino
radiation to the matter behind a stalled shock wave, and this energy
transfer may be necessary to propel
the shock through the outer layers in an explosion 
\cite{colgate66,bethe85}. 
But whether or not such neutrino heating is the
proximate cause of explosion, the fact that neutrinos dominate the
energetics implies that accurate neutrino transport
is integral to any realistic and comprehensive study of the 
explosion mechanism.

The importance of accurate neutrino transport is 
a lesson learned from experience with
supernova simulations. Parametrized studies
\cite{janka96}
highlight the sensitivity of explosions to neutrino luminosities
and to conditions in the semi-transparent region near the 
nascent neutron star's surface (see also Ref. 
\cite{janka01})---precisely the region where
neutrino energy and angle dependence must be tracked carefully. 
Moreover, there remains a nagging qualititative uncertainty in simulations
with multidimensional hydrodynamics: Those with neutrino transport
that depends on direction in configuration space but is
averaged over energy and angles exhibit explosions 
\cite{herant94,burrows95,fryer02}, while those with neutrino transport
that depends on energy but is averaged over angles in both configuration and
momentum space do not show explosions \cite{mezzacappa98,mezzacappa98b}.
It may be that these differing outcomes are due to the different 
neutrino transport schemes, both of which are ultimately inadequate.

Moving beyond these general arguments for the necessity of accurate
neutrino transport, 
quantitative consideration of the energetics shows how severe
the requirements are on 
one aspect of accuracy:
energy conservation. 
As mentioned above, virtually all of the gravitational
potential energy (\inlineSFinmath${\Tilde {{10}^{53}}\multsp\ \Mvariable{\rm erg}}$) 
released during collapse is eventually converted into intense neutrino 
fluxes lasting several seconds. 
However, 
supernova explosion energies 
(the kinetic energy of the ejecta)
are observed to be only
\inlineSFinmath${\Tilde {{10}^{51}}\Mvariable\ {\rm erg}}$. 
Now, because it is difficult to argue with any rigor about the physical impact of any energy lost
or gained due to numerical error, 
the total energy should be conserved to a precision corresponding to the 
phenomena of interest in the problem.
Hence a simulation's result for explosion energy accurate to, say, \inlineSFinmath${10\InvisibleSpace \%}$ would require
total energy conservation to an accuracy of about \inlineSFinmath${0.1\InvisibleSpace \%}$. Allowing for systematic error accrual, the total energy
would have to be conserved at a level of \inlineSFinmath${0.1\InvisibleSpace \%/N}$ per time step, where \inlineSFinmath${N\Tilde {{10}^5}}$ is the total
number of time steps in the simulation.

Conservative formulations of kinetic theory would help to meet the
numerical challenge of particle number and energy conservation in 
core-collapse supernova
simulations.
To give an idea of the kind of formulation of kinetic theory that we seek,
we first review familiar descriptions of the dynamics of a fluid medium. The dynamics can be described in two different ways,
which might be called {\em elemental} and {\em conservative}.

The elemental formulation expresses the evolution of the fluid in terms of equations of motion for the velocity and some independent set of 
quantities (e.g. temperature, densities of various species comprising
the fluid, etc.) measured by an observer moving along with the fluid
(``comoving observer'').
For example, 
consider
a spacetime with metric components $\{g_{\mu\nu}\}$
and metric determinant $g\equiv \det(g_{\mu\nu})$, 
containing
a perfect fluid with  4-velocity components
$\{u^\mu\}$  
and comoving frame total energy density $\rho$, pressure $p$, and 
baryon density $n$. 
Iin the absence of radiative transfer and significant energy
input from nuclear reactions, the perfect fluid
 evolves  according to
\begin{eqnarray}
(\rho +p){u^{\mu }}\Big(\frac{\partial {u^{i }}}{\partial {x^{\mu }}}+{{{{\Gamma }^{i }}\InvisibleSpace }_{\Mvariable{\rho
\mu }}}{u^{\rho }}\Big)+({g^{\Mvariable{i \mu }}}+{u^{i }}{u^{\mu }})\frac{\partial p}{\partial {x^{\mu }}}&=&0, \label{euler} \\
\dispSFNumberedEquationmath{u^{\mu }}\frac{\partial \rho }{\partial {x^{\mu }}}+\frac{(\rho +p)}{{\sqrt{-g}}}\frac{\partial }{\partial {x^{\mu }}}\big({\sqrt{-g}}{u^{\mu
}}\big)&=&0, \label{energy} \\
\dispSFNumberedEquationmath{u^{\mu }}\frac{\partial n}{\partial {x^{\mu }}}+\frac{n}{{\sqrt{-g}}}\frac{\partial }{\partial {x^{\mu }}}\big({\sqrt{-g}}{u^{\mu
}}\big)&=&0.\label{baryon}
\end{eqnarray}
(Greek and latin letters are spacetime and space indices respectively.)
Supplementary relations between \inlineSFinmath${\rho }$, \inlineSFinmath${p}$, and \inlineSFinmath${n}$---referred to as the {\em equation of state}---are
determined by the microphysics of the fluid. The name {\em elemental} 
denotes the fact that 
by writing down separate equations of motion for the velocity
and 
comoving-frame
quantities, the kinetic and ``intrinsic'' fluid 
energies---two ``elements'' of the system---are analytically separated. 

The conservative approach
expresses the evolution of the system in terms of the divergence of the stress-energy tensor \inlineSFinmath${{T^{\Mvariable{\mu
\nu }}}}$. For a perfect fluid, $T^{\mu\nu}\inlineSFinmath{=(\rho +p){u^{\mu }}{u^{\nu }}+p {g^{\Mvariable{\mu \nu }}}}$, and 
Eqs. (\ref{euler}) and (\ref{energy}) 
are replaced by
\begin{equation}
\dispSFNumberedEquationmath{\frac{1}{{\sqrt{-g}}}\frac{\partial }{\partial {x^{\mu }}}\big({\sqrt{-g}}{T^{\Mvariable{\nu \mu }}}\big)=-{{{{\Gamma
}^{\nu }}\InvisibleSpace }_{\Mvariable{\rho \mu }}}{T^{\Mvariable{\rho \mu}}},}\label{divergence}
\end{equation}
while Eq. (\ref{baryon}) is replaced by
\begin{equation}
\dispSFNumberedEquationmath{\frac{1}{{\sqrt{-g}}}\frac{\partial }{\partial {x^{\mu }}}\big({\sqrt{-g}}n\multsp {u^{\mu }}\big)=0.}\label{baryon2}
\end{equation}
Volume integrals of the left-hand sides of Eqs. (\ref{divergence}) 
and (\ref{baryon2})---obtained by multiplying by the invariant
spacetime volume element $\sqrt{-g}\,d^4x$ and integrating---are
related to surface integrals in an obvious manner. Physically, this
relates the time rates of change of 4-momentum and baryon number 
in a volume to fluxes
through a surface surrounding that volume; hence the 
labeling
of
Eqs. (\ref{divergence}) 
and (\ref{baryon2}) as {\em conservative}. 

The right-hand side of Eq. (\ref{divergence}) deserves note.
After discussing the 
reasons for 
this term's existence,
we 
comment on what it means for conservation issues.

There are several 
important cases where
the connection coefficients
${{{{\Gamma
}^{\nu }}\InvisibleSpace }_{\Mvariable{\rho \mu }}}$
on the right-hand side of Eq. (\ref{divergence})
might not vanish. They are present
in curved spacetime, where (at least in part) they embody 
gravitational forces. But even in flat spacetime, 
coordinates employed by accelerated observers give rise to 
connection coefficients.
And even without spacetime curvature or accelerated reference
frames, connection coefficients arise from the use of 
curvilinear coordinates.

What does the right-hand side of Eq. (\ref{divergence}) mean for
4-momentum conservation? Only when it vanishes---that is, only
for inertial observers in flat spacetime employing rectilinear 
coordinates---are the components of total 4-momentum
constant in time (``conserved''). For only in this case do the coordinates
reflect the translation invariance of flat spacetime, the physical origin
of 4-momentum conservation. (Curved spacetime lacks translation
invariance, so there is no 4-momentum conservation.) 
Because the presence, for whatever reason, of a source term like the
right-hand side of Eq. (\ref{divergence}) 
means that the 4-momentum components in such a basis
are not conserved, it 
might more properly be called a ``balance equation'' than a
``conservation law''. But because the volume integral of the left-hand side  
easily translates into a surface integral, we still call it a 
``conservative'' formulation. 

There are some special cases in which a conserved quantity
associated with the time coordinate $t$ can be found, however.
For unaccelerated observers in flat spacetime, 
${{{{\Gamma
}^{t }}\InvisibleSpace }_{\Mvariable{\rho \mu }}}=0$, even in
curvilinear coordinates. This means that energy is conserved, though
the 3-momentum components in cuvilinear coordinates are not. Another
special case is the restriction to spherical symmetry in general relativity:
Here certain coordinate choices allow the non-vanishing ${{{{\Gamma
}^{t }}\InvisibleSpace }_{\Mvariable{\rho \mu }}}$ terms to be absorbed into
the left-hand aside, leading to the identification of a conserved energy-like
quantity (see e.g. Ref. \cite{liebendoerfer01}). 


Having used the familiar example of a perfect fluid to discuss what we
mean by ``elemental'' and ``conservative'' formulations, 
we now consider kinetic theory in terms of these categories.
The evolution of a particle type described by kinetic theory is often expressed as an equation of motion for the 
distribution function \inlineSFinmath${f}$, 
the ensemble-averaged density of a given particle type in phase space. 
(These concepts will be defined with greater precision in subsequent sections;
for the present discussion it is sufficient to understand that phase space 
is the combination of configuration space and momentum space, and that
multiplying $f$ by the volume of an infinitesimal cell in phase space gives
the number of particles having positions and momenta within the ranges 
defined by that cell.)
The distribution function evolves due to advection through phase space and particle interactions. 

Advection through phase space gives rise to derivatives of $f$ with respect to the components
of the position vector $x$ and the momentum vector $p$. 
For numerical evolution 
it is convenient to parametrize the 
distribution function in terms of $\{x^\mu\}$, the
components of $x$ in a global ``coordinate basis'' 
\footnote{In a ``coordinate basis'' the basis vectors are
$\vec{e}_\mu = {\partial\over \partial x^\mu}$. These basis vectors
``commute'' because the order of partial derivatives can
be interchanged freely. 
The connection coefficients in a coordinate basis are given
by 
${{{{\Gamma }^{\mu }}\InvisibleSpace }_{\Mvariable{\nu \rho }}}=\frac{1}{2}{g^{\Mvariable{\mu \sigma }}}\Big(\frac{\partial
{g_{\Mvariable{\sigma \nu }}}}{\partial {x^{\rho }}}+\frac{\partial {g_{\Mvariable{\sigma \rho }}}}{\partial {x^{\nu }}}-\frac{\partial {g_{\Mvariable{\nu
\rho }}}}{\partial {x^{\sigma }}}\Big).$
In a ``non-coordinate basis'' obtained from the coordinate basis
by a transformation 
$\vec{e}_{\mu'} = {L^\mu}_{\mu'}\vec{e}_\mu$,
the basis vectors
$\vec{e}_{\mu'} = {L^\mu}_{\mu'}{\partial\over \partial x^\mu}$ do
not commute if ${L^\mu}_{\mu'}$ depends on position in spacetime. The
connection coefficients in a non-coordinate basis have extra terms
associated with the non-vanishing commutators of the basis vectors.
In this paper, the connection coefficients in a non-coordinate basis
are obtained from those of a coordinate basis via the transformation
${{{{\Gamma }^{{\mu' }}}}_{{\nu' }{\rho' }}}={{{{L
}^{{\mu' }}}}_{\mu }}{{{{L }^{\nu }}}_{{\nu' }}}{{{{L }^{\rho }}}_{{\rho'
}}}{{{{\Gamma }^{\mu }}}_{\Mvariable{\nu \rho }}}+{{{{L }^{{\mu' }}}}_{\mu }}{{{{L }^{\rho
}}}_{{\rho' }}}\frac{\partial {{{{L }^{\mu }}}_{{\nu' }}}}{\partial {x^{\rho }}}$. 
}. 
(Here and throughout
this paper, quantities
defined with respect to the coordinate basis have 
indices without accents.)
For 
momentum 
we make a different choice of basis, 
because interactions with a fluid
are most easily described---and best handled numerically---in terms 
of momentum components measured by a comoving observer. 
The change from the coordinate basis to an 
orthonormal basis associated with the 
comoving observer (a ``non-coordinate basis'') has two parts. 
First there is a transformation 
\inlineSFinmath${{{\Mvariable{dx}}^{\overvar{\mu
}{\_}}}={{{e^{\overvar{\mu }{\_}}}\InvisibleSpace }_{\mu }}{{\Mvariable{dx}}^{\mu }}}$ to an (in general non-comoving) orthonormal basis. 
(Here and throughout
this paper, quantities
defined with respect to the non-comoving orthonormal basis have indices
accented with a bar.)
This is followed by
a Lorentz transformation \inlineSFinmath${{{\Mvariable{dx}}^{\overvar{\mu }{\RawWedge }}}={{{{\Lambda }^{\overvar{\mu }{\RawWedge }}}\InvisibleSpace }_{\overvar{\mu
}{\_}}}{{\Mvariable{dx}}^{\overvar{\mu }{\_}}}}$ 
to a comoving orthonormal basis. (Here and throughout
this paper, quantities
defined with respect to the comoving orthonormal basis have indices
accented with a hat.)
Hence advection through 
phase space will involve derivatives of $f$
with respect to the 
coordinate basis position components
$\{x^\mu\}$ and
the comoving orthonormal basis 
momentum components $\{p^{\overvar{\mu}{\RawWedge}}\}$.

Turning from advection to particle interactions, we consider the case
where the particle species are sufficiently dilute that the interactions
can be described in terms of a collision integral \inlineSFinmath${\mathbb{C}[f]}$ depending only on 
the distribution functions $f$ of the individual particle species. 
(This approximation ignores correlations between particles; that is,
the number of instances of finding particles at the same position is obtained
from the product of their distribution functions.)
In this case, the
equation of motion for the distribution function \inlineSFinmath${f}$ is 
\cite{lindquist66,mezzacappa89}
\begin{equation}
\dispSFNumberedEquationmath{{p^{\overvar{\mu }{\RawWedge }}}\bigg({{{{\Lambda }^{\overvar{\mu }{\_}}}\InvisibleSpace }_{\overvar{\mu }{\RawWedge
}}}{{{e^{\mu }}\InvisibleSpace }_{\overvar{\mu }{\_}}}\frac{\partial f}{\partial {x^{\mu }}}-{{{{\Gamma }^{\overvar{\nu }{\RawWedge }}}\InvisibleSpace
}_{\overvar{\rho }{\RawWedge }\overvar{\mu }{\RawWedge }}}{p^{\overvar{\rho }{\RawWedge }}}\frac{\partial f}{\partial {p^{\overvar{\nu }{\RawWedge
}}}}\bigg)=\mathbb{C}[f].} \label{boltzmann}
\end{equation}
This is the {\em Boltzmann equation}.


In terms of the categories described above in connection with fluid evolution, the Boltzmann equation is ``elemental'': The most fundamental quantity---the
distribution function---is the 
evolved variable,
and volume integrals of the equation in both configuration space and momentum space
are not obviously related to surface integrals. This ``non-conservative'' character is present even in flat spacetime and rectangular coordinates. 
The first term of Eq. (\ref{boltzmann}) is non-conservative even in flat
spacetime and rectilinear coordinates because the boost \inlineSFinmath${{{{{\Lambda }^{\overvar{\mu }{\_}}}\InvisibleSpace
}_{\overvar{\mu }{\RawWedge }}}}$---which depends on the coordinates
$\{x^\mu\}$---sits outside
the derivative $\partial f/\partial x^\mu$.
The spatial dependence of the boost also gives rise to non-vanishing ${{{{\Gamma }^{\overvar{\nu }{\RawWedge }}}\InvisibleSpace
}_{\overvar{\rho }{\RawWedge }\overvar{\mu }{\RawWedge }}}$, even in flat
spacetime and rectilinear coordinates; therefore the 
second term of Eq. (\ref{boltzmann}) does not vanish. This second term is 
non-conservative because the factor $p^{\overvar{\rho }{\RawWedge }}$
sits outside the derivative ${\partial f}/{\partial {p^{\overvar{\nu }{\RawWedge
}}}}$.

While the Boltzmann equation is ``elemental'' or ``non-conservative'', it is well known (e.g., see Refs. \cite{lindquist66,ehlers71,israel72}) 
the first two momentum moments of the distribution function
(integrated over a suitable invariant momentum space volume element \inlineSFinmath${\Mvariable{dP}}$) constitute a particle number current \inlineSFinmath${{N^{\mu
}}}$ and particle stress-energy tensor \inlineSFinmath${{T^{\Mvariable{\mu \nu }}}}$,
\begin{eqnarray}
\dispSFNumberedEquationmath{N^{\mu }}&=&\int f\multsp {p^{\mu }}\Mvariable{dP}
=\int f\multsp 
\,{p^{\overvar{\mu }{\RawWedge }}}{{{{\Lambda }^{\overvar{\mu }{\_}}}\InvisibleSpace }_{\overvar{\mu }{\RawWedge
}}}{{{e^{\mu }}\InvisibleSpace }_{\overvar{\mu }{\_}}}\,
\Mvariable{dP}, 
\label{particleNumberCurrent}
\\
\dispSFNumberedEquationmath{T^{\Mvariable{\mu \nu }}}&=&\int f\multsp {p^{\mu }}{p^{\nu }}\Mvariable{dP}
= \int f\multsp \,
{p^{\overvar{\mu }{\RawWedge }}}{{{{\Lambda }^{\overvar{\mu }{\_}}}\InvisibleSpace }_{\overvar{\mu }{\RawWedge
}}}{{{e^{\mu }}\InvisibleSpace }_{\overvar{\mu }{\_}}}\,
{p^{\overvar{\nu }{\RawWedge }}}{{{{\Lambda }^{\overvar{\nu }{\_}}}\InvisibleSpace }_{\overvar{\nu }{\RawWedge
}}}{{{e^{\nu }}\InvisibleSpace }_{\overvar{\nu }{\_}}}\,
\Mvariable{dP},
\label{particleStressEnergy}
\end{eqnarray}
which (for electrically neutral particles) obey the balance equations 
\begin{eqnarray}
\dispSFNumberedEquationmath\frac{1}{{\sqrt{-g}}}\frac{\partial }{\partial {x^{\mu }}}\big({\sqrt{-g}}{N^{\mu }}\big)&=&\int \mathbb{C}[f]\Mvariable{dP},
\label{particleNumberConservation} \\
\dispSFNumberedEquationmath\frac{1}{{\sqrt{-g}}}\frac{\partial }{\partial {x^{\mu }}}\big({\sqrt{-g}}{T^{\Mvariable{\nu \mu }}}\big)
&=&
-{{{{\Gamma
}^{\nu }}\InvisibleSpace }_{\Mvariable{\rho \mu }}}{T^{\Mvariable{\rho \mu }}}
+ \int \mathbb{C}[f]{p^{\nu }}\Mvariable{dP}. 
\label{particleEnergyConservation}
\end{eqnarray}

While this result is often stated in the literature
(e.g., see Refs. \cite{lindquist66,ehlers71,israel72}), 
because of the non-conservative character of Eq. (\ref{boltzmann}) 
it is not obvious how its momentum moments give rise to
Eqs. (\ref{particleNumberConservation}) and 
(\ref{particleEnergyConservation}). 
Equation (\ref{boltzmann}) contains factors
${{{{\Lambda }^{\overvar{\mu }{\_}}}\InvisibleSpace }_{\overvar{\mu }{\RawWedge
}}}{{{e^{\mu }}\InvisibleSpace }_{\overvar{\mu }{\_}}}$ outside
${\partial f}/{\partial {x^{\mu }}}$; but from Eqs. (\ref{particleNumberCurrent})-(\ref{particleEnergyConservation}) we see that these factors have come inside the derivative with
respect to $x^\mu$. What happens to the spacetime derivatives of 
${{{{\Lambda }^{\overvar{\mu }{\_}}}\InvisibleSpace }_{\overvar{\mu }{\RawWedge
}}}{{{e^{\mu }}\InvisibleSpace }_{\overvar{\mu }{\_}}}$ that are 
generated in taking these factors inside the derivative? 
Furthermore, according to Eqs. (\ref{particleNumberCurrent}) and 
(\ref{particleStressEnergy}), the quantities \inlineSFinmath${{N^{\mu
}}}$ and \inlineSFinmath${{T^{\Mvariable{\mu \nu }}}}$
involve no momentum derivatives of $f$. But the
second term of Eq. (\ref{boltzmann}) contains a factor
$p^{\overvar{\rho }{\RawWedge }}({\partial f}/{\partial {p^{\overvar{\nu }{\RawWedge
}}}})$. Because of the momentum factor
outside the momentum derivative it is not obvious how 
this term will contribute to Eqs. (\ref{particleNumberConservation})
and (\ref{particleEnergyConservation}) when integrated
over momentum space.

How, then, is the connection between Eq. (\ref{boltzmann}) and
Eqs. (\ref{particleNumberConservation}) and 
(\ref{particleEnergyConservation}) established in detail?
The reviews of relativistic kinetic theory by Lindquist \cite{lindquist66} and Israel \cite{israel72} do not provide detailed proofs. 
As discussed in Sec. \ref{sec:distribution}, an explicit proof 
by Ehlers \cite{ehlers71}
relies on the fact that \inlineSFinmath${{N^{\mu }}}$ and \inlineSFinmath${{T^{\Mvariable{\mu
\nu }}}}$ are momentum-integrated quantities, and no direct
insight is gained into what happens to the momentum derivatives of $f$ in the 
integration over momentum space.


For those interested in computer models of radiative transfer
problems, these are not idle academic questions; they are issues that must
be faced in order to build simulations capable of making meaningful
scientific statements about the core-collapse supernova explosion mechanism. 
Experience with 
supernova simulations in spherical
symmetry shows that understanding the detailed connection between
the Boltzmann equation and the particle number and 4-momentum balance
equations has important consequences for how well these quantities are
conserved in a simulation \cite{mezzacappa93,liebendoerfer02}.  
While a discretization of the Boltzmann equation is
a natural numerical method of evolving the neutrino species, na\"{\i}ve differencings of the various terms in 
Eq. (\ref{boltzmann}) generally will not be consistent with 
a straightfoward differencing of
Eqs. (\ref{particleNumberConservation}) and (\ref{particleEnergyConservation}), leading to unacceptable numerical errors in 
particle number and 
energy conservation. 
In Lagrangian (or ``comoving'') 
coordinates in spherical symmetry, Mezzacappa and Bruenn
\cite{mezzacappa93} derive a conservative formulation of the Boltzmann
equation transparently related to particle number balance as expressed
in Eq. (\ref{particleNumberConservation}). They also devise methods of
handling momentum derivatives that are consistent with both number and
energy conservation \cite{mezzacappa93b}.
Liebend\"orfer et al. \cite{liebendoerfer02} went a step further in
this spherically symmetric case, deducing the connnection between 
the Mezzacappa and Bruenn ``number conservative'' Boltzmann equation
in comoving coordinates
and energy conservation 
as represented by an Eulerian (or ``lab frame'') version of
Eq. (\ref{particleEnergyConservation}).
Using complicated, non-intuitive differencings of hydrodynamic and 
gravitational variables, 
they construct a numerical implementation of 
the Mezzacappa and Bruenn ``number conservative'' Boltzmann equation
that is stable and 
faithful to its analytic connection to 
the lab frame version of 
Eq. (\ref{particleEnergyConservation}) 
to the accuracy necessary to make solid scientific statements about the
neutrino-driven explosion mechanism in spherical symmetry 
\cite{mezzacappa01,liebendoerfer01b,liebendoerfer02}.

Because spherically symmetric models with Boltzmann transport fail to reproduce some important observable characteristics of core-collapse supernovae (e.g. the launch of an explosion
\cite{mezzacappa01,liebendoerfer01b,liebendoerfer02,rampp00,rampp02,thompson02}), this 
work must be followed up in multiple spatial dimensions
(see Refs. \cite{janka02,janka02b} for some early efforts).
In this paper we develop---allowing for full relativity and 
multiple spatial dimensions---conservative formulations of kinetic theory, 
such that volume integrals in both configuration and momentum space
are {\em trivially related} to surface integrals. 
These conservative expressions 
{\em make transparent} the connection between Eq. (\ref{boltzmann}) and Eqs. 
(\ref{particleNumberConservation}) and (\ref{particleEnergyConservation}).
They can be used to deduce the term-by-term cancellations involved in
this connection, thereby illuminating the complicated differencings
required to achieve the cancellations numerically.
Our conservative formulations also
suggest new primary variables of radiation transport: particle number
and energy variables describing the contribution of each {\em comoving orthonormal
basis} momentum bin to the {\em coordinate basis} number and energy
densities.
It may be that the use of these new radiation variables
could
provide a simpler path to an accurate accounting of particle number and energy in simulations of radiative transfer
problems.

The organization of this paper is as follows. Differential forms and exterior calculus are natural mathematical tools for handling the volume elements
and integrations needed in relativistic kinetic theory. We closely follow (and slightly extend) Ehlers' work \cite{ehlers71}
in reviewing their use in the description
of phase space for particles of definite mass (Sec. \ref{sec:phaseSpace})
and the derivation of the Boltzmann equation (Sec. \ref{sec:distribution}). 
The centerpieces of this paper---two conservative reformulations of the
Boltzmann equation, which provide transparent connections to particle number and energy-momentum balance as expressed in Eqs. (\ref{particleNumberConservation}) and (\ref{particleEnergyConservation})---are presented in 
Sec. \ref{sec:conservative}.  
Because differential forms and exterior calculus may be unfamiliar to
those whose primary interests are radiation transport, in general,
or supernova science, in particular,
Secs. \ref{sec:phaseSpace}-\ref{sec:conservative} each will contain two
subsections: one containing a general derivation, and a second that 
explicitly demonstrates aspects of 
the derivation in the familiar case of the $O(v)$ limit in
Lagrangian coordinates in spherical symmetry.
(While we review some aspects of exterior calculus in our presentation,
these are mostly in endnotes, and are 
more in the character of reminders than a 
self-contained introduction. For the latter we refer the reader to
Refs. \cite{ehlers71,mtw}.) 
A conclusion (Sec. \ref{sec:conclusion})
summarizes our results,
comments on their connection to moment formalisms, and discusses the
utility of these formulations for supernova simulations. 
As an application of our formalism, an appendix contains \inlineSFinmath{$O
(v)$} equations in flat spacetime, but in coordinates sufficiently general to represent rectangular, spherical, and cylindrical coordinate systems.

\section{Phase space for particles of definite mass}
\label{sec:phaseSpace}

In this section we consider phase space for particles of definite
mass. We specify particle trajectories---the ``states'' whose average
occupation is specified by the distribution function $f$---and 
the volume elements needed to derive macroscopic observables from
$f$ and operators that act upon it.
In discussing the general case (Subsec. \ref{subsec:phaseSpaceGeneral})
we follow Ehlers \cite{ehlers71} closely, and refer the reader to his work for more detailed discussions and proofs of some of the assertions
made here. We extend his discussion to the use of momentum components measured in a frame comoving with the fluid that interacts with the particle species
treated by kinetic theory. The specific case of Lagrangian coordinates in
spherical symmetry to $O(v)$ is treated in Subsec. 
\ref{subsec:phaseSpaceSpecific}.

\subsection{Phase space for particles of definite mass: the general case}
\label{subsec:phaseSpaceGeneral}

A study of kinetic theory begins with consideration of the trajectories of individual particles. The worldline 
$x(\lambda)$
of a particle of mass \inlineSFinmath${m}$ with 4-momentum 
$p$
and electric charge \inlineSFinmath${e}$, moving through 
a spacetime
with metric components \inlineSFinmath$\{g_{\Mvariable{\mu \nu }}\}$ and 
electromagnetic field tensor components 
\inlineSFinmath$\{{F_{\Mvariable{\mu \nu }}}\}$, 
is determined
by
\begin{eqnarray}
\dispSFNumberedEquationmath\frac{{{\Mvariable{dx}}^{\mu }}}{\Mvariable{d\lambda }}&=&{p^{\mu }},\label{momentum}\\
\dispSFNumberedEquationmath\frac{{{\Mvariable{dp}}^{\mu }}}{\Mvariable{d\lambda }}&=&e\multsp {{{F^{\mu }}\InvisibleSpace }_{\nu }}{p^{\nu }}-{{{{\Gamma
}^{\mu }}\InvisibleSpace }_{\Mvariable{\nu \rho }}}{p^{\nu }}{p^{\rho }}.
\label{geodesic}
\end{eqnarray}
In the coordinate basis associated with 
the position vector components
$\{x^\mu\}$, the connection coefficients are
\begin{equation}
\dispSFNumberedEquationmath{{{{{\Gamma }^{\mu }}\InvisibleSpace }_{\Mvariable{\nu \rho }}}=\frac{1}{2}{g^{\Mvariable{\mu \sigma }}}\Big(\frac{\partial
{g_{\Mvariable{\sigma \nu }}}}{\partial {x^{\rho }}}+\frac{\partial {g_{\Mvariable{\sigma \rho }}}}{\partial {x^{\nu }}}-\frac{\partial {g_{\Mvariable{\nu
\rho }}}}{\partial {x^{\sigma }}}\Big).}\label{connectionCoordinate}
\end{equation}
As mentioned in Sec. \ref{sec:introduction}, we seek results expressed in terms of (orthonormal) momentum components $\{p^{\hat\mu}\}$ 
measured by an observer comoving with the
fluid with which the particles interact. Combining the transformation \inlineSFinmath${{{{e^{\overvar{\mu }{\_}}}\InvisibleSpace }_{\mu }}}$ to an
orthonormal tetrad and  boost \inlineSFinmath${{{{{\Lambda }^{\overvar{\mu }{\RawWedge }}}\InvisibleSpace }_{\overvar{\mu }{\_}}}}$ to the comoving
frame into the composite transformation 
\begin{equation}
\dispSFNumberedEquationmath{{{{{\ScriptCapitalL }^{\overvar{\mu }{\RawWedge }}}\InvisibleSpace }_{\mu }}={{{{\Lambda }^{\overvar{\mu }{\RawWedge
}}}\InvisibleSpace }_{\overvar{\mu }{\_}}}{{{e^{\overvar{\mu }{\_}}}\InvisibleSpace }_{\mu }},}\label{compositeTransformation}
\end{equation}
Eqs. (\ref{momentum}) and (\ref{geodesic}) are replaced by 
\begin{eqnarray}
\dispSFNumberedEquationmath\frac{{{\Mvariable{dx}}^{\mu }}}{\Mvariable{d\lambda }}&=&{{{{\ScriptCapitalL }^{\mu }}\InvisibleSpace }_{\overvar{\mu
}{\RawWedge }}}{p^{\overvar{\mu }{\RawWedge }}},\label{momentum2}\\
\dispSFNumberedEquationmath\frac{{{\Mvariable{dp}}^{\overvar{\mu }{\RawWedge }}}}{\Mvariable{d\lambda }}&=&e\multsp {{{F^{\overvar{\mu }{\RawWedge
}}}\InvisibleSpace }_{\overvar{\nu }{\RawWedge }}}{p^{\overvar{\nu }{\RawWedge }}}-{{{{\Gamma }^{\overvar{\mu }{\RawWedge }}}\InvisibleSpace }_{\overvar{\nu
}{\RawWedge }\overvar{\rho }{\RawWedge }}}{p^{\overvar{\nu }{\RawWedge }}}{p^{\overvar{\rho }{\RawWedge }}}.\label{geodesic2}
\end{eqnarray}
The connection coefficients do not transform exactly as tensor components, but as
\begin{equation}
\dispSFNumberedEquationmath{{{{{\Gamma }^{\overvar{\mu }{\RawWedge }}}}_{\overvar{\nu }{\RawWedge }\overvar{\rho }{\RawWedge }}}={{{{\ScriptCapitalL
}^{\overvar{\mu }{\RawWedge }}}}_{\mu }}{{{{\ScriptCapitalL }^{\nu }}}_{\overvar{\nu }{\RawWedge }}}{{{{\ScriptCapitalL }^{\rho }}}_{\overvar{\rho
}{\RawWedge }}}{{{{\Gamma }^{\mu }}}_{\Mvariable{\nu \rho }}}+{{{{\ScriptCapitalL }^{\overvar{\mu }{\RawWedge }}}}_{\mu }}{{{{\ScriptCapitalL }^{\rho
}}}_{\overvar{\rho }{\RawWedge }}}\frac{\partial {{{{\ScriptCapitalL }^{\mu }}}_{\overvar{\nu }{\RawWedge }}}}{\partial {x^{\rho }}}.}
\label{connectionCoefficients}
\end{equation}
This completes the desired specification of particle trajectories.

The set of position vectors \inlineSFinmath${x}$ and momentum vectors \inlineSFinmath${p}$ comprise an 8-dimensional manifold \inlineSFinmath${M}$, the
{\itshape one-particle phase space for particles of arbitrary rest masses. }The positions \inlineSFinmath${x}$ are points in spacetime, and the momenta
\inlineSFinmath${p}$ are points in that portion of the tangent space at \inlineSFinmath${x}$ characterized by \inlineSFinmath${{p^2}\leq 0}$ and \inlineSFinmath${p}$
future-directed. The curves 
$\left(x(\lambda),p(\lambda)\right)$
obtained from the particle equations
of motion comprise the {\itshape phase flow }in \inlineSFinmath${M}$. The
phase flow is generated by the {\itshape Liouville vector}
\begin{equation}
\dispSFNumberedEquationmath{L={p^{\overvar{\mu }{\RawWedge }}}{{{{\ScriptCapitalL }^{\mu }}\InvisibleSpace }_{\overvar{\mu }{\RawWedge }}}\frac{\partial
}{\partial {x^{\mu }}}+\big(e\multsp {{{F^{\overvar{\mu }{\RawWedge }}}\InvisibleSpace }_{\overvar{\nu }{\RawWedge }}}{p^{\overvar{\nu }{\RawWedge
}}}-{{{{\Gamma }^{\overvar{\mu }{\RawWedge }}}\InvisibleSpace }_{\overvar{\nu }{\RawWedge }\overvar{\rho }{\RawWedge }}}{p^{\overvar{\nu }{\RawWedge
}}}{p^{\overvar{\rho }{\RawWedge }}}\big)\frac{\partial }{\partial {p^{\overvar{\mu }{\RawWedge }}}}.}
\end{equation}
Here \inlineSFinmath${\bigg(\left\{\frac{\partial }{\partial {x^{\mu }}}\right\},\left\{\frac{\partial }{\partial {p^{\overvar{\mu }{\RawWedge }}}}\right\}\bigg)}$ have been chosen as 
basis vectors
on \inlineSFinmath${M}$.

Specification of a particle mass \inlineSFinmath${m}$ defines a hypersurface of \inlineSFinmath${M}$, a 7-dimensional manifold \inlineSFinmath${{M_m}}$
called the {\itshape one-particle phase space for particles of definite mass }\inlineSFinmath${m}$. The mass satisifies
\begin{equation}
\dispSFNumberedEquationmath{{m^2}=-{g_{\Mvariable{\mu \nu }}}{p^{\mu }}{p^{\nu }}=-{g_{\Mvariable{\mu \nu }}}{{{{\ScriptCapitalL }^{\mu }}\InvisibleSpace
}_{\overvar{\mu }{\RawWedge }}}{{{{\ScriptCapitalL }^{\nu }}\InvisibleSpace }_{\overvar{\nu }{\RawWedge }}}{p^{\overvar{\mu }{\RawWedge }}}{p^{\overvar{\nu
}{\RawWedge }}}.}\label{massShell}
\end{equation}
Considered as a scalar function 
$m(x,p)$ on $M$,
the particle mass
satisfies
\begin{equation}
\dispSFNumberedEquationmath{L[m]=0},\label{constantMass}
\end{equation}
which expresses the constancy of \inlineSFinmath${m}$ on each phase space trajectory. Hence the 7-dimensional manifold \inlineSFinmath${{M_m}}$ is generated
by all the phase space trajectories of particles of mass \inlineSFinmath${m}$. Equation (\ref{constantMass}) shows that \inlineSFinmath${L}$ is tangent to \inlineSFinmath${{M_m}}$,
and Eq. (\ref{massShell}) is a constraint indicating that it is one of the momentum dimensions of \inlineSFinmath${M}$ that has been lost in going over to
\inlineSFinmath${{M_m}}$. Hence the Liouville operator restricted to \inlineSFinmath${{M_m}}$ can be expressed  
\begin{equation}
\dispSFNumberedEquationmath{{L_m}={p^{\overvar{\mu }{\RawWedge }}}{{{{\ScriptCapitalL }^{\mu }}\InvisibleSpace }_{\overvar{\mu }{\RawWedge }}}\frac{\partial
}{\partial {x^{\mu }}}+\big(e\multsp {{{F^{\overvar{i}{\RawWedge }}}\InvisibleSpace }_{\overvar{\nu }{\RawWedge }}}{p^{\overvar{\nu }{\RawWedge }}}-{{{{\Gamma
}^{\overvar{i}{\RawWedge }}}\InvisibleSpace }_{\overvar{\nu }{\RawWedge }\overvar{\rho }{\RawWedge }}}{p^{\overvar{\nu }{\RawWedge }}}{p^{\overvar{\rho
}{\RawWedge }}}\big)\frac{\partial }{\partial {p^{\overvar{i}{\RawWedge }}}},}
\label{liouville}
\end{equation}
where \inlineSFinmath${\bigg(\left\{\frac{\partial }{\partial {x^{\mu }}}\right\},\left\{\frac{\partial }{\partial {p^{\overvar{i}{\RawWedge }}}}\right\}\bigg)}$ have been chosen
as basis vectors in \inlineSFinmath${{M_m}}$. To allow for a change of coordinates in momentum space from \inlineSFinmath${\big\{{p^{\overvar{i}{\RawWedge
}}}\big\}}$ to \inlineSFinmath${\big\{{u^{\overvar{i}{\RawWedge }}}\big\}}$ (to momentum space spherical coordinates, for example), we rewrite Eq. (\ref{liouville}) as
\begin{equation}
\dispSFNumberedEquationmath{{L_m}={p^{\overvar{\mu }{\RawWedge }}}{{{{\ScriptCapitalL }^{\mu }}\InvisibleSpace }_{\overvar{\mu }{\RawWedge }}}\frac{\partial
}{\partial {x^{\mu }}}+\big(e\multsp {{{F^{\overvar{j}{\RawWedge }}}\InvisibleSpace }_{\overvar{\nu }{\RawWedge }}}{p^{\overvar{\nu }{\RawWedge }}}-{{{{\Gamma
}^{\overvar{j}{\RawWedge }}}\InvisibleSpace }_{\overvar{\nu }{\RawWedge }\overvar{\rho }{\RawWedge }}}{p^{\overvar{\nu }{\RawWedge }}}{p^{\overvar{\rho
}{\RawWedge }}}\big)\frac{\partial {u^{\overvar{i}{\RawWedge }}}}{\partial {p^{\overvar{j}{\RawWedge }}}}\frac{\partial }{\partial {u^{\overvar{i}{\RawWedge
}}}},} \label{liouvilleFinal}
\end{equation}
where \inlineSFinmath${\Big(\left\{\frac{\partial }{\partial {x^{\mu }}}\right\},\left\{\frac{\partial }{\partial {u^{\overvar{i}{\RawWedge }}}}\right\}\Big)}$ are now the chosen
basis vectors on \inlineSFinmath${{M_m}}$.

Before discussing volume and ``surface'' elements on $M_m$, we give a 
brief reminder of some properties 
of differential forms and their exterior derivatives (e.g., see Refs.
 \cite{ehlers71,mtw}). If \inlineSFinmath${\{{{\Mvariable{dz}}^a}\}}$
are basis 1-forms on some manifold, an \inlineSFinmath${m}$-form \inlineSFinmath${\psi }$ can be expanded as
\begin{equation}
\dispSFNumberedEquationmath{\psi =\frac{1}{m!}{{\psi }_{{a_1}{{\Mvariable{\ldots a}}_m}}}{{\Mvariable{dz}}^{{a_1}{{\Mvariable{\ldots a}}_m}}},}
\end{equation}
where $dz^{a_1\ldots a_m}$ denotes the wedge product $dz^{a_1}\wedge
\ldots\wedge dz^{a_m}$.
The quantities \inlineSFinmath${{{\psi }_{{a_1}{{\Mvariable{\ldots a}}_m}}}}$ are the {\itshape components} of \inlineSFinmath${\psi }$. The components are
completely antisymmetric (\inlineSFinmath${{{\psi }_{{a_1}{{\Mvariable{\ldots a}}_m}}}}=\inlineSFinmath{{{\psi }_{[{a_1}{{\Mvariable{\ldots a}}_m}]}}}$), as
is the wedge product,
so that \inlineSFinmath${\psi }$ can also be expressed
\begin{equation}
\dispSFNumberedEquationmath{\psi ={{\psi }_{|{a_1}{{\Mvariable{\ldots a}}_m}|}}{{\Mvariable{dz}}^{{a_1}{{\Mvariable{\ldots a}}_m}}},}\label{bars}
\end{equation}
where the vertical bars indicate that summation is taken only over \inlineSFinmath${{a_1}<{a_2}<\ldots <{a_m}}$. The {\itshape exterior derivative}
takes an \inlineSFinmath${m}$-form into an \inlineSFinmath${(m+1)}$-form as follows:
\begin{equation}
\dispSFNumberedEquationmath{\Mvariable{d\psi }=\frac{\partial {{\psi }_{|{a_1}{{\Mvariable{\ldots a}}_m}|}}}{\partial {z^{{a_{m+1}}}}}{{\Mvariable{dz}}^{{a_{m+1}}{a_1}{{\Mvariable{\ldots
a}}_m}}}.}\label{exteriorDerivative}
\end{equation}
Two properties of the exterior derivative will be used later in this paper:
\begin{eqnarray}
\dispSFNumberedEquationmath d(\phi +\psi )&=&\Mvariable{d\phi }+\Mvariable{d\psi }, \label{exteriorSum}\\
\dispSFNumberedEquationmath d(\phi \wedge \psi )&=&\Mvariable{d\phi }\wedge \psi +{{(-1)}^p}\phi \wedge \Mvariable{d\psi },\label{exteriorProduct}
\end{eqnarray}
where \inlineSFinmath${\phi }$ is a \inlineSFinmath${p}$-form.

Now we turn to a discussion of
volume and ``surface'' elements on \inlineSFinmath${{M_m}}$.
The invariant volume element in spacetime is the 4-form
\begin{equation}
\dispSFNumberedEquationmath{\eta ={1\over 4!}{\sqrt{-g}}\,
{{\epsilon }_{\Mvariable{\mu \nu \rho \sigma }}}\,
{{\Mvariable{dx}}^{\mu\nu\rho\sigma}},}
\label{spacetimeElement}
\end{equation}
where $g$ is the determinant of the metric,
\inlineSFinmath${{{\epsilon }_{\Mvariable{\mu
\nu \rho \sigma }}}}$ is the completely antisymmetric symbol with \inlineSFinmath${{{\epsilon }_{0123}}=+1}$, and
the wedge product of basis 1-forms is abbreviated by the notation $\inlineSFinmath{{{\Mvariable{dx}}^{\mu\nu\rho\sigma}}\equiv {{\Mvariable{dx}}^\mu}\wedge
{{\Mvariable{dx}}^\nu}\wedge {{\Mvariable{dx}}^\rho}\wedge {{\Mvariable{dx}}^\sigma}}$. 
Shifting attention to the momentum space at a
given spacetime point, an
invariant volume element on a mass shell corresponding to mass \inlineSFinmath${m}$ in the tangent space is 
\begin{equation}
\dispSFNumberedEquationmath{{{\pi }_m}={1\over 3!}\frac{{\sqrt{-g}}}{|{p_0}|}
\epsilon_{0ijk}{{\Mvariable{dp}}^{ijk}}.}
\end{equation}
Expressed in terms of basis 1-forms conjugate to orthonormal comoving frame momentum components, it becomes
\begin{equation}
\dispSFNumberedEquationmath{{{\pi }_m}=\frac{1}{E({\bf p})}{{\Mvariable{dp}}^{\overvar{1}{\RawWedge }\overvar{2}{\RawWedge }\overvar{3}{\RawWedge
}}},}
\end{equation}
where the bold character denotes a 3-vector, and
$E({\bf p}) \equiv \sqrt{|{\bf p}|^2 + m^2}$.
After a final change of momentum coordinates from
${\bf p}$ to ${\bf u}$,
the momentum space volume element can also be expressed as
\begin{equation}
\dispSFNumberedEquationmath{{{\pi }_m}=\frac{1}{E({\bf p})}\Bigg|\det \big[\frac{\partial {\bf p}}{\partial {\bf u}}\big]\Bigg|{{\Mvariable{du}}^{\overvar{1}{\RawWedge
}\overvar{2}{\RawWedge }\overvar{3}{\RawWedge }}}.}\label{momentumElement}
\end{equation}
A volume element on \inlineSFinmath${{M_m}}$ is a 7-form, which can be expressed as
\begin{eqnarray}
\dispSFNumberedEquationmath\Omega& =&\eta \wedge {{\pi }_m} \label{volume}\\
 &=&\frac{{\sqrt{-g}}}{E({\bf p})}\Bigg|\det \big[\frac{\partial {\bf p}}{\partial {\bf u}}\big]\Bigg|{{\Mvariable{dx}}^{0123}}\wedge
{{\Mvariable{du}}^{\overvar{1}{\RawWedge }\overvar{2}{\RawWedge }\overvar{3}{\RawWedge }}}.\label{volume2}
\end{eqnarray}
The ``surface element'' normal to the phase flow 
in \inlineSFinmath${{M_m}}$ is an important quantity; it is  
a 6-form, obtained by contracting 
\footnote{The contraction of a basis vector ${\partial \over \partial z^a}$ 
with a 1-form
$dz^b$ is ${\partial \over\partial z^a}
 \cdot dz^b = \delta^b_a$. In the contraction
of a vector with a {\em direct} product of 1-forms, the contraction is with 
the first 1-form in the direct product. This means that the contraction of
a vector with a {\em wedge} product of $n$ 1-forms gives $n$ terms, because
the wedge product is the 
completely antisymmetrized sum of $n$ direct products.}
\inlineSFinmath${\Omega }$ with the Liouville vector
\inlineSFinmath${{L_m}}$: 
\begin{eqnarray}
\omega &=&{L_m}\cdot \Omega  \nonumber \\
 & =&\frac{{\sqrt{-g}}}{3!}{p^{\overvar{\mu }{\RawWedge }}}{{{{\ScriptCapitalL }^{\mu }}\InvisibleSpace }_{\overvar{\mu }{\RawWedge }}}{{\epsilon
}_{\Mvariable{\mu \nu \rho \sigma }}}({{\Mvariable{dx}}^{\Mvariable{\nu \rho \sigma }}}\wedge {{\pi }_m})+\IndentingNewLine \nonumber  \\
 & & \frac{1}{2!{E({\bf p})}}\Bigg|\det \big[\frac{\partial {\bf p}}{\partial {\bf u}}\big]\Bigg|\big(e\multsp {{{F^{\overvar{j}{\RawWedge
}}}\InvisibleSpace }_{\overvar{\nu }{\RawWedge }}}{p^{\overvar{\nu }{\RawWedge }}}-{{{{\Gamma }^{\overvar{j}{\RawWedge }}}\InvisibleSpace }_{\overvar{\nu
}{\RawWedge }\overvar{\rho }{\RawWedge }}}{p^{\overvar{\nu }{\RawWedge }}}{p^{\overvar{\rho }{\RawWedge }}}\big)\frac{\partial {u^{\overvar{i}{\RawWedge
}}}}{\partial {p^{\overvar{j}{\RawWedge }}}}{{\epsilon }_{\overvar{0}{\RawWedge }\overvar{i}{\RawWedge }\overvar{k}{\RawWedge }\overvar{n}{\RawWedge
}}}\big({{\Mvariable{du}}^{\overvar{k}{\RawWedge }\overvar{n}{\RawWedge }}}\wedge \eta \big). \label{surface}
\end{eqnarray}
The first term in Eq. (\ref{surface}) can be written in terms of 
surface elements in spacetime. A surface element in spacetime 
with normal in the $x^\mu$ direction
is given by the contraction of the vector $\partial/\partial x^\mu$ with
the spacetime volume element $\eta$:
\begin{eqnarray}
{{\sigma }_{\mu }}&=& {\partial\over\partial x^\mu}\cdot\eta \nonumber
\\
&=&\frac{{\sqrt{-g}}}{3!}{{\epsilon }_{\Mvariable{\mu \nu \rho \sigma }}}{{\Mvariable{dx}}^{\Mvariable{\nu
\rho \sigma }}}.\label{spacetimeSurface}
\end{eqnarray}
In terms of these spacetime surface elements, 
the first term in Eq. (\ref{surface}) can be expressed
\begin{equation}
\dispSFNumberedEquationmath{\frac{{\sqrt{-g}}}{3!}{p^{\overvar{\mu }{\RawWedge }}}{{{{\ScriptCapitalL }^{\mu }}\InvisibleSpace }_{\overvar{\mu }{\RawWedge
}}}{{\epsilon }_{\Mvariable{\mu \nu \rho \sigma }}}({{\Mvariable{dx}}^{\Mvariable{\nu \rho \sigma }}}\wedge {{\pi }_m})={p^{\overvar{\mu }{\RawWedge
}}}{{{{\ScriptCapitalL }^{\mu }}\InvisibleSpace }_{\overvar{\mu }{\RawWedge }}}({{\sigma }_{\mu }}\wedge {{\pi }_m}).}\label{surfaceFirst}
\end{equation}
This first term of the ``surface element'' $\omega$ is all that contributes
to an integration over momentum space 
\footnote{The
computational procedure for integrals over differential forms
involves (1) forming an infinitesimal parallelpiped---a wedge
product of the vectors forming the edges of the infinitesimal 
region---at each point in the region over which the integration
is performed; (2) contracting this infinitesimal parallelpiped
with the differential form at each point; (3) calculating the
integral with the usual rules of elementary calculus. An 
infinitesimal parallelpiped in momentum space is  
a wedge product of displacement vectors in
each of the three directions in momentum space. 
In the contraction of this parallelpiped
with $\omega$, the only terms that do not vanish are those
in Eq. (\ref{surfaceFirst}); the other terms of Eq. (\ref{surface})
have only two momentum space 1-forms.}; it corresponds 
to a cell in the familiar 6-dimensional Newtonian phase space.
The surface element \inlineSFinmath${\omega }$ also shares an important
property with a Newtonian phase space volume element: It is invariant under phase flow. This result
follows from the vanishing
exterior derivative of \inlineSFinmath${\omega }$,
\begin{equation}
\dispSFNumberedEquationmath{\Mvariable{d\omega }=0,\label{domega}}
\end{equation}
and the use of Stokes' theorem on a small tube of worldlines \cite{ehlers71}.

\subsection{Phase space for particles of definite mass: Lagrangian coordinates in spherical symmetry to $O(v)$}
\label{subsec:phaseSpaceSpecific}

Here we specialize certain results of the previous subsection to massless,
electrically
neutral particles propagating through flat spacetime, employing Lagrangian
coordinates in spherical symmetry and retaining only terms to $O(v)$.

First we introduce these coordinates. 
In flat spacetime and Eulerian spherical coordinates, the 
metric components are specified by the line element
\begin{equation}
ds^2 = -d{\tilde t}^2 + dr^2 + r^2\, d\theta^2 + r^2\sin^2\theta\,d\phi^2.
\label{eulerianMetric}
\end{equation}
We seek Lagrangian coordinates $t$ and $m$ to replace the Eulerian
coordinates $\tilde t$ and $r$. The enclosed baryon mass
$m$ is defined by 
\begin{equation}
m=\int^r 4\pi r^2\rho\,dr,\label{massCoordinate}
\end{equation}
where 
$\rho$ is the baryon mass density. 
The mass density
obeys the
$O(v)$ conservation law
\begin{equation}
{\partial \rho\over\partial \tilde t} + {1\over r^2}{\partial\over\partial r}
\left(r^2 \rho v\right)=0,\label{baryonConservation}
\end{equation}
where
\begin{equation}
v={dr\over d\tilde t}
\end{equation}
is the radial velocity measured by an Eulerian observer.
We choose the Lagrangian time coordinate such that the metric components are
specified by a line element of the form 
\begin{equation}
ds^2 = -dt^2 + g_{mm}\,dm^2 + r^2\, d\theta^2 + 
r^2\sin^2\theta\,d\phi^2.\label{comovingMetricForm}
\end{equation}
The transformations between the Eulerian and Lagrangian 
coordinates involve eight nontrivial quantities, the elements of
the Jacobian matrices
\begin{eqnarray}
J_{\{\tilde t, r\} \rightarrow \{t, m\}} &=&
\pmatrix{\partial \tilde t\over\partial t & \partial \tilde t\over \partial m \cr
\partial r\over\partial t & \partial r\over\partial m},\\
J_{\{t, m\} \rightarrow \{\tilde t, r\}} &=&
\pmatrix{\partial t\over\partial\tilde t & \partial t\over \partial r \cr
\partial m\over\partial\tilde t & \partial m\over\partial r}.
\end{eqnarray}
Two of these elements, $\partial m/\partial r$ and $\partial m/\partial
\tilde t$, are determined by Eqs. (\ref{massCoordinate}) and
(\ref{baryonConservation}). The relationship between Eqs. 
(\ref{eulerianMetric}) and (\ref{comovingMetricForm})---specifically,
the requirements $g_{tt}=1$ and $g_{tm}=0$---provide two more
equations. The identity $J_{\{\tilde t, r\} \rightarrow \{t, m\}} =
\left(J_{\{t, m\} \rightarrow \{\tilde t, r\}}\right)^{-1}$
provides the remaining four equations, and we find that to $O(v)$
\begin{eqnarray}
\pmatrix{\partial \tilde t\over\partial t & \partial \tilde t\over \partial m \cr
\partial r\over\partial t & \partial r\over\partial m}
&=&
\pmatrix{ 1 & v\over 4\pi r^2 \rho \cr  v & 1\over 4\pi r^2 \rho},
\label{inverseCoordinateTransformation}
\\
\pmatrix{\partial t\over\partial\tilde t & \partial t\over \partial r \cr
\partial m\over\partial\tilde t & \partial m\over\partial r}
&=&
\pmatrix{ 1 & -v \cr -4\pi r^2\rho v & 4\pi r^2 \rho}.
\label{coordinateTransformation}
\end{eqnarray}
From these relations we see that to $O(v)$, the line element in
our Lagrangian coordinates is
\begin{eqnarray}
ds^2& =& g_{\mu\nu}dx^\mu dx^\nu \nonumber\\ 
&=& -dt^2 + \left(1\over 4\pi r^2\rho\right)^2 dm^2 + 
  r^2\, d\theta^2 + r^2\sin^2\theta\,d\phi^2,\label{comovingMetric}
\end{eqnarray}
whose determinant is
\begin{equation}
g = -\left(\sin\theta \over 4\pi\rho\right)^2.\label{comovingDeterminant}
\end{equation}
These metric components can be used to obtain results 
valid to $O(v)$ \cite{castor72}.

Now we consider the equations of motion that determine 
the particle trajectories. The rates of change, along a
particle worldline, of
the coordinate basis position components $\{x^\mu\}=\{t,m,\theta,\phi\}$ 
and comoving
orthornomal basis momentum components $\{p^{\hat\mu}\}=\{p^{\hat 0},
p^{\hat 1},p^{\hat 2},p^{\hat 3}\}$ are given by
Eqs. (\ref{momentum2}) and (\ref{geodesic2}):
\begin{eqnarray}
\dispSFNumberedEquationmath\frac{{{\Mvariable{dx}}^{\mu }}}{\Mvariable{d\lambda }}&=&{{{{\ScriptCapitalL }^{\mu }}\InvisibleSpace }_{\overvar{\mu
}{\RawWedge }}}{p^{\overvar{\mu }{\RawWedge }}},\label{trajectory1}\\
\dispSFNumberedEquationmath\frac{{{\Mvariable{dp}}^{\overvar{\mu }{\RawWedge }}}}{\Mvariable{d\lambda }}&=&-{{{{\Gamma }^{\overvar{\mu }{\RawWedge }}}\InvisibleSpace }_{\overvar{\nu
}{\RawWedge }\overvar{\rho }{\RawWedge }}}{p^{\overvar{\nu }{\RawWedge }}}{p^{\overvar{\rho }{\RawWedge }}}.\label{trajectory2}
\end{eqnarray}
We now consider the quantities 
${{\cal L}^{\mu}}_{\hat\mu}$ and ${{{{\Gamma }^{\overvar{\mu }{\RawWedge }}}\InvisibleSpace }_{\overvar{\nu
}{\RawWedge }\overvar{\rho }{\RawWedge }}}$ appearing in these equations.
As discussed in the previous subsection, 
${{{{\ScriptCapitalL }^{\overvar{\mu }{\RawWedge }}}\InvisibleSpace }_{\mu }}={{{{\Lambda }^{\overvar{\mu }{\RawWedge
}}}\InvisibleSpace }_{\overvar{\mu }{\_}}}{{{e^{\overvar{\mu }{\_}}}\InvisibleSpace }_{\mu }}$ is a composite transformation from the coordinate basis
(denoted by unadorned indices) to an orthonormal comoving basis (denoted
by hatted indices). In the present case the coordinate basis is already
comoving, so that the boost 
${{{\Lambda }^{\overvar{\mu }{\RawWedge
}}}\InvisibleSpace }_{\overvar{\mu }{\_}}$ 
is simply the identity transformation. Under the remaining transformation
${{{e^{\overvar{\mu }{\_}}}\InvisibleSpace }_{\mu }}$ to an orthonormal
basis, the metric of Eq. (\ref{comovingMetric}) must become the Lorentz 
metric. We make an obvious choice and take the non-zero transformation
elements 
${{{{\ScriptCapitalL }^{\overvar{\mu }{\RawWedge }}}\InvisibleSpace }_{\mu }}$
to be
\begin{eqnarray}
{{{{\ScriptCapitalL }^{\overvar{0 }{\RawWedge }}}\InvisibleSpace }_{t }}
&=& 1, \label{transformationFirst} \\
{{{{\ScriptCapitalL }^{\overvar{1 }{\RawWedge }}}\InvisibleSpace }_{m }}
&=& {1\over 4\pi r^2\rho}, \\
{{{{\ScriptCapitalL }^{\overvar{2 }{\RawWedge }}}\InvisibleSpace }_{\theta }}
&=& r, \\
{{{{\ScriptCapitalL }^{\overvar{3 }{\RawWedge }}}\InvisibleSpace }_{\phi }}
&=& r\sin\theta.\label{transformationLast}
\end{eqnarray}
The elements  
${{{{\ScriptCapitalL }^{\mu }}_{\overvar{\mu }{\RawWedge }}}\InvisibleSpace }$
of the inverse transformation are obvious. Turning to the connection coefficients, we employ Eqs. 
(\ref{connectionCoordinate}), (\ref{comovingMetric}), 
and (\ref{inverseCoordinateTransformation}) 
to find the 
non-zero 
coordinate basis connection coefficients
${{{{\Gamma }^{\mu }}\InvisibleSpace }_{\Mvariable{\nu \rho }}}$
to be
\begin{eqnarray}
{{{{\Gamma }^{t}}\InvisibleSpace }_{\Mvariable{m m }}}
&=& -\left(1\over 4\pi r^2\rho\right)^2\left({2 v\over r}+
 {\partial\ln\rho\over\partial t }\right),\label{connectionCoordinateFirst}\\
{{{{\Gamma }^{t}}\InvisibleSpace }_{\Mvariable{\theta\theta }}}
&=& r v,\\
{{{{\Gamma }^{t}}\InvisibleSpace }_{\Mvariable{\phi\phi }}}
&=& r v\sin^2\theta,\\
{{{{\Gamma }^{m}}\InvisibleSpace }_{\Mvariable{t m }}}=
{{{{\Gamma }^{m}}\InvisibleSpace }_{\Mvariable{m t }}}
&=& -\left({2 v\over r}+
 {\partial\ln\rho\over\partial t }\right),\\
{{{{\Gamma }^{m}}\InvisibleSpace }_{\Mvariable{m m }}}
&=& -\left({1\over 2\pi r^3\rho}+{\partial\ln\rho\over\partial m}\right),\\
{{{{\Gamma }^{m}}\InvisibleSpace }_{\Mvariable{\theta\theta}}}
&=& -4\pi r^3\rho,\\
{{{{\Gamma }^{m}}\InvisibleSpace }_{\Mvariable{\phi\phi}}}
&=& -4\pi r^3\rho\sin^2\theta,\\
{{{{\Gamma }^{\theta}}\InvisibleSpace }_{\Mvariable{t \theta }}}=
{{{{\Gamma }^{\theta}}\InvisibleSpace }_{\Mvariable{\theta t }}}=
{{{{\Gamma }^{\phi}}\InvisibleSpace }_{\Mvariable{t \phi }}}=
{{{{\Gamma }^{\phi}}\InvisibleSpace }_{\Mvariable{\phi t }}}
&=& { v\over r},\\
{{{{\Gamma }^{\theta}}\InvisibleSpace }_{\Mvariable{m \theta }}}=
{{{{\Gamma }^{\theta}}\InvisibleSpace }_{\Mvariable{\theta m }}}=
{{{{\Gamma }^{\phi}}\InvisibleSpace }_{\Mvariable{m \phi }}}=
{{{{\Gamma }^{\phi}}\InvisibleSpace }_{\Mvariable{\phi m }}}
&=& { 1\over 4\pi r^3\rho},\\
{{{{\Gamma }^{\theta}}\InvisibleSpace }_{\Mvariable{\phi\phi}}}
&=& -\sin\theta \cos\theta,\\
{{{{\Gamma }^{\phi}}\InvisibleSpace }_{\Mvariable{\theta \phi }}}=
{{{{\Gamma }^{\phi}}\InvisibleSpace }_{\Mvariable{\phi \theta }}}
&=& \cot\theta.\label{connectionCoordinateLast}
\end{eqnarray}
Now we can obtain 
the connection coefficients in the orthonormal 
comoving frame 
by employing Eqs. (\ref{connectionCoefficients}),  
(\ref{transformationFirst})-(\ref{transformationLast})
and (\ref{connectionCoordinateFirst})-(\ref{connectionCoordinateLast}).
We find that the non-zero 
${{{{\Gamma }^{\overvar{\mu }{\RawWedge }}}\InvisibleSpace }_{\overvar{\nu
}{\RawWedge }\overvar{\rho }{\RawWedge }}}$
are
\begin{eqnarray}
{{{{\Gamma }^{\hat 0}}\InvisibleSpace }_{\Mvariable{\hat 1 \hat 1 }}}=
{{{{\Gamma }^{\hat 1}}\InvisibleSpace }_{\Mvariable{\hat 1 \hat 0 }}}
&=& -\left({2 v\over r}+
 {\partial\ln\rho\over\partial t }\right),\label{connectionComovingFirst}\\
{{{{\Gamma }^{\hat 0}}\InvisibleSpace }_{\Mvariable{\hat 2 \hat 2 }}}=
{{{{\Gamma }^{\hat 0}}\InvisibleSpace }_{\Mvariable{\hat 3 \hat 3 }}}=
{{{{\Gamma }^{\hat 2}}\InvisibleSpace }_{\Mvariable{\hat 2 \hat 0 }}}=
{{{{\Gamma }^{\hat 3}}\InvisibleSpace }_{\Mvariable{\hat 3 \hat 0 }}}
&=&{v\over r}, \\
{{{{\Gamma }^{\hat 1}}\InvisibleSpace }_{\Mvariable{\hat 2 \hat 2 }}}=
{{{{\Gamma }^{\hat 1}}\InvisibleSpace }_{\Mvariable{\hat 3 \hat 3 }}}=
-{{{{\Gamma }^{\hat 2}}\InvisibleSpace }_{\Mvariable{\hat 2 \hat 1 }}}
-{{{{\Gamma }^{\hat 3}}\InvisibleSpace }_{\Mvariable{\hat 3 \hat 1 }}}
&=&-{1\over r},\\
{{{{\Gamma }^{\hat 2}}\InvisibleSpace }_{\Mvariable{\hat 3 \hat 3 }}}=
-{{{{\Gamma }^{\hat 3}}\InvisibleSpace }_{\Mvariable{\hat 3 \hat 2 }}}
&=&-{\cot\theta\over r}. \label{connectionComovingLast}
\end{eqnarray}
Note that unlike the
${{{{\Gamma }^{\mu }}\InvisibleSpace }_{\Mvariable{\nu \rho }}}$,
the
${{{{\Gamma }^{\overvar{\mu }{\RawWedge }}}\InvisibleSpace }_{\overvar{\nu
}{\RawWedge }\overvar{\rho }{\RawWedge }}}$
are not symmetric in their lower indices
${\overvar{\nu
}{\RawWedge }\overvar{\rho }{\RawWedge }}$
because the comoving orthonormal basis is not a coordinate basis.
With ${{\cal L}^{\mu}}_{\hat\mu}$ and ${{{{\Gamma }^{\overvar{\mu }{\RawWedge }}}\InvisibleSpace }_{\overvar{\nu
}{\RawWedge }\overvar{\rho }{\RawWedge }}}$ now explicitly specified, 
our consideration of the particle equations of motion is complete.

The next topic to consider from the previous subsection is the
Liouville vector. This vector generates the phase flow---the 
set of trajectories 
$\left(x(\lambda),p(\lambda)\right)$
obtained from the particle equations
of motion, Eqs. (\ref{trajectory1}) and (\ref{trajectory2}). These
are trajectories through the 8-dimensional manifold comprised of position
vectors $x$ and momentum vectors $p$. But a 7-dimensional submanifold,
the phase space for particles of definite mass, is of more immediate
interest: The physical fact that the particles have
definite mass---as embodied in the ``mass shell'' constraint of Eq.  
(\ref{massShell})---implies that there are only three independent
momentum variables. We choose these to be $\{u^{\hat i}\}=\{\epsilon,\mu,
\varphi\}$, related to the comoving orthonormal basis 
momentum components $\{p^{\hat i}\}$ by
\begin{eqnarray}
{p^{\overvar{1}{\RawWedge }}}&=&\epsilon\, \multsp \mu,\label{momentum1}  \\
{p^{\overvar{2}{\RawWedge }}}&=&\epsilon\, \multsp \sqrt{1-\mu^2}\, \Mvariable{\cos}\varphi ,\IndentingNewLine  
\\
{p^{\overvar{3}{\RawWedge }}}&=&\epsilon\, \multsp \sqrt{1-\mu^2}\, \Mvariable{\sin}\varphi .\label{momentum3}
\end{eqnarray}
In the present case of massless particles, the mass shell condition
gives 
\begin{equation}
p^{\hat 0}=\epsilon. \label{momentum0}
\end{equation}
From these expressions one can compute the Jacobian $(\partial p^{\hat i} /
\partial u^{\hat j})$ and its inverse $(\partial u^{\hat i} /
\partial p^{\hat j})$; this inverse is 
\begin{equation}
\pmatrix{
\partial \epsilon\over\partial p^{\hat 1} & \partial\epsilon
\over \partial p^{\hat 2} & \partial\epsilon\over \partial p^{\hat 3} \cr
\partial \mu\over\partial p^{\hat 1} & \partial\mu
\over \partial p^{\hat 2} & \partial\mu\over \partial p^{\hat 3} \cr
\partial \varphi\over\partial p^{\hat 1} & \partial\varphi
\over \partial p^{\hat 2} & \partial\varphi\over \partial p^{\hat 3}
}
=
\pmatrix{
\mu &\sqrt{1-\mu^2}\,\cos\varphi  & \sqrt{1-\mu^2}\,\sin\varphi \cr
1-\mu^2\over \epsilon & -\mu \sqrt{1-\mu^2}\,\cos\varphi \over \epsilon & 
 -\mu \sqrt{1-\mu^2}\,\sin\varphi \over \epsilon \cr
0 & -\sin\varphi\over\epsilon\sqrt{1-\mu^2} &
\cos\varphi\over\epsilon\sqrt{1-\mu^2} 
}.\label{momentumTransformation}
\end{equation}
We are now ready to write down the Liouville vector on the 
phase space for particles of definite mass, obtained with the help of 
Eqs. (\ref{liouvilleFinal}), 
(\ref{transformationFirst})-(\ref{transformationLast}), 
(\ref{connectionComovingFirst})-(\ref{connectionComovingLast}), and
(\ref{momentum1})-(\ref{momentumTransformation}):
\begin{eqnarray}
L_m& =& \epsilon {\partial \over \partial t} + 4\pi r^2\rho \epsilon\mu
{\partial\over\partial m} + 
\epsilon^2\left[\mu^2\left({3 v\over r} +
{\partial\ln\rho\over\partial t}\right)-{v\over r}\right]
{\partial\over\partial\epsilon} +\nonumber \\
& &\epsilon(1-\mu^2)\left[{1\over r}+\mu\left({3 v\over r}+
{\partial\ln\rho\over\partial t}\right)
\right]{\partial\over\partial\mu}+\ldots,\label{comovingLiouville}
\end{eqnarray}
where ``\ldots'' represents $\partial/\partial\theta$, $\partial/\partial\phi$,
and $\partial/\partial\varphi$ terms whose explicit form we will not need
in this spherically symmetric case.

Finally we consider volume and ``surface'' elements in phase space.
From Eqs. (\ref{spacetimeElement}) and (\ref{comovingDeterminant}),
the spacetime volume element is
\begin{equation}
\eta = {\sin\theta \over 4\pi\rho}\, dt\wedge dm\wedge d\theta\wedge d\phi.
\label{sphericalEta}
\end{equation} 
Contraction of $\eta$ with the coordinate basis vectors 
$\partial/\partial t$, $\partial/\partial m$, $\partial/\partial \theta$, 
and $\partial/\partial \phi$
produces (see Eq. (\ref{spacetimeSurface})) the spacetime surface elements
\begin{eqnarray}
\sigma_t &=& {\sin\theta \over 4\pi\rho}\, dm\wedge d\theta\wedge d\phi, \\
\sigma_m &=& - {\sin\theta \over 4\pi\rho}\, dt\wedge d\theta\wedge d\phi, \\
\sigma_\theta &=& {\sin\theta \over 4\pi\rho}\, dt\wedge dm\wedge d\phi, \\
\sigma_\phi &=& - {\sin\theta \over 4\pi\rho}\, dt\wedge dm\wedge d\theta.
\end{eqnarray}
Turning to momentum space, from Eqs. (\ref{momentumElement}) and (\ref{momentum1})-(\ref{momentum0})
the invariant volume element in momentum space is found to be
\begin{equation}
\pi_m = \epsilon\,d\epsilon\wedge d\mu\wedge d\varphi.
\label{sphericalPi}
\end{equation}
We form the volume element on phase space, using Eqs. (\ref{volume}),
(\ref{sphericalEta}), and (\ref{sphericalPi}):
\begin{eqnarray}
\Omega &=& \eta \wedge \pi_m \\
&= & {\sin\theta\; \epsilon\over 4\pi \rho}\, 
dt\wedge dm\wedge d\theta\wedge d\phi\wedge d\epsilon\wedge d\mu\wedge 
d\varphi.
\label{comovingVolume}
\end{eqnarray}
With the help of Eqs. (\ref{comovingLiouville}) and (\ref{comovingVolume}),
we also construct the surface element in phase space that is normal to 
the phase flow,
the 6-form $\omega$ of Eq. (\ref{surface}):
\begin{eqnarray}
\omega &=&  L_m \cdot {\Omega} 
\\
&=& {\sin\theta\;\epsilon\over 4\pi\rho} 
\left[ (L_m)^t\, dm\wedge d\theta\wedge d\phi\wedge d\epsilon\wedge d\mu 
       \wedge d\varphi -
       (L_m)^m\, dt\wedge d\theta\wedge d\phi\wedge d\epsilon\wedge d\mu 
       \wedge d\varphi + \right. \nonumber
\\
& & \left.      (L_m)^\epsilon\, dt\wedge d\theta\wedge d\phi\wedge 
       dm\wedge d\mu \wedge d\varphi -
       (L_m)^\mu\, dt\wedge dm\wedge d\theta\wedge d\phi\wedge d\epsilon
       \wedge d\varphi + \ldots
\right] \\
&=& {\sin\theta\;\epsilon^2\over 4\pi\rho}  \, 
dm\wedge d\theta\wedge d\phi\wedge d\epsilon\wedge d\mu 
       \wedge d\varphi - 
      r^2\sin\theta\;\epsilon^2\mu\, 
    dt\wedge d\theta\wedge d\phi\wedge d\epsilon\wedge d\mu 
       \wedge d\varphi +\nonumber
\\
  & &   {\sin\theta\;\epsilon^3\over 4\pi\rho}\left[\mu^2\left({3 v\over r} +
{\partial\ln\rho\over\partial t}\right)-{v\over r}\right] \,
dt\wedge d\theta\wedge d\phi\wedge 
       dm\wedge d\mu \wedge d\varphi -\nonumber 
\\
& &     {\sin\theta\;\epsilon^2(1-\mu^2)\over 4\pi\rho}
\left[{1\over r}+\mu\left({3 v\over r}+
{\partial\ln\rho\over\partial t}\right)
\right]\, dt\wedge dm\wedge d\theta\wedge d\phi\wedge d\epsilon
       \wedge d\varphi + \ldots,
\label{surfaceSpecific}
\end{eqnarray} 
where ``\ldots'' represents terms arising from the
$\partial/\partial\theta$, $\partial/\partial\phi$,
and $\partial/\partial\varphi$ terms of the Liouville vector.  
The validity of Eq. (\ref{domega}) can be verified (see Eq. 
(\ref{exteriorDerivative}) for the definition of the exterior
derivative):
\begin{eqnarray}
d\omega &=&\left(
{\partial\over\partial t}\left({\sin\theta\;\epsilon^2\over 4\pi\rho}\right) +
{\partial\over\partial m}\left(r^2\sin\theta\;\epsilon^2\mu \right) + \right.
\nonumber
\\
& &\left. {\partial\over\partial \epsilon}\left\{ 
  {\sin\theta\;\epsilon^3\over 4\pi\rho}\left[\mu^2\left({3 v\over r} +
{\partial\ln\rho\over\partial t}\right)-{v\over r}\right]
\right\} +\right. \nonumber
\\
& &\left. {\partial\over\partial\mu}\left\{
  {\sin\theta\;\epsilon^2(1-\mu^2)\over 4\pi\rho}
\left[{1\over r}+\mu\left({3 v\over r}+
{\partial\ln\rho\over\partial t}\right)
\right]+\ldots
\right\}
\right)\times \nonumber
\\
& & dt\wedge dm\wedge d\theta\wedge d\phi\wedge d\epsilon\wedge d\mu\wedge 
d\varphi  \label{domegaSpherical}
\\
&=&0,
\end{eqnarray}
where the expression for $\partial r/\partial m$ is taken from
Eq. (\ref{inverseCoordinateTransformation}). The terms explicitly
displayed in Eq. (\ref{domegaSpherical}) do in fact sum to zero by 
themselves, and it can be shown 
that those represented by ``\ldots'' do as well.

\section{The distribution function, Boltzmann equation, and balance equations}
\label{sec:distribution}

In this section---again following Ehlers \cite{ehlers71} for the 
general case---we define the distribution function $f$, derive the
Boltzmann equation, and present the equations 
describing the balance of particle number and 4-momentum. 
In the general case we reproduce Ehlers' derivation 
of the momentum-integrated
number balance equation, and explain why it does not yield a conservative
reformulation of the Boltzmann equation.
We also present the balance equation for 4-momentum without
derivation. In the special case of
Lagrangian coordinates in spherical symmetry to $O(v)$, 
we verify part of the general derivation of the Boltzmann
equation by explicit calculation. We also display the
Boltzmann equation and the momentum-integrated number and
4-momentum balance equations pertaining to this special case, including
the transformation to the lab frame needed to obtain a 
``conserved'' energy.

\subsection{The distribution function, Boltzmann equation, and balance equations: the general case}

The {\itshape distribution function} $f$ for a particle of a given type represents the density of 
particles
in phase space. The particle type's
7-dimensional phase space for particles of definite mass, \inlineSFinmath${{M_m}}$, is filled with trajectories 
$\left(x(\lambda),p(\lambda)\right)$,
or ``states''. As a collection of particles evolves, the number of particles in each state changes due to collisions. If
one considers a 6-dimensional hypersurface \inlineSFinmath${\Sigma }$ in \inlineSFinmath${{M_m}}$, the ensemble-averaged number \inlineSFinmath${\Mfunction{N}[\Sigma
]}$ of occupied states crossing \inlineSFinmath${\Sigma }$ is 
\begin{equation}
\dispSFNumberedEquationmath{\Mfunction{N}[\Sigma ]={\int_\Sigma }f\multsp \omega ,}\label{worldlines}
\end{equation}
where the surface element $\omega$ is given by Eq. (\ref{surface}). 
(Hypersurfaces like $\Sigma$ are oriented, and the particle trajectories have a direction associated with them as well. The crossing of a hypersurface
by an occupied trajectory can give a positive or negative contribution to
Eq. (\ref{worldlines}), depending on their relative orientation.)
With this definition, the distribution function \inlineSFinmath${f(x,p)}$ is a scalar \cite{ehlers71}.

An equation governing the evolution of the distribution function is obtained by considering a closed 6-dimensional hypersurface \inlineSFinmath${\partial
D}$ bounding a region \inlineSFinmath${D}$ in \inlineSFinmath${{M_m}}$. The net number of occupied states emerging from \inlineSFinmath${D}$ is, from Eq.
(\ref{worldlines}) and the generalized Stokes' theorem,
\begin{equation}
\dispSFNumberedEquationmath{\Mfunction{N}[\partial D]=\int_{\partial D} f\multsp \omega =\int_D d(f\multsp \omega ).}\label{stokes}
\end{equation}
(We note in passing---and discuss in more detail in Sec. \ref{sec:conservative}---that
this expression, which relates a volume integral to a surface integral,
is key to obtaining conservative formulations of kinetic theory.)
Using the ``product rule'' of Eq. (\ref{exteriorProduct}), 
the vanishing exterior derivative of \inlineSFinmath${\omega }$ (Eq. (\ref{domega})), and the definition of \inlineSFinmath${\omega }$ as the contraction
of the Liouville vector \inlineSFinmath${{L_m}}$ with the volume element \inlineSFinmath${\Omega }$ (Eq. (\ref{surface})), we have
\begin{equation}
\dispSFNumberedEquationmath{d(f\multsp \omega )=\Mvariable{df}\wedge \omega =\Mvariable{df}\wedge ({L_m}\cdot \Omega ).}\label{almostLiouville}
\end{equation}
Any scalar function \inlineSFinmath${f}$,  vector field \inlineSFinmath${L}$, and \inlineSFinmath${n}$-form field \inlineSFinmath${\Omega }$ on an \inlineSFinmath${n}$-dimensional
manifold obey the identity \cite{ehlers71} 
\begin{equation}
\dispSFNumberedEquationmath{\Mvariable{df}\wedge (L\cdot \Omega )=L[f]\Omega .}
\label{almostLiouville2}
\end{equation}
Hence the net number of occupied states emerging from \inlineSFinmath${D}$ is
\begin{equation}
\dispSFNumberedEquationmath{\Mfunction{N}[\partial D]=\int_D {L_m}[f]\Omega .}
\label{almostBoltzmann}
\end{equation}
(Equations (\ref{stokes})-(\ref{almostBoltzmann}) amount to the Liouville
theorem in a relativistic context.)
The domain $D$ in phase space consists of a region $H$ in spacetime,
together with regions $K_x$ in the momentum space (the mass shell in 
the tangent space) at each spacetime point $x$. Recall also that the
volume element $\Omega$ in phase space is the product of the spacetime
and momentum space volume elements, $\Omega = \eta\wedge\pi_m$.
Hence 
the integral over $D$ in Eq. (\ref{almostBoltzmann}) 
can be expressed as the iterated integral
\begin{equation}
\dispSFNumberedEquationmath{\Mfunction{N}[\partial D]=\int_H \eta \Bigg(\int_{K_x}{L_m}[f]{{\pi }_m}\Bigg).}\label{boltzmannLeft}
\end{equation}
Because $\partial D$ is a closed surface, 
the net number of 
particles in trajectories
emerging from \inlineSFinmath${D}$ must equal the net number of collisions in \inlineSFinmath${D}$. If correlations
between particles can be neglected, the spacetime density of collisions can be expressed in terms of a {\itshape collision integral }\inlineSFinmath${\mathbb{C}[f]}$
that depends only on one-particle distribution functions. Therefore, if 
$\int_{K_x}\mathbb{C}[f]{{\pi }_m}$ denotes the spacetime density of 
collisions, we have
\begin{equation}
\dispSFNumberedEquationmath{\Mfunction{N}[\partial D]=\int_H \eta \Bigg(\int_{K_x}\mathbb{C}[f]{{\pi }_m}\Bigg).}\label{boltzmannRight}
\end{equation}
Because the regions $H$ and $K_x$ are arbitrary, comparison of
Eqs. (\ref{boltzmannLeft}) and (\ref{boltzmannRight}) shows that
the evolution of \inlineSFinmath${f}$ is determined by 
\begin{equation}
\dispSFNumberedEquationmath{{L_m}[f]=\mathbb{C}[f],}\label{formalBoltzmann}
\end{equation}
or, using Eq. (\ref{liouvilleFinal}),
\begin{equation}
\dispSFNumberedEquationmath{{p^{\overvar{\mu }{\RawWedge }}}{{{{\ScriptCapitalL }^{\mu }}\InvisibleSpace }_{\overvar{\mu }{\RawWedge }}}\frac{\partial
f}{\partial {x^{\mu }}}+\big(e\multsp {{{F^{\overvar{j}{\RawWedge }}}\InvisibleSpace }_{\overvar{\nu }{\RawWedge }}}{p^{\overvar{\nu }{\RawWedge
}}}-{{{{\Gamma }^{\overvar{j}{\RawWedge }}}\InvisibleSpace }_{\overvar{\nu }{\RawWedge }\overvar{\rho }{\RawWedge }}}{p^{\overvar{\nu }{\RawWedge
}}}{p^{\overvar{\rho }{\RawWedge }}}\big)\frac{\partial {u^{\overvar{i}{\RawWedge }}}}{\partial {p^{\overvar{j}{\RawWedge }}}}\frac{\partial f}{\partial
{u^{\overvar{i}{\RawWedge }}}}=\mathbb{C}[f].}\label{fullBoltzmann}
\end{equation}
This is the Boltzmann equation.

Next we consider the particle number 4-current, electromagnetic 4-current, and stress-energy tensor, 
and present 
the balance equations they satisfy. 
In Eq. (\ref{worldlines}), specialize the hypersurface of \inlineSFinmath${{M_m}}$ to be \inlineSFinmath${\Sigma =G\times {P_m}(x)}$, where
\inlineSFinmath${G}$ is an infinitesimal spacelike hypersurface in spacetime at point \inlineSFinmath${x}$, and \inlineSFinmath${{P_m}(x)}$ is the 
entire momentum space at point \inlineSFinmath${x}$ \footnote{Other 6-dimensional hypersurfaces can be formed, which contain a 
4-dimensional region in spacetime and a 2-dimensional surface in momentum
space or a timelike region in spacetime together with a 3-dimensional region
in momentum space; such hypersurfaces are not of use in defining the particle number 4-current and stress-energy tensor.}.
As explained in the endnote to the discussion following 
Eq. (\ref{surfaceFirst}), on such a hypersurface the only relevant term of \inlineSFinmath${\omega }$ 
is its first term in Eq. (\ref{surface})
(given also by Eq. (\ref{surfaceFirst})),
so that Eq. (\ref{worldlines}) becomes
\begin{equation}
\dispSFNumberedEquationmath{N[G\times P_m(x)]={\int_G }{{\sigma }_{\mu }}\Bigg(\int_{P_m(x)}{{{f {\ScriptCapitalL }^{\mu
}}\InvisibleSpace }_{\overvar{\mu }{\RawWedge }}}{p^{\overvar{\mu }{\RawWedge }}}\multsp \multsp {{\pi }_m}\Bigg).}\label{worldlines2}
\end{equation}
This is the number of particles whose worldlines cross \inlineSFinmath${G}$ (at \inlineSFinmath${x}$); hence the 4-vector
\begin{equation}
\dispSFNumberedEquationmath{{N^{\mu }(x)}=\int_{P_m(x)}{{{{f \ScriptCapitalL }^{\mu }}\InvisibleSpace }_{\overvar{\mu
}{\RawWedge }}}{p^{\overvar{\mu }{\RawWedge }}}\multsp \multsp {{\pi }_m}=\int_{P_m(x)}f{p^{\mu }}\multsp \multsp {{\pi }_m}} \label{numberVector}
\end{equation}
is the particle 4-current density. 
The electromagnetic 4-current is obtained by multiplying by the electric charge: \inlineSFinmath${{J^{\mu }}=e\multsp
{N^{\mu }}}$. The particle 4-current satisfies the 
balance equation
\begin{equation}
\dispSFNumberedEquationmath{\frac{1}{{\sqrt{-g}}}\frac{\partial }{\partial {x^{\mu }}}\big({\sqrt{-g}}{N^{\mu }}\big)=\int \mathbb{C}[f]{{\pi
}_m},}\label{numberConservation}
\end{equation}
and similarly for the electromagnetic 4-current. (Of course the collisions will be such that the sum of the divergences of the electromagnetic 4-currents
of all particle species will vanish.)
 The stress-energy tensor is
\begin{equation}
\dispSFNumberedEquationmath{{T^{\Mvariable{\mu \nu }}}=
\int_{P_m(x)}{{{{f \ScriptCapitalL }^{\mu }}\InvisibleSpace }_{\overvar{\mu
}{\RawWedge }}}
{{{ \ScriptCapitalL }^{\nu }}\InvisibleSpace }_{\overvar{\nu
}{\RawWedge }}
{p^{\overvar{\mu }{\RawWedge }}}{p^{\overvar{\nu }{\RawWedge }}}\multsp \multsp {{\pi }_m} =
\int_{P_m(x)}f{p^{\mu }}\multsp {p^{\nu }}\multsp {{\pi }_m},}
\label{stressEnergy}
\end{equation}
which obeys the 
balance equation
\begin{equation}
\dispSFNumberedEquationmath{\frac{1}{{\sqrt{-g}}}\frac{\partial }{\partial {x^{\mu }}}\big({\sqrt{-g}}{T^{\Mvariable{\nu \mu }}}\big)={{{F^{\nu }}\InvisibleSpace }_{\mu }}{J^{\mu }}
-{{{{\Gamma
}^{\nu }}\InvisibleSpace }_{\Mvariable{\rho \mu }}}{T^{\Mvariable{\rho \mu }}}
+\int \mathbb{C}[f]
{{{ \ScriptCapitalL }^{\nu }}\InvisibleSpace }_{\overvar{\nu
}{\RawWedge }}{p^{\hat\nu
}}{{\pi }_m}.}\label{energyConservation}
\end{equation}

To conclude this subsection, we reproduce Ehlers' derivation 
of Eq. (\ref{numberConservation}) for 
number balance, and explain why this proof does not yield a conservative
reformulation of the Boltzmann equation. The proof involves
forming an integral over a suitable hypersurface in phase space, evaluating
that integral in two different ways, and comparing the results.

First we specialize the integral in Eq. (\ref{worldlines}) 
to a specific hypersurface of integration.
Consider an arbitrary region in spacetime, $\hat D$, whose boundary is
$\partial\hat D$. 
Form a hypersurface $\partial D$ in the seven-dimensional phase space
$M_m$,
composed of the boundary region $\partial\hat D$ in spacetime 
together with the entire momentum space $P_m(x)$ at each 
point of $\partial\hat D$. Equation (\ref{worldlines}) becomes
\begin{equation}
{N}[\partial D]=\int_{\partial D} f\multsp \omega,
\label{numberProof1}
\end{equation}
the integral we will evaluate in two different ways.

On the one hand, as derived in Eqs. (\ref{stokes})-(\ref{almostBoltzmann}), 
the integral in Eq. (\ref{numberProof1}) can be written
\begin{equation}
{N}[\partial D]=\int_{D} L_m[f]\Omega.
\end{equation}
With the Boltzmann equation, Eq. (\ref{formalBoltzmann}),
\begin{equation}
{N}[\partial D]=\int_{D} \mathbb{C}[f]\Omega 
=\int_{\hat D}\eta\left(\int_{P_m(x)}\mathbb{C}[f]\pi_m \right).
\label{numberProof15}
\end{equation}

On the other hand, the integral in Eq. (\ref{numberProof1}) 
can be evaluated directly. As a first step, integrate
over momentum space at each point of $\partial\hat D$. 
As discussed in Subsec. \ref{subsec:phaseSpaceSpecific},
in such integrals over 3-dimensional regions in momentum space 
only the first term of $\omega$ in 
Eq. (\ref{surface}) contributes (see in particular the endnote
in the sentence following Eq. (\ref{surfaceFirst})). 
Using Eq. (\ref{surfaceFirst}) to express the first term of $\omega$,
Eq. (\ref{numberProof1}) can be expressed as 
\begin{eqnarray}
{N}[\partial D]&=&\int_{\partial\hat D}\sigma_\mu 
\left(\int_{P_m(x)}{{\cal L}^\mu}_{\hat\mu} p^{\hat\mu} f\multsp\, \pi_m\right)
\\
&=& \int_{\partial\hat D}\sigma_\mu N^\mu(x),
\label{numberProof2}
\end{eqnarray}
where the definition (Eq. (\ref{numberVector})) of the particle number current 
has been employed in the second step. Equation (\ref{numberProof2}) is a 
closed surface integral in spacetime; by the divergence theorem (a special
case of the generalized Stokes theorem), it can be converted into a
spacetime volume integral:
\begin{eqnarray}
{N}[\partial D]&=&\int_{\hat D}\eta\; {N^\mu}_{;\mu} \\
&=&\int_{\hat D}\eta \;\frac{1}{{\sqrt{-g}}}\frac{\partial }{\partial {x^{\mu }}}\big({\sqrt{-g}}{N^{\mu }}\big).\label{numberProof3}
\end{eqnarray} 
Because the spacetime region $\hat D$ is arbitrary, 
Eq. (\ref{numberConservation}) follows from comparison of Eqs. 
(\ref{numberProof15}) and (\ref{numberProof3}).

Now we can see why this type of proof does not yield a
conservative reformulation of the Boltzmann equation. In particular,
we can see why no direct insight is gained into the fate of
the non-conservative momentum derivative terms in the Boltzmann equation 
(Eq. (\ref{fullBoltzmann})) upon
integration over momentum space. The left-hand side of 
Eq. (\ref{fullBoltzmann}) arises from the action of the Liouville vector
on $f$. Similarly, the phase space surface element
$\omega$ of Eq. (\ref{surface}) is given by the contraction of the 
Liouville vector with the phase space volume element $\Omega$. Inspection
shows that the second term of Eq. (\ref{surface}) is closely related to
the momentum derivative terms in the Boltzmann equation. But in the 
above derivation of the number balance equation,
an expression for ${{N^{\mu }}}$ is obtained at an early stage
by integration over momentum space; this integration kills 
these terms of $\omega$,
so that they are no longer present by the time the spacetime divergence
theorem is applied. The power and elegance of exterior
calculus has hidden the algebraic details of the connection between the
Boltzmann equation and the number balance equation!

This is remedied in 
Section \ref{sec:conservative}. Instead of choosing a specialized 
hypersurface involving all of momentum space and evaluating the surface
integral of the left-hand side of Eq. (\ref{stokes}), the path to
conservative reformulations of the Boltzmann equations involves 
analysis of the volume integral on the right-hand side of Eq. (\ref{stokes});
for here {\em all} the terms of $\omega$ have been brought inside the
exterior derivative, including the ones related to the momentum
derivative terms appearing in the Boltzmann equation.

\subsection{The distribution function, Boltzmann equation, and balance equations: Lagrangian coordinates in spherical symmetry to $O(v)$}

The first result to specialize from the previous subsection is the 
Boltzmann equation. An aspect of the derivation that can be demonstrated
by explicit calculation is the assertion 
\begin{equation}
df\wedge\omega = L_m[f]\Omega
\end{equation}
contained in Eqs. (\ref{almostLiouville})-(\ref{almostLiouville2}). 
In spherical symmetry $f=f(t,m,\epsilon,\mu)$, so that its gradient is
\begin{equation}
 d f = {\partial f\over\partial t}dt +{\partial f\over\partial m}dm
+{\partial f\over\partial\epsilon}d\epsilon + {\partial f\over\partial\mu}d\mu.
\end{equation}
Forming the wedge product
with $\omega$ of Eq. (\ref{surfaceSpecific}) results in
\begin{eqnarray}
 d f\wedge\omega
&=&\left\{ {\sin\theta\;\epsilon^2\over 4\pi\rho}\,
{\partial f\over\partial t} +
      r^2 \sin\theta\;\epsilon^2\mu\, {\partial f\over\partial m} +
{\sin\theta\;\epsilon^3\over 4\pi\rho}\left[\mu^2\left({3 v\over r} +
{\partial\ln\rho\over\partial t}\right)-{v\over r}\right] \,
{\partial f\over\partial\epsilon} +\right.\nonumber 
\\
& &     
\left. {\sin\theta\;\epsilon^2(1-\mu^2)\over 4\pi\rho}
\left[{1\over r}+\mu\left({3 v\over r}+
{\partial\ln\rho\over\partial t}\right)
\right]\,{\partial f\over\partial\mu}\right\} \times\nonumber
\\
& &
dt\wedge dm\wedge d\theta\wedge d\phi\wedge d\epsilon\wedge d\mu\wedge 
d\varphi,
\end{eqnarray}
where all other terms vanish because the the wedge product of a 1-form 
with itself vanishes due to antisymmetry. Comparison with Eqs. 
(\ref{comovingLiouville})  and (\ref{comovingVolume}) for $L_m$
and $\Omega$ then shows that
$ d f\wedge\omega =  L_m[f]\Omega$, as was to
be demonstrated. The Boltzmann equation, Eq. (\ref{fullBoltzmann}),
specializes to
\begin{eqnarray}
 \epsilon {\partial f\over \partial t} + 4\pi r^2\rho \epsilon\mu
{\partial f\over\partial m} + 
\epsilon^2\left[\mu^2\left({3 v\over r} +
{\partial\ln\rho\over\partial t}\right)-{v\over r}\right]
{\partial f\over\partial\epsilon} + & &\nonumber \\
\epsilon(1-\mu^2)\left[{1\over r}+\mu\left({3 v\over r}+
{\partial\ln\rho\over\partial t}\right)
\right]{\partial f\over\partial\mu}&=&\mathbb{C}[f].\label{comovingBoltzmann}
\end{eqnarray}
This agrees with Eq. (20) of Ref. \cite{castor72}.

Next we specialize Eq. (\ref{numberConservation}) for particle number
balance. Equations (\ref{numberVector}), 
(\ref{transformationFirst})-(\ref{transformationLast}), 
(\ref{momentum1})-(\ref{momentum0}), and (\ref{sphericalPi}),
together with the fact that $f=f(t,m,\epsilon,\mu)$ in spherical symmetry,
imply that 
\begin{eqnarray}
N^t &=& \int f\; 2\pi\,\epsilon^2\, d\epsilon \, d\mu \equiv \rho H^N,
\\
N^m &=& 4\pi r^2\rho \int f\;2\pi \epsilon^2 \mu\, d\mu \equiv
4\pi r^2\rho^2 G^N
\end{eqnarray} 
are the only nonvanishing components of the
particle number vector. With these expressions and Eqs. 
(\ref{comovingDeterminant}) and (\ref{sphericalPi}), 
Eq. (\ref{numberConservation}) becomes
\begin{equation}
{\partial H^N \over\partial t} +
{\partial\over\partial m}\left(4\pi r^2\rho G^N \right) =
{1\over \rho}\int \mathbb{C}[f]\; \epsilon \, d\epsilon \, d\mu \, d\varphi.
\label{numberComoving}
\end{equation}
This agrees with Eq. (28) (with lapse function $\alpha$ set to 1 for
flat spacetime) of Ref. \cite{liebendoerfer01}.

Finally we specialize Eq. (\ref{energyConservation}) for particle number
4-momentum balance. Equations (\ref{numberVector}), 
(\ref{transformationFirst})-(\ref{transformationLast}), 
(\ref{momentum1})-(\ref{momentum0}), and (\ref{sphericalPi}),
together with the fact that $f=f(t,m,\epsilon,\mu)$ in spherical symmetry,
imply that 
\begin{eqnarray}
T^{tt} &=& \int f\; 2\pi\,\epsilon^3\, d\epsilon \, d\mu \equiv \rho H^E,
\\
T^{tm} = T^{mt} &=& 4\pi r^2\rho \int f\;2\pi \epsilon^3 \mu\, d\mu \equiv
4\pi r^2\rho^2 G^E, 
\\
T^{mm} &=& (4\pi r^2\rho)^2 \int f\;2\pi \epsilon^3 \mu^2\, d\mu
\equiv (4\pi r^2\rho)^2 \rho P^E,
\\ 
T^{\theta\theta} &=& {1\over r^2} \int f\;\pi \epsilon^3 (1-\mu^2)\, d\mu
\equiv {\rho\over 2 r^2}(H^E - P^E), 
\\ 
T^{\phi\phi} &=& {1\over r^2\sin^2\theta} 
\int f\;\pi \epsilon^3 (1-\mu^2)\, d\mu
\equiv {\rho\over 2 r^2\sin^2\theta}(H^E - P^E) 
\end{eqnarray} 
are the only nonvanishing components of the
particle stress-energy tensor. With these expressions and Eqs. 
(\ref{comovingDeterminant}),
(\ref{connectionCoordinateFirst})-(\ref{connectionCoordinateLast}), 
(\ref{transformationFirst})-(\ref{transformationLast}), 
(\ref{momentum1})-(\ref{momentum0}), and (\ref{sphericalPi}), 
the $t$ and $m$ components of Eq. (\ref{energyConservation}) become
\begin{eqnarray}
{\partial H^E \over\partial t} +
{\partial\over\partial m}\left(4\pi r^2\rho G^E \right) +
{v\over r} (H^E - P^E) &-&\left({2v\over r}
+{\partial\ln\rho \over \partial t}\right) P^E
\nonumber\\
&=& {1\over\rho}
\int \mathbb{C}[f]\; \epsilon^2 \, d\epsilon \, d\mu \, d\varphi,
\label{energyComoving}
\\
{1\over r^2 \rho}{\partial  \over\partial t}(r^2 \rho G^E) + {1\over 4\pi r^2\rho}
{\partial\over\partial m}\left((4\pi r^2\rho)^2 P^E \right) -
{1\over r} (H^E - P^E) &-& 2 
\left({2v\over r}+{\partial\ln\rho \over \partial t}\right) G^E 
- 
\nonumber\\ 
(4\pi r^2\rho)
\left({1\over 2\pi r^3\rho}+{\partial\ln\rho \over \partial m}\right) P^E
&=& {1\over\rho}
\int \mathbb{C}[f]\; \epsilon^2\mu \, d\epsilon \, d\mu \, d\varphi.
\end{eqnarray}
With the help of Eq. (\ref{inverseCoordinateTransformation}), the
$m$ component equation can be expressed
\begin{eqnarray}
{\partial G^E \over\partial t} +
{\partial\over\partial m}\left(4\pi r^2\rho P^E \right) -
{1\over r} (H^E - P^E) &-&  
\left({2v\over r}+{\partial\ln\rho \over \partial t}\right) G^E 
\nonumber\\ 
&=& {1\over\rho}
\int \mathbb{C}[f]\; \epsilon^2\mu \, d\epsilon \, d\mu \, d\varphi.
\label{momentumComoving}
\end{eqnarray}
Equations (\ref{energyComoving}) and (\ref{momentumComoving}) 
agree with the expressions following Eq. (31) of 
Ref. \cite{liebendoerfer01}. As expected from the discussion in 
Sec. \ref{sec:introduction}, these 4-momentum balance equations
have non-conservative terms arising from the non-vanishing connection
coefficients of Eqs. 
(\ref{connectionCoordinateFirst})-(\ref{connectionCoordinateLast}).

It was also noted in Sec. \ref{sec:introduction} that in Eulerian
coordinates in flat spacetime, the connection coefficients 
${\Gamma^{\tilde t}}_{\mu\nu}$ vanish, so that a conserved energy
can be defined. Equation (\ref{energyConservation}) for particle
4-momentum balance is a vector equation; the $\tilde t$ component
of a vector $V$ 
is obtained from the $t$ and $m$ components by the transformation
\begin{equation}
V^{\tilde t} = {\partial\tilde t \over \partial t}\, V^t +
{\partial t\over \partial m}\, V^m,
\label{vectorTransformation}
\end{equation}
where the transformation coefficients are given by
Eq. (\ref{inverseCoordinateTransformation}).  Equation (\ref{energyComoving})
is actually $\rho^{-1}$ times the $t$ component of 4-momentum balance,
and Eq. (\ref{momentumComoving}) is $(4\pi r^2\rho^2)^{-1}$ times the
$m$ component. These 
factors must be restored before plugging into Eq. (\ref{vectorTransformation});
the final result for the $\tilde t$ component of 4-momentum balance is
\begin{eqnarray}
{\partial \over \partial t}\left(H^E + v G^E\right) + 
{\partial \over \partial m}\left[4\pi r^2\rho\left(G^E+v P^E\right)\right] &-&
{1\over r^2\rho}{\partial\over\partial t}\left(r^2 \rho v\right) G^E
\nonumber \\
&=&{1\over\rho}
\int \mathbb{C}[f]\; (1+ v\mu) \epsilon^2 \, d\epsilon \, d\mu \, d\varphi.
\label{energyEulerian}
\end{eqnarray}
(The $O(v)$ baryon conservation expression
\begin{equation}
{\partial\ln\rho \over \partial t} + {2v\over r} + 4\pi r^2\rho
{\partial v\over \partial m} = 0,
\label{baryonConservationComoving}
\end{equation}
obtained from Eqs. (\ref{baryonConservation}) and 
(\ref{inverseCoordinateTransformation})-(\ref{coordinateTransformation}),
has been employed.) 
Equation (\ref{energyEulerian}) agrees with Eq. (32) of
Ref. \cite{liebendoerfer01}. The non-conservative third term on the
left-hand side is $O(v^2)$ and ought to be dropped at the
level of approximation we are using. However, it is retained in
Ref. \cite{liebendoerfer01} for a practical reason. In supernova simulations
velocities can exceed values for which the $O(v)$ approximation is 
strictly valid; and while use of the $O(v)$ formalism might be questioned
on physical grounds, the $O(v^2)$ non-conservative term can still be
used to check that the numerical implementation of the $O(v)$
Boltzmann equation is consistent with Eq. (\ref{energyEulerian}).

\section{Conservative formulations of particle kinetics}
\label{sec:conservative}

We are now in a position to present conservative formulations of kinetic theory. We seek expressions closely tied to Eq. (\ref{numberConservation}) 
for particle number
balance and Eq. (\ref{energyConservation}) for particle 4-momentum balance. 
In addition to deriving these conservative formulations in the general
case, we show the relationship between them and explain the care that
must be exercised in finite differencing the number balance equation in
order to make it consistent with the energy balance equation.
We also specialize these results to Lagrangian coordinates in spherical
symmetry to $O(v)$, and see that the ``number conservative'' Boltzmann
equation---arrived at in the past by guesswork \cite{mezzacappa93}---emerges 
naturally from our formalism.

\subsection{Conservative formulations of particle kinetics:
the general case}
\label{subsec:conservativeGeneral}

For an expression related to particle number balance, the derivation of the Boltzmann equation in the previous section had us closer to the
desired result than might first be realized. In rushing headlong towards an equation for \inlineSFinmath${f}$, the key relation is easily overlooked:
It is Eq. (\ref{stokes}), the result of the generalized Stokes' theorem. 
The integrand on the right-hand side of Eq. (\ref{stokes}), $d(f \omega)$,
is conservative: Having been obtained from Stokes' theorem 
(the generalization) of the divergence theorem, it has {\em everything} 
inside the exterior derivative. Being an exterior derivative, it is too
abstract to be directly useful; 
but massaging it just enough to bring it into the form
$d(f \omega) = \mathbb{N}[f]\Omega$, where $\Omega$ is
the volume element in phase space, we shall see that $\mathbb{N}[f]$
is in fact a conservative differential operator in the familiar, elementary
sense.

Now we take a detailed look at Eq. (\ref{stokes}), in particular the exterior derivative
\begin{eqnarray}
d(f\multsp \omega )&=&d\big({\sqrt{-g}}{p^{\overvar{\mu }{\RawWedge }}}{{{{\ScriptCapitalL }^{\mu }}\InvisibleSpace }_{\overvar{\mu }{\RawWedge }}}f\multsp
{{\epsilon }_{\mu |\Mvariable{\nu \rho \sigma }|}}{{\Mvariable{dx}}^{\Mvariable{\nu \rho \sigma }}}\big)\wedge {{\pi }_m}-\nonumber \\
& &\big({\sqrt{-g}}{p^{\overvar{\mu
}{\RawWedge }}}{{{{\ScriptCapitalL }^{\mu }}\InvisibleSpace }_{\overvar{\mu }{\RawWedge }}}f\multsp {{\sigma }_{\mu }}\big)\wedge {{\Mvariable{d\pi
}}_m}+\nonumber  \\
& &  d\Bigg(\frac{1}{E({\bf p})}\Bigg|\det \big[\frac{\partial {\bf p}}{\partial {\bf u}}\big]\Bigg|\big(e\multsp {{{F^{\overvar{j}{\RawWedge
}}}\InvisibleSpace }_{\overvar{\nu }{\RawWedge }}}{p^{\overvar{\nu }{\RawWedge }}}-{{{{\Gamma }^{\overvar{j}{\RawWedge }}}\InvisibleSpace }_{\overvar{\nu
}{\RawWedge }\overvar{\rho }{\RawWedge }}}{p^{\overvar{\nu }{\RawWedge }}}{p^{\overvar{\rho }{\RawWedge }}}\big)\frac{\partial {u^{\overvar{i}{\RawWedge
}}}}{\partial {p^{\overvar{j}{\RawWedge }}}}f\multsp {{\epsilon }_{\overvar{0}{\RawWedge }\overvar{i}{\RawWedge }\big|\overvar{k}{\RawWedge }\overvar{n}{\RawWedge
}\big|}}{{\Mvariable{du}}^{\overvar{k}{\RawWedge }\overvar{n}{\RawWedge }}}\Bigg)\wedge \eta +\nonumber   \\
& & \Bigg(\frac{1}{E({\bf p})}\Bigg|\det \big[\frac{\partial {\bf p}}{\partial {\bf u}}\big]\Bigg|\big(e\multsp {{{F^{\overvar{j}{\RawWedge
}}}\InvisibleSpace }_{\overvar{\nu }{\RawWedge }}}{p^{\overvar{\nu }{\RawWedge }}}-{{{{\Gamma }^{\overvar{j}{\RawWedge }}}\InvisibleSpace }_{\overvar{\nu
}{\RawWedge }\overvar{\rho }{\RawWedge }}}{p^{\overvar{\nu }{\RawWedge }}}{p^{\overvar{\rho }{\RawWedge }}}\big)\frac{\partial {u^{\overvar{i}{\RawWedge
}}}}{\partial {p^{\overvar{j}{\RawWedge }}}}f\multsp {{\epsilon }_{\overvar{0}{\RawWedge }\overvar{i}{\RawWedge }\big|\overvar{k}{\RawWedge }\overvar{n}{\RawWedge
}\big|}}{{\Mvariable{du}}^{\overvar{k}{\RawWedge }\overvar{n}{\RawWedge }}}\Bigg)\wedge \Mvariable{d\eta },\label{exteriorNumber}
\end{eqnarray}
where we have used Eqs. (\ref{surface}), (\ref{spacetimeSurface}), (\ref{exteriorSum}), and (\ref{exteriorProduct}) and employed the vertical bar notation introduced in Eq. (\ref{bars}). The exterior derivatives
will be expressed in terms of the basis \inlineSFinmath${\big\{{{\Mvariable{dx}}^{\mu }},{{\Mvariable{du}}^{\overvar{i}{\RawWedge }}}\big\}}$ on \inlineSFinmath${{M_m}}$.
First we note that in our chosen momentum coordinates, the second term in Eq. (\ref{exteriorNumber}) vanishes because \inlineSFinmath${{{\Mvariable{d\pi }}_m}=0}$:
This is because \inlineSFinmath${{{\pi }_m}}$ as expressed in Eq. 
(\ref{momentumElement}) has no dependence on \inlineSFinmath${\{{x^{\mu }}\}}$; and while it depends
on \inlineSFinmath${\big\{{u^{\overvar{i}{\RawWedge }}}\big\}}$, adding another momentum \inlineSFinmath${1}$-form to the wedge product \inlineSFinmath${{{\Mvariable{du}}^{\overvar{1}{\RawWedge
}\overvar{2}{\RawWedge }\overvar{3}{\RawWedge }}}}$ would cause it to vanish due to its antisymmetry. Similarly, the fourth term vanishes because
\inlineSFinmath${\Mvariable{d\eta }=0}$. To understand what happens to the first term, consider one of the terms in the sum over \inlineSFinmath${\mu
}$:
\begin{equation}
\dispSFNumberedEquationmath{d\big({\sqrt{-g}}{p^{\overvar{\mu }{\RawWedge }}}{{{{\ScriptCapitalL }^0}\InvisibleSpace }_{\overvar{\mu }{\RawWedge
}}}f\multsp {{\epsilon }_{0|123|}}{{\Mvariable{dx}}^{123}}\big)\wedge {{\pi }_m}=\frac{\partial }{\partial {x^0}}\big({\sqrt{-g}}{p^{\overvar{\mu
}{\RawWedge }}}{{{{\ScriptCapitalL }^0}\InvisibleSpace }_{\overvar{\mu }{\RawWedge }}}f\multsp \big){{\Mvariable{dx}}^{0123}}\wedge {{\pi }_m}.}
\end{equation}
While derivatives with respect to $\{x^i\}$ and \inlineSFinmath${\big\{{u^{\overvar{i}{\RawWedge }}}\big\}}$ are nonvanishing, only the wedge product of $dx^0$ with $dx^{123}$ is non-vanishing, and the wedge product in \inlineSFinmath${{{\pi
}_m}}$ does not admit another momentum \inlineSFinmath${1}$-form. The other terms in the sum over \inlineSFinmath${\mu }$ are similar. To understand the
third term in Eq. (\ref{exteriorNumber}), consider a particular term in the sum over \inlineSFinmath${\overvar{i}{\RawWedge }}$:
\begin{eqnarray}
d\Bigg(\frac{1}{E({\bf p})}\Bigg|\det \big[\frac{\partial {\bf p}}{\partial {\bf u}}\big]\Bigg|\big(e\multsp {{{F^{\overvar{j}{\RawWedge }}}\InvisibleSpace
}_{\overvar{\nu }{\RawWedge }}}{p^{\overvar{\nu }{\RawWedge }}}-{{{{\Gamma }^{\overvar{j}{\RawWedge }}}\InvisibleSpace }_{\overvar{\nu }{\RawWedge
}\overvar{\rho }{\RawWedge }}}{p^{\overvar{\nu }{\RawWedge }}}{p^{\overvar{\rho }{\RawWedge }}}\big)\frac{\partial {u^{\overvar{1}{\RawWedge }}}}{\partial
{p^{\overvar{j}{\RawWedge }}}}f\multsp {{\epsilon }_{\overvar{0}{\RawWedge }\overvar{1}{\RawWedge }\big|\overvar{2}{\RawWedge }\overvar{3}{\RawWedge
}\big|}}{{\Mvariable{du}}^{\overvar{2}{\RawWedge }\overvar{3}{\RawWedge }}}\Bigg)\wedge \eta =\nonumber  \\
 \frac{\partial }{\partial {u^{\overvar{1}{\RawWedge }}}}\Bigg(\frac{1}{E({\bf p})}\Bigg|\det \big[\frac{\partial {\bf p}}{\partial
{\bf u}}\big]\Bigg|\big(e\multsp {{{F^{\overvar{j}{\RawWedge }}}\InvisibleSpace }_{\overvar{\nu }{\RawWedge }}}{p^{\overvar{\nu }{\RawWedge }}}-{{{{\Gamma
}^{\overvar{j}{\RawWedge }}}\InvisibleSpace }_{\overvar{\nu }{\RawWedge }\overvar{\rho }{\RawWedge }}}{p^{\overvar{\nu }{\RawWedge }}}{p^{\overvar{\rho
}{\RawWedge }}}\big)\frac{\partial {u^{\overvar{1}{\RawWedge }}}}{\partial {p^{\overvar{j}{\RawWedge }}}}f\multsp \Bigg){{\Mvariable{du}}^{\overvar{1}{\RawWedge
}\overvar{2}{\RawWedge }\overvar{3}{\RawWedge }}}\wedge \eta ,
\end{eqnarray}
and similarly for the other terms in the sum over \inlineSFinmath${\overvar{i}{\RawWedge }}$. With Eq. (\ref{volume2}), all terms in Eq. (\ref{exteriorNumber}) can be assembled to form
the expression
\begin{equation}
\dispSFNumberedEquationmath{d(f\multsp \omega )=\mathbb{N}[f]\Omega ,}
\label{conservativeNumberOperator}
\end{equation}
where 
\begin{eqnarray}
\mathbb{N}[f]&\equiv&\frac{1}{{\sqrt{-g}}}\frac{\partial }{\partial {x^{\mu }}}\big({\sqrt{-g}}{{{{\ScriptCapitalL
}^{\mu }}\InvisibleSpace }_{\overvar{\mu }{\RawWedge }}}
{p^{\overvar{\mu }{\RawWedge }}}f\multsp \big)+\nonumber  \\
& & E({\bf p})\Bigg|\det \big[\frac{\partial {\bf p}}{\partial {\bf u}}\big]{{\Bigg|}^{-1}}  \frac{\partial }{\partial {u^{\overvar{i}{\RawWedge }}}}\Bigg(\frac{1}{E({\bf p})}\Bigg|\det \big[\frac{\partial {\bf p}}{\partial
{\bf u}}\big]\Bigg|\big(e\multsp {{{F^{\overvar{j}{\RawWedge }}}\InvisibleSpace }_{\overvar{\nu }{\RawWedge }}}{p^{\overvar{\nu }{\RawWedge }}}-{{{{\Gamma
}^{\overvar{j}{\RawWedge }}}\InvisibleSpace }_{\overvar{\nu }{\RawWedge }\overvar{\rho }{\RawWedge }}}{p^{\overvar{\nu }{\RawWedge }}}{p^{\overvar{\rho
}{\RawWedge }}}\big)\frac{\partial {u^{\overvar{i}{\RawWedge }}}}{\partial {p^{\overvar{j}{\RawWedge }}}}f\Bigg).\label{conservativeNumberOperator2}
\end{eqnarray}
In the derivation of the Boltzmann equation, Eq. (\ref{conservativeNumberOperator}) can be used to replace Eq. (\ref{almostBoltzmann}) with a similar expression in which \inlineSFinmath${{L_m}[f]}$
is replaced by \inlineSFinmath${\mathbb{N}[f]}$, with the result that
\begin{equation}
\dispSFNumberedEquationmath{\mathbb{N}[f]=\mathbb{C}[f].}
\label{conservativeNumberKinetics}
\end{equation}
This is our conservative formulation of particle number kinetics.

Next we seek a formulation related to particle 4-momentum balance. With arbitrary \inlineSFinmath${{v_{\mu }}}$, we evaluate the exterior derivative
\inlineSFinmath${d({v_{\mu }}{p^{\mu }}f\multsp \omega )}$ in two separate ways. (To simplify formulae, we often write \inlineSFinmath${{p^{\mu }}}$
instead of \inlineSFinmath${{{{{\ScriptCapitalL }^{\mu }}\InvisibleSpace }_{\overvar{\mu }{\RawWedge }}}{p^{\overvar{\mu }{\RawWedge }}}}$, but because
of our chosen momentum coordinates on \inlineSFinmath${{M_m}}$ the latter expression must often be used in computational steps.) First, employing
Eqs. (\ref{conservativeNumberOperator}) 
and (\ref{conservativeNumberKinetics}), 
\begin{equation}
\dispSFNumberedEquationmath{d({v_{\mu }}{p^{\mu }}f\multsp \omega )=d({v_{\mu }}{p^{\mu }})\wedge f\multsp \omega +{v_{\mu }}{p^{\mu }}\mathbb{C}[f]\Omega
.}\label{energyTwoTerms}
\end{equation}
Computation shows that
\begin{eqnarray}
\dispSFNumberedEquationmath
d({v_{\mu }}{p^{\mu }})\wedge f\multsp \omega& =&\bigg({p^{\mu }}{p^{\nu }}f\frac{\partial {v_{\mu }}}{\partial {x^{\nu }}}+{v_{\mu }}f\multsp {p^{\nu
}}{p^{\overvar{\mu }{\RawWedge }}}\frac{\partial {{{{\ScriptCapitalL }^{\mu }}\InvisibleSpace }_{\overvar{\mu }{\RawWedge }}}}{\partial {x^{\nu }}}\bigg)\Omega
+ \nonumber \\
& & {v_{\mu }}\Bigg({{{{\ScriptCapitalL }^{\mu }}\InvisibleSpace }_{\overvar{0}{\RawWedge }}}\frac{\partial {p^{\overvar{0}{\RawWedge }}}}{\partial
{p^{\overvar{j}{\RawWedge }}}}+{{{{\ScriptCapitalL }^{\mu }}\InvisibleSpace }_{\overvar{i}{\RawWedge }}}\frac{\partial {p^{\overvar{i}{\RawWedge
}}}}{\partial {p^{\overvar{j}{\RawWedge }}}}\Bigg)\big(e\multsp {{{F^{\overvar{j}{\RawWedge }}}\InvisibleSpace }_{\overvar{\nu }{\RawWedge }}}{p^{\overvar{\nu
}{\RawWedge }}}-{{{{\Gamma }^{\overvar{j}{\RawWedge }}}\InvisibleSpace }_{\overvar{\nu }{\RawWedge }\overvar{\rho }{\RawWedge }}}{p^{\overvar{\nu
}{\RawWedge }}}{p^{\overvar{\rho }{\RawWedge }}}\big)f\multsp \Omega.\label{energyStep}
\end{eqnarray}
Because of the mass shell constraint (Eq. (\ref{massShell})), \inlineSFinmath${{p^{\overvar{0}{\RawWedge }}}}$ is considered a function of the \inlineSFinmath${\big\{{p^{\overvar{i}{\RawWedge
}}}\big\}}$. The geodesic equation (equation (\ref{geodesic2})) can be used to show that
\begin{equation}
\dispSFNumberedEquationmath{\frac{\partial {p^{\overvar{0}{\RawWedge }}}}{\partial {p^{\overvar{j}{\RawWedge }}}}\big(e\multsp {{{F^{\overvar{j}{\RawWedge
}}}\InvisibleSpace }_{\overvar{\nu }{\RawWedge }}}{p^{\overvar{\nu }{\RawWedge }}}-{{{{\Gamma }^{\overvar{j}{\RawWedge }}}\InvisibleSpace }_{\overvar{\nu
}{\RawWedge }\overvar{\rho }{\RawWedge }}}{p^{\overvar{\nu }{\RawWedge }}}{p^{\overvar{\rho }{\RawWedge }}}\big)=e\multsp {{{F^{\overvar{0}{\RawWedge
}}}\InvisibleSpace }_{\overvar{\nu }{\RawWedge }}}{p^{\overvar{\nu }{\RawWedge }}}-{{{{\Gamma }^{\overvar{0}{\RawWedge }}}\InvisibleSpace }_{\overvar{\nu
}{\RawWedge }\overvar{\rho }{\RawWedge }}}{p^{\overvar{\nu }{\RawWedge }}}{p^{\overvar{\rho }{\RawWedge }}}.}
\label{timeGeodesic}
\end{equation}
This, together with Eq. (\ref{connectionCoefficients}) for the transformation of the connection coefficients, allows Eq. (\ref{energyStep}) to be written
\begin{equation}
\dispSFNumberedEquationmath{d({v_{\mu }}{p^{\mu }})\wedge f\multsp \omega =\Big({p^{\mu }}{p^{\nu }}f\frac{\partial {v_{\mu }}}{\partial {x^{\nu
}}}+{v_{\mu }}(e\multsp {{{F^{\mu }}\InvisibleSpace }_{\nu }}{p^{\nu }}-{{{{\Gamma }^{\mu }}\InvisibleSpace }_{\Mvariable{\nu \rho }}}{p^{\nu }}{p^{\rho
}})f\multsp \Big)\Omega .}\label{energyTerm1}
\end{equation}
Second, we evaluate \inlineSFinmath${d({v_{\mu }}{p^{\mu }}f\multsp \omega )}$ directly. A computation similar to that leading to Eqs. (\ref{conservativeNumberOperator})-(\ref{conservativeNumberOperator2})
yields
\begin{equation}
\dispSFNumberedEquationmath{d({v_{\mu }}{p^{\mu }}f\multsp \omega )=\Big({p^{\mu }}{p^{\nu }}f\frac{\partial {v_{\mu }}}{\partial {x^{\nu }}}+{v_{\mu
}}{\mathbb{T}^{\mu}}[f]\Big)\Omega ,}
\end{equation}
where
\begin{eqnarray}
{\mathbb{T}^{\mu}}[f]&\equiv&\frac{1}{{\sqrt{-g}}}\frac{\partial }{\partial {x^{\nu }}}\big({\sqrt{-g}}{{{{\ScriptCapitalL }^{\mu }}\InvisibleSpace
}_{\overvar{\mu }{\RawWedge }}}{{{{\ScriptCapitalL }^{\nu }}\InvisibleSpace }_{\overvar{\nu }{\RawWedge }}}{p^{\overvar{\mu }{\RawWedge }}}{p^{\overvar{\nu
}{\RawWedge }}}f\multsp \big)+  \nonumber \\
& & E({\bf p})\Bigg|\det \big[\frac{\partial{\bf p}}{\partial{\bf u}}\big]{{\Bigg|}^{-1}}  \frac{\partial }{\partial {u^{\overvar{i}{\RawWedge }}}}\Bigg(\frac{1}{E({\bf p})}\Bigg|\det \big[\frac{\partial{\bf p}}{\partial
{\bf u}}\big]\Bigg|\big(e\multsp {{{F^{\overvar{j}{\RawWedge }}}\InvisibleSpace }_{\overvar{\nu }{\RawWedge }}}{p^{\overvar{\nu }{\RawWedge }}}-{{{{\Gamma
}^{\overvar{j}{\RawWedge }}}\InvisibleSpace }_{\overvar{\nu }{\RawWedge }\overvar{\rho }{\RawWedge }}}{p^{\overvar{\nu }{\RawWedge }}}{p^{\overvar{\rho
}{\RawWedge }}}\big)\frac{\partial {u^{\overvar{i}{\RawWedge }}}}{\partial {p^{\overvar{j}{\RawWedge }}}}{{{{\ScriptCapitalL }^{\mu }}\InvisibleSpace
}_{\overvar{\mu }{\RawWedge }}}{p^{\overvar{\mu }{\RawWedge }}}f\Bigg).
\label{energyLastFactor}
\end{eqnarray}
Recalling that \inlineSFinmath${{v_{\mu }}}$ is arbitrary, Eqs. (\ref{energyTwoTerms}) and (\ref{energyTerm1})-(\ref{energyLastFactor}) can be combined into
\begin{equation}
\dispSFNumberedEquationmath{{\mathbb{T}^{\mu }}[f]
=
\big(e\multsp {{{F^{\mu }}\InvisibleSpace }_{\nu }}{{{{\ScriptCapitalL }^{\nu }}\InvisibleSpace
}_{\overvar{\nu }{\RawWedge }}}{p^{\overvar{\nu }{\RawWedge }}}
-{{{{\Gamma }^{\mu }}\InvisibleSpace }_{\Mvariable{\nu \rho }}}{{{{\ScriptCapitalL
}^{\nu }}\InvisibleSpace }_{\overvar{\nu }{\RawWedge }}}{{{{\ScriptCapitalL }^{\rho }}\InvisibleSpace }_{\overvar{\rho }{\RawWedge }}}{p^{\overvar{\nu
}{\RawWedge }}}{p^{\overvar{\rho }{\RawWedge }}}\big)f +
{{{{\ScriptCapitalL }^{\mu }}\InvisibleSpace }_{\overvar{\mu }{\RawWedge }}}{p^{\overvar{\mu
}{\RawWedge }}}\mathbb{C}[f],}
\label{conservativeMomentumKinetics}
\end{equation}
a ``conservative'' formulation of particle 4-momentum kinetics.

To illuminate the practical problem of accurately accounting for both
particle number and energy, we consider the relationship between 
Eq. (\ref{conservativeNumberKinetics}) for particle number kinetics
and the $t$ component of Eq. (\ref{conservativeMomentumKinetics}) for
particle 4-momentum kinetics. 
(This is an issue when one follows the general approach of, for example, Refs.
\cite{mezzacappa01,liebendoerfer01b,liebendoerfer02}, in which
a particle number distribution equation is solved and a particle
energy equation is used as a consistency check.)
Specifically, we need to relate the
spatial and momentum divergence terms of these equations to each other
and identify terms that cancel.

First we relate the
spatial and momentum divergence terms of Eq. 
(\ref{conservativeNumberKinetics}) and the $t$ component of
Eq. (\ref{conservativeMomentumKinetics}).
Their spatial divergence terms,
\begin{eqnarray}
\mathbb{N}_x[f]&\equiv&\frac{1}{{\sqrt{-g}}}\frac{\partial }{\partial {x^{\mu }}}\big({\sqrt{-g}}{{{{\ScriptCapitalL
}^{\mu }}\InvisibleSpace }_{\overvar{\mu }{\RawWedge }}}
{p^{\overvar{\mu }{\RawWedge }}}f\multsp \big), 
\\
\left(\mathbb{T}^t\right)_x[f]&\equiv&\frac{1}{{\sqrt{-g}}}\frac{\partial }{\partial {x^{\mu }}}\big({\sqrt{-g}}{{{{\ScriptCapitalL }^{t }}\InvisibleSpace
}_{\overvar{\nu }{\RawWedge }}}{{{{\ScriptCapitalL }^{\mu }}\InvisibleSpace }_{\overvar{\mu }{\RawWedge }}}{p^{\overvar{\nu }{\RawWedge }}}{p^{\overvar{\mu
}{\RawWedge }}}f\multsp \big),
\end{eqnarray}
are related by
\begin{equation}
{{\ScriptCapitalL }^{t }\InvisibleSpace
}_{\overvar{\nu }{\RawWedge }}{p^{\overvar{\nu }{\RawWedge }}}\; 
\mathbb{N}_x[f]
=
\left(\mathbb{T}^t\right)_x[f] - 
f \; 
{p^{\overvar{\mu}{\RawWedge }}} {p^{\overvar{\nu }{\RawWedge }}}
{{{{\ScriptCapitalL }^{\mu }}\InvisibleSpace }_{\overvar{\mu }{\RawWedge }}}
{\partial\over\partial x^\mu}\left(
{{{{\ScriptCapitalL }^{t }}\InvisibleSpace }_{\overvar{\nu }{\RawWedge }}}
\right).\label{numberRelation}
\end{equation}
Their momentum divergence terms,
\begin{eqnarray}
\mathbb{N}_p[f]&\equiv&
E({\bf p})\Bigg|\det \big[\frac{\partial {\bf p}}{\partial {\bf u}}\big]{{\Bigg|}^{-1}}  \frac{\partial }{\partial {u^{\overvar{i}{\RawWedge }}}}\Bigg(\frac{1}{E({\bf p})}\Bigg|\det \big[\frac{\partial {\bf p}}{\partial
{\bf u}}\big]\Bigg|\big(e\multsp {{{F^{\overvar{j}{\RawWedge }}}\InvisibleSpace }_{\overvar{\mu }{\RawWedge }}}{p^{\overvar{\mu }{\RawWedge }}}-{{{{\Gamma
}^{\overvar{j}{\RawWedge }}}\InvisibleSpace }_{\overvar{\mu }{\RawWedge }\overvar{\rho }{\RawWedge }}}{p^{\overvar{\mu }{\RawWedge }}}{p^{\overvar{\rho
}{\RawWedge }}}\big)\frac{\partial {u^{\overvar{i}{\RawWedge }}}}{\partial {p^{\overvar{j}{\RawWedge }}}}f\Bigg),
\\
\left(\mathbb{T}^t\right)_p[f]&\equiv&
E({\bf p})\Bigg|\det \big[\frac{\partial{\bf p}}{\partial{\bf u}}\big]{{\Bigg|}^{-1}}  \frac{\partial }{\partial {u^{\overvar{i}{\RawWedge }}}}\Bigg(\frac{1}{E({\bf p})}\Bigg|\det \big[\frac{\partial{\bf p}}{\partial
{\bf u}}\big]\Bigg|\big(e\multsp {{{F^{\overvar{j}{\RawWedge }}}\InvisibleSpace }_{\overvar{\mu }{\RawWedge }}}{p^{\overvar{\mu }{\RawWedge }}}-{{{{\Gamma
}^{\overvar{j}{\RawWedge }}}\InvisibleSpace }_{\overvar{\mu }{\RawWedge }\overvar{\rho }{\RawWedge }}}{p^{\overvar{\mu }{\RawWedge }}}{p^{\overvar{\rho
}{\RawWedge }}}\big)\frac{\partial {u^{\overvar{i}{\RawWedge }}}}{\partial {p^{\overvar{j}{\RawWedge }}}}{{{{\ScriptCapitalL }^{t }}\InvisibleSpace
}_{\overvar{\nu }{\RawWedge }}}{p^{\overvar{\nu }{\RawWedge }}}f\Bigg),
\end{eqnarray}
are related by
\begin{equation}
{{\ScriptCapitalL }^{t }\InvisibleSpace
}_{\overvar{\nu }{\RawWedge }}{p^{\overvar{\nu }{\RawWedge }}}\; 
\mathbb{N}_p[f]
=
\left(\mathbb{T}^t\right)_p[f] - 
f\;
\big(e\multsp {{{F^{\overvar{j}{\RawWedge }}}\InvisibleSpace }_{\overvar{\mu }{\RawWedge }}}{p^{\overvar{\mu }{\RawWedge }}}-{{{{\Gamma
}^{\overvar{j}{\RawWedge }}}\InvisibleSpace }_{\overvar{\mu }{\RawWedge }\overvar{\rho }{\RawWedge }}}{p^{\overvar{\mu }{\RawWedge }}}{p^{\overvar{\rho
}{\RawWedge }}}\big)
\frac{\partial {u^{\overvar{i}{\RawWedge }}}}{\partial {p^{\overvar{j}{\RawWedge }}}} \;
{{{{\ScriptCapitalL }^{t }}\InvisibleSpace }_{\overvar{\nu }{\RawWedge }}}\;
\frac{\partial {p^{\overvar{\nu}{\RawWedge }}}}{\partial {u^{\overvar{i}{\RawWedge }}}}
.\label{energyRelation}
\end{equation}

Comparison of Eqs. (\ref{numberRelation}) and (\ref{energyRelation}) with
Eq. (\ref{conservativeNumberKinetics}) and the $t$ component of
Eq. (\ref{conservativeMomentumKinetics}) shows that the following 
equation must be valid:
\begin{equation}
E_S + E_M
=
e\multsp {{{F^{t }}\InvisibleSpace }_{\mu }}{{{{\ScriptCapitalL }^{\mu }}\InvisibleSpace
}_{\overvar{\mu }{\RawWedge }}}{p^{\overvar{\mu }{\RawWedge }}}
-{{{{\Gamma }^{t }}\InvisibleSpace }_{\Mvariable{\mu \rho }}}{{{{\ScriptCapitalL
}^{\mu }}\InvisibleSpace }_{\overvar{\mu }{\RawWedge }}}{{{{\ScriptCapitalL }^{\rho }}\InvisibleSpace }_{\overvar{\rho }{\RawWedge }}}{p^{\overvar{\mu
}{\RawWedge }}}{p^{\overvar{\rho }{\RawWedge }}},
\label{matching}
\end{equation}
where 
\begin{equation}
E_S \equiv 
f\,{p^{\overvar{\mu}{\RawWedge }}} {p^{\overvar{\nu }{\RawWedge }}}
{{{{\ScriptCapitalL }^{\mu }}\InvisibleSpace }_{\overvar{\mu }{\RawWedge }}}
{\partial\over\partial x^\mu}\left(
{{{{\ScriptCapitalL }^{t }}\InvisibleSpace }_{\overvar{\nu }{\RawWedge }}}
\right)
\label{spaceExtra}
\end{equation}
is the ``extra'' term of Eq. (\ref{numberRelation})
relating the spatial divergence terms of the number and energy balance
equations, and
\begin{equation}
E_M \equiv f\,
\big(e\multsp {{{F^{\overvar{j}{\RawWedge }}}\InvisibleSpace }_{\overvar{\mu }{\RawWedge }}}{p^{\overvar{\mu }{\RawWedge }}}-{{{{\Gamma
}^{\overvar{j}{\RawWedge }}}\InvisibleSpace }_{\overvar{\mu }{\RawWedge }\overvar{\rho }{\RawWedge }}}{p^{\overvar{\mu }{\RawWedge }}}{p^{\overvar{\rho
}{\RawWedge }}}\big)
\frac{\partial {u^{\overvar{i}{\RawWedge }}}}{\partial {p^{\overvar{j}{\RawWedge }}}} \;
{{{{\ScriptCapitalL }^{t }}\InvisibleSpace }_{\overvar{\nu }{\RawWedge }}}\;
\frac{\partial {p^{\overvar{\nu}{\RawWedge }}}}{\partial {u^{\overvar{i}{\RawWedge }}}}
\label{momentumExtra}
\end{equation}
is the ``extra'' term in Eq. (\ref{energyRelation})
relating the momentum divergence terms of the number and energy balance
equations. Noting that 
\begin{equation}
\frac{\partial {u^{\overvar{i}{\RawWedge }}}}{\partial {p^{\overvar{j}{\RawWedge }}}} \;
{{{{\ScriptCapitalL }^{t }}\InvisibleSpace }_{\overvar{\nu }{\RawWedge }}}\;
\frac{\partial {p^{\overvar{\nu}{\RawWedge }}}}{\partial {u^{\overvar{i}{\RawWedge }}}}
=
\frac{\partial {p^{\overvar{0}{\RawWedge }}}}{\partial {p^{\overvar{j}{\RawWedge }}}}
{{{{\ScriptCapitalL }^{t }}\InvisibleSpace }_{\overvar{0 }{\RawWedge }}}\; 
+
\delta^{\overvar{k}{\RawWedge }}_{\overvar{j}{\RawWedge }}
{{{{\ScriptCapitalL }^{t }}\InvisibleSpace }_{\overvar{k }{\RawWedge }}}, 
\end{equation}
Eq. (\ref{timeGeodesic}) can be used to rewrite Eq. (\ref{momentumExtra})
as
\begin{equation}
E_M 
= f\,
\big(e\multsp {{{F^{\overvar{\nu}{\RawWedge }}}\InvisibleSpace }_{\overvar{\mu }{\RawWedge }}}{p^{\overvar{\mu }{\RawWedge }}}-{{{{\Gamma
}^{\overvar{\nu}{\RawWedge }}}\InvisibleSpace }_{\overvar{\mu }{\RawWedge }\overvar{\rho }{\RawWedge }}}{p^{\overvar{\mu }{\RawWedge }}}{p^{\overvar{\rho
}{\RawWedge }}}\big)
{{{{\ScriptCapitalL }^{t }}\InvisibleSpace }_{\overvar{\nu }{\RawWedge }}}.
\end{equation}
Use of Eq. (\ref{connectionCoefficients}) and two applications of
the identity
\begin{equation}
{{{{\ScriptCapitalL }^{\mu }}\InvisibleSpace }_{\overvar{\rho }{\RawWedge }}}
{\partial \over \partial x^\sigma}\left({{{{\ScriptCapitalL }^{\overvar{\rho }{\RawWedge }}}\InvisibleSpace }_\nu}\right) 
= -
{{{{\ScriptCapitalL }^{\overvar{\rho }{\RawWedge }}}\InvisibleSpace }_\nu}
{\partial \over \partial x^\sigma}\left({{{{\ScriptCapitalL }^{\mu }}\InvisibleSpace }_{\overvar{\rho }{\RawWedge }}}\right)
\end{equation}
lead finally to
\begin{equation}
E_M 
= f \left[
e\multsp {{{F^{t }}\InvisibleSpace }_{\mu }}{{{{\ScriptCapitalL }^{\mu }}\InvisibleSpace
}_{\overvar{\mu }{\RawWedge }}}{p^{\overvar{\mu }{\RawWedge }}}
-{{{{\Gamma }^{t }}\InvisibleSpace }_{\Mvariable{\mu \rho }}}{{{{\ScriptCapitalL
}^{\mu }}\InvisibleSpace }_{\overvar{\mu }{\RawWedge }}}{{{{\ScriptCapitalL }^{\rho }}\InvisibleSpace }_{\overvar{\rho }{\RawWedge }}}{p^{\overvar{\mu
}{\RawWedge }}}{p^{\overvar{\rho }{\RawWedge }}} 
-
{p^{\overvar{\mu}{\RawWedge }}} {p^{\overvar{\nu }{\RawWedge }}}
{{{{\ScriptCapitalL }^{\mu }}\InvisibleSpace }_{\overvar{\mu }{\RawWedge }}}
{\partial\over\partial x^\mu}\left(
{{{{\ScriptCapitalL }^{t }}\InvisibleSpace }_{\overvar{\nu }{\RawWedge }}}
\right)\right].
\label{momentumExtraFinal}
\end{equation}
Equations (\ref{spaceExtra}) and (\ref{momentumExtraFinal}) make it 
obvious that Eq. (\ref{matching}) is satisfied.

In a computational approach to transport in which
a particle number distribution equation is solved and a particle
energy equation is used as a consistency check, care must be taken
that Eq. (\ref{matching}) is satisfied numerically.
In particular, a given finite difference representation 
of the spacetime divergence
$\mathbb{N}_x[f]$ implies a corresponding finite difference
representation of position-dependent quantities in $E_S$, 
and the position-dependent
quantities in terms of $E_M$ that
cancel with $E_S$ must have a matching finite difference 
representation. Similarly,
a given finite difference representation 
of the momentum divergence
$\mathbb{N}_p[f]$ implies a corresponding finite difference
representation of momentum variables in $E_M$, and the momentum variables
in terms of $E_S$ that
cancel with terms of $E_M$ must have a matching finite difference 
representation.

\subsection{Conservative formulations of particle kinetics:
Lagrangian coordinates in spherical symmetry to $O(v)$}
\label{subsec:conservativeComoving}

Using Eqs. (\ref{comovingDeterminant}), 
(\ref{transformationFirst})-(\ref{transformationLast}), 
(\ref{connectionComovingFirst})-(\ref{connectionComovingLast}), 
and (\ref{momentum1})-(\ref{momentum0}), the ``number conservative''
formulation of 
Eq. (\ref{conservativeNumberKinetics}) specializes to
\begin{eqnarray}
{\partial \over \partial t}\left(f\over\rho\right)
+ 
 {\partial\over\partial m}\left(4\pi r^2\rho\mu\, {f\over\rho}\right)
+
{1\over\epsilon^2}{\partial\over\partial\epsilon}
\left(\epsilon^3\left[\mu^2\left({3 v\over r} +
{\partial\ln\rho\over\partial t}\right)-{v\over r}\right]{f\over\rho}\right)
 + & &\nonumber \\
{\partial \over\partial\mu}\left(
(1-\mu^2)\left[{1\over r}+\mu\left({3 v\over r}+
{\partial\ln\rho\over\partial t}\right)
\right]{f\over\rho}\right)
&=&{1\over\rho\, \epsilon}\,\mathbb{C}[f],
\label{numberConservativeBoltzmann}
\end{eqnarray}
which agrees with Eq. (23) of Ref. \cite{mezzacappa93}.
In Ref. \cite{mezzacappa93}, the necessity of making $f/\rho$ the
evolved variable in order to get a conservative particle number
equation is left unexplained. In the present formalism, we see that
the factor $1/\rho$ comes from the factor $\sqrt{-g}=\sin\theta/4\pi\rho$ 
in Eq. (\ref{numberConservation}). Multiplication of Eq. 
(\ref{numberConservativeBoltzmann}) by $\epsilon^2\,d\epsilon\,d\mu\,d\varphi$
and integrating immediately yields Eq. (\ref{numberComoving}), expressing
particle number ``conservation''.

Similarly, the $t$ and $m$ components of the ``momentum conservative''
formulation of Eq. (\ref{conservativeMomentumKinetics}) are
\begin{eqnarray}
{\partial \over \partial t}\left(\epsilon f\over\rho\right)
&+& 
 {\partial\over\partial m}\left(4\pi r^2\rho \epsilon\mu\, {f\over\rho}\right)
+
{1\over\epsilon^2}{\partial\over\partial\epsilon}
\left(\epsilon^4\left[\mu^2\left({3 v\over r} +
{\partial\ln\rho\over\partial t}\right)-{v\over r}\right]{f\over\rho}\right)
 + \nonumber 
\\
& &{\partial \over\partial\mu}\left(
(1-\mu^2)\left[{1\over r}+\mu\left({3 v\over r}+
{\partial\ln\rho\over\partial t}\right)
\right]{\epsilon f\over\rho}\right)\nonumber
\\
&=&
\left({2v\over r}
+
{\partial\ln\rho \over \partial t}\right)\epsilon\mu^2\,f
-
{v\over r}\,\epsilon (1-\mu^2)f
+
{1\over\rho}\,\mathbb{C}[f]
\label{energyConservativeBoltzmann}
\end{eqnarray}
and
\begin{eqnarray}
{1\over r^2\rho}{\partial \over \partial t}\left(r^2\rho\,\epsilon\mu f\over\rho\right)
&+& 
{1\over 4\pi r^2\rho} {\partial\over\partial m}\left((4\pi r^2\rho)^2 
\epsilon\mu^2\, {f\over\rho}\right)
+
{1\over\epsilon^2}{\partial\over\partial\epsilon}
\left(\epsilon^4\mu\left[\mu^2\left({3 v\over r} +
{\partial\ln\rho\over\partial t}\right)-{v\over r}\right]{f\over\rho}\right)
 + \nonumber 
\\
& &{\partial \over\partial\mu}\left(
(1-\mu^2)\left[{1\over r}+\mu\left({3 v\over r}+
{\partial\ln\rho\over\partial t}\right)
\right]{\epsilon\mu f\over\rho}\right)\nonumber
\\
&=&
2 \left({2v\over r}+
{\partial\ln\rho \over \partial t}\right)\epsilon\mu\,f
+
(4\pi r^2\rho)
\left({1\over 2\pi r^3\rho}+{\partial\ln\rho \over \partial m}\right)
\epsilon\mu^2\,f\nonumber
\\
& &+
{1\over r}\,\epsilon (1-\mu^2)\,f
+
{\mu\over\rho}\,\mathbb{C}[f].
\label{momentumConservativeBoltzmann}
\end{eqnarray}
With the help of Eq. (\ref{inverseCoordinateTransformation}), the
$m$ component equation can be expressed
\begin{eqnarray}
{\partial \over \partial t}\left(\epsilon\mu f\over\rho\right)
&+& 
 {\partial\over\partial m}\left(4\pi r^2\rho 
\epsilon\mu^2\, {f\over\rho}\right)
+
{1\over\epsilon^2}{\partial\over\partial\epsilon}
\left(\epsilon^4\mu\left[\mu^2\left({3 v\over r} +
{\partial\ln\rho\over\partial t}\right)-{v\over r}\right]{f\over\rho}\right)
 + \nonumber 
\\
& &{\partial \over\partial\mu}\left(
(1-\mu^2)\left[{1\over r}+\mu\left({3 v\over r}+
{\partial\ln\rho\over\partial t}\right)
\right]{\epsilon\mu f\over\rho}\right)\nonumber
\\
&=&
\left({2v\over r}+
{\partial\ln\rho \over \partial t}\right)\epsilon\mu\,f
+
{1\over r}\,\epsilon (1-\mu^2)\,f
+
{\mu\over\rho}\,\mathbb{C}[f].
\label{momentumConservativeBoltzmann2}
\end{eqnarray}
Multiplication of Eqs. 
(\ref{energyConservativeBoltzmann})-(\ref{momentumConservativeBoltzmann2}) 
by $\epsilon^2\,d\epsilon\,d\mu\,d\varphi$
and integrating immediately yields Eqs. 
(\ref{energyComoving})-(\ref{momentumComoving}), expressions for 
particle 4-momentum balance in the comoving frame.

Just as the comoving frame $t$ and $m$ momentum-integrated balance
equations---Eqs. 
(\ref{energyComoving}) and (\ref{momentumComoving})---can be combined
to form the lab frame $\tilde t$ ``conservation'' equation, Eq. (\ref{energyEulerian}),
 Eqs. (\ref{energyConservativeBoltzmann}) and 
(\ref{momentumConservativeBoltzmann2}) can be combined as
\begin{eqnarray}
{\partial \over \partial t}\left[\epsilon (1+v\mu) f\over\rho\right]
&+& 
 {\partial\over\partial m}\left(4\pi r^2\rho \epsilon\mu(1+v\mu) 
{f\over\rho}\right)
-{1\over r^2\rho}{\partial\over\partial t}\left(r^2 \rho v\right)
\epsilon\mu\, {f\over\rho}\nonumber
\\
& &{1\over\epsilon^2}{\partial\over\partial\epsilon}
\left(\epsilon^4(1+v\mu)\left[\mu^2\left({3 v\over r} +
{\partial\ln\rho\over\partial t}\right)-{v\over r}\right]{f\over\rho}\right)
 + \nonumber 
\\
& &{\partial \over\partial\mu}\left(
(1-\mu^2)\left[{1\over r}+\mu\left({3 v\over r}+
{\partial\ln\rho\over\partial t}\right)
\right]\epsilon(1+v\mu) {f\over\rho}\right)
=
{1\over\rho}(1+v\mu)\mathbb{C}[f].\nonumber \\
& &
\label{energyConservativeBoltzmannEulerian}
\end{eqnarray}
Multiplication of this equation  
by $\epsilon^2\,d\epsilon\,d\mu\,d\varphi$
and integrating immediately yields Eq. 
(\ref{energyEulerian}) for lab frame energy ``conservation''.

We now examine the relationship between Eq. 
(\ref{numberConservativeBoltzmann}), a conservative
formulation of particle number kinetics, and Eq. 
(\ref{energyConservativeBoltzmannEulerian}), a conservative formulation of
particle energy kinetics. We multiply Eq. 
(\ref{numberConservativeBoltzmann}) by $(1+v\mu)$ and
consider what it takes to get the terms in Eq.
(\ref{energyConservativeBoltzmannEulerian}). In doing so, we
examine carefully only $O(v)$ terms.
First the time derivative terms:
\begin{equation}
\epsilon(1+v\mu){\partial \over \partial t}\left(f\over\rho\right)
=
{\partial \over \partial t}\left[\epsilon (1+v\mu) f\over\rho\right] 
+ O(v^2).
\end{equation}
Next we compare the mass derivative terms:
\begin{equation} 
\epsilon(1+v\mu){\partial\over\partial m}\left(4\pi r^2\rho\mu\, {f\over\rho}\right)
=
 {\partial\over\partial m}\left(4\pi r^2\rho \epsilon\mu(1+v\mu) 
{f\over\rho}\right) -  4\pi r^2\epsilon\mu^2\,{\partial v
\over \partial m}\,{f\over\rho}.
\end{equation}
In the notation of the previous subsection, the ``extra terms'' from 
the spacetime divergence are
\begin{equation}
E_S = 4\pi r^2\epsilon\mu^2\,{\partial v
\over \partial m}\,{f \over\rho} + O(v^2).
\label{spaceExtraComoving}
\end{equation}
Next we relate the momentum divergence terms in the number and
energy equations, looking first at the energy derivatives, 
\begin{eqnarray}
\epsilon(1+v\mu){1\over\epsilon^2}{\partial\over\partial\epsilon}
\left(\epsilon^3\left[\mu^2\left({3 v\over r} +
{\partial\ln\rho\over\partial t}\right)-{v\over r}\right]{f\over\rho}\right)
 &=&\nonumber
\\
{1\over\epsilon^2}{\partial\over\partial\epsilon}
\left(\epsilon^4(1+v\mu)\left[\mu^2\left({3 v\over r} +
{\partial\ln\rho\over\partial t}\right)-{v\over r}\right]{f\over\rho}\right)
&-&  \epsilon\left[\mu^2\left({3 v\over r} +
{\partial\ln\rho\over\partial t}\right)-{v\over r}\right]{f\over\rho}
+O(v^2),\nonumber\\
& &
\end{eqnarray}
and then at the angle derivatives:
\begin{eqnarray}
\epsilon(1+v\mu){\partial \over\partial\mu}\left(
(1-\mu^2)\left[{1\over r}+\mu\left({3 v\over r}+
{\partial\ln\rho\over\partial t}\right)
\right]{f\over\rho}\right)
&=&\nonumber
\\
{\partial \over\partial\mu}\left(
(1-\mu^2)\left[{1\over r}+\mu\left({3 v\over r}+
{\partial\ln\rho\over\partial t}\right)
\right]\epsilon(1+v\mu) {f\over\rho}\right)
&-&
{\epsilon v(1-\mu^2)\over r}\,{f\over \rho}
+ O(v^2).
\end{eqnarray}
In the notation of the previous subsection, the ``extra terms'' from 
the momentum divergence are
\begin{equation}
E_M = \epsilon\left[\mu^2\left({3 v\over r} +
{\partial\ln\rho\over\partial t}\right)-{v\over r}\right]{f\over\rho}
+
{\epsilon v(1-\mu^2)\over r}\,{f\over \rho}
+ 
O(v^2).
\label{momentumExtraComoving}
\end{equation}
Having $E_S$ and $E_M$, we are ready to verify Eq. (\ref{matching}),
which must be satisfied for consistency between the particle 
number and energy balance equations. Having specified electrically neutral
particles, and having chosen to work with the lab frame energy 
expression of Eq. (\ref{energyConservativeBoltzmannEulerian})
(in which the connection coefficients vanish), the right-hand side of
Eq. (\ref{matching}) vanishes. Employing Eq. 
(\ref{baryonConservationComoving}) for baryon conservation,
from Eqs. (\ref{spaceExtraComoving}) and (\ref{momentumExtraComoving})
it is easy to see that
\begin{equation}
E_S + E_M = 0
\label{matchingComoving}
\end{equation}
is indeed satisfied analytically. 
But in solving the ``number conservative'' Eq. 
(\ref{numberConservativeBoltzmann}) numerically, consistency with
with the ``energy conservative'' Eq. 
(\ref{energyConservativeBoltzmannEulerian}) requires that Eq. 
(\ref{matchingComoving}) be satisfied {\em numerically} as well.
Ref. \cite{liebendoerfer02} provides an example of a finite difference
representation of Eq. (\ref{numberConservativeBoltzmann}) that 
satisfies this criterion.

\section{Conclusion}
\label{sec:conclusion}

In this section we summarize our conservative formulations of kinetic theory, comment on their relation to moment formalisms, and discuss their possible
application in the core-collapse supernova environment.

Having in mind computational radiative transfer in astrophysical environments, we have sought formulations of relativistic kinetic theory with the
following properties: (1) they are expressed in terms of global,
Eulerian (or ``lab-frame'') spacetime coordinates \inlineSFinmath${\{{x^{\mu }}\}}$; (2) they are expressed
in terms of convenient three-momentum coordinates \inlineSFinmath${\big\{{u^{\overvar{i}{\RawWedge }}}\big\}}$ (e.g. spherical polar), which are taken
from the orthonormal momentum components \inlineSFinmath${\big\{{p^{\overvar{i}{\RawWedge }}}\big\}}$ measured by an observer comoving with the medium;
and (3) they are 
``conservative'', having transparent connections to total particle number and 4-momentum balance as expressed in Eqs. (\ref{numberConservation})
and (\ref{energyConservation}). 

To express our formulations having these properties, we here introduce the {\itshape specific particle number flux vector }
\begin{equation}
\dispSFNumberedEquationmath{{{\ScriptCapitalN }^{\mu }}\equiv {{{{\ScriptCapitalL }^{\mu }}\InvisibleSpace }_{\overvar{\mu }{\RawWedge }}}{p^{\overvar{\mu
}{\RawWedge }}}f}
\end{equation}
and the {\itshape specific particle stress-energy tensor}
\begin{equation}
\dispSFNumberedEquationmath{{{\ScriptCapitalT }^{\Mvariable{\mu \nu }}}\equiv {{{{\ScriptCapitalL }^{\mu }}\InvisibleSpace }_{\overvar{\mu }{\RawWedge
}}}{{{{\ScriptCapitalL }^{\nu }}\InvisibleSpace }_{\overvar{\nu }{\RawWedge }}}{p^{\overvar{\mu }{\RawWedge }}}{p^{\overvar{\nu }{\RawWedge }}}f,}
\end{equation}
where the transformation to the comoving frame \inlineSFinmath${{{{{\ScriptCapitalL }^{\overvar{\mu }{\RawWedge }}}\InvisibleSpace }_{\mu }}}$ is given
by equation (\ref{compositeTransformation}). (While the adjective ``specific'' often denotes a quantity measured per unit mass, in this context we use it to denote the particle
flux and stress-energy in a given invariant momentum space volume element.) While the distribution function \inlineSFinmath${f}$ of a given particle type of mass
\inlineSFinmath${m}$ and charge \inlineSFinmath${e}$ obeys the Boltzmann equation (Eq. (\ref{fullBoltzmann})), the specific particle number flux and stress-energy satisfy
the  conservative equations
\begin{eqnarray}
\frac{1}{{\sqrt{-g}}}\frac{\partial }{\partial {x^{\mu }}}\big({\sqrt{-g}}{{\ScriptCapitalN }^{\mu }}\multsp \big)+\nonumber \\
 E({\bf p})\Bigg|\det \big[\frac{\partial {\bf p}}{\partial {\bf u}}\big]{{\Bigg|}^{-1}}  \frac{\partial }{\partial {u^{\overvar{i}{\RawWedge }}}}\Bigg(\frac{1}{{E({\bf p})}}\Bigg|\det \big[\frac{\partial {\bf p}}{\partial
{\bf u}}\big]\Bigg|\big(e\multsp {{{F^{\overvar{j}{\RawWedge }}}\InvisibleSpace }_{\overvar{\mu }{\RawWedge }}}-{{{{\Gamma }^{\overvar{j}{\RawWedge }}}\InvisibleSpace
}_{\overvar{\mu }{\RawWedge }\overvar{\nu }{\RawWedge }}}{p^{\overvar{\nu }{\RawWedge }}}\big)\frac{\partial {u^{\overvar{i}{\RawWedge }}}}{\partial
{p^{\overvar{j}{\RawWedge }}}}{{{{\ScriptCapitalL }^{\overvar{\mu }{\RawWedge }}}\InvisibleSpace }_{\mu }}{{\ScriptCapitalN }^{\mu }}\Bigg)\IndentingNewLine
\nonumber\\ = \mathbb{C}[f], \label{conservative1} \\
\frac{1}{{\sqrt{-g}}}\frac{\partial }{\partial {x^{\nu }}}\big({\sqrt{-g}}{{\ScriptCapitalT }^{\Mvariable{\mu \nu }}}\big)+ \nonumber \\
 {E({\bf p})}\Bigg|\det \big[\frac{\partial {\bf p}}{\partial {\bf u}}\big]{{\Bigg|}^{-1}} \frac{\partial }{\partial {u^{\overvar{i}{\RawWedge }}}}\Bigg(\frac{1}{{E({\bf p})}}\Bigg|\det \big[\frac{\partial {\bf p}}{\partial
{\bf u}}\big]\Bigg|\big(e\multsp {{{F^{\overvar{j}{\RawWedge }}}\InvisibleSpace }_{\overvar{\nu }{\RawWedge }}}-{{{{\Gamma }^{\overvar{j}{\RawWedge }}}\InvisibleSpace
}_{\overvar{\nu }{\RawWedge }\overvar{\rho }{\RawWedge }}}{p^{\overvar{\rho }{\RawWedge }}}\big)\frac{\partial {u^{\overvar{i}{\RawWedge }}}}{\partial
{p^{\overvar{j}{\RawWedge }}}}{{{{\ScriptCapitalL }^{\overvar{\nu }{\RawWedge }}}\InvisibleSpace }_{\nu }}{{\ScriptCapitalT }^{\Mvariable{\mu \nu
}}}\Bigg)\IndentingNewLine \nonumber\\  = {{{F^{\mu }}\InvisibleSpace }_{\nu }}{{\ScriptCapitalN }^{\nu }}
-
{{{{\Gamma }^{\mu }}\InvisibleSpace
}_{\Mvariable{\nu \rho }}}{{\ScriptCapitalT }^{\Mvariable{\nu \rho }}}
+
{{{{\ScriptCapitalL }^{\mu }}\InvisibleSpace }_{\overvar{\mu }{\RawWedge
}}}{p^{\overvar{\mu }{\RawWedge }}}\mathbb{C}[f].\label{conservative2}
\end{eqnarray}
In stating that Eqs. (\ref{conservative1}) and (\ref{conservative2}) constitute  conservative formulations of particle kinetics, we mean that the connection to the balance equations of Eqs. (\ref{numberConservation}) and (\ref{energyConservation}) is transparent, in the following sense. We can 
use elementary calculus to form the familiar invariant momentum space volume element 
\begin{equation}
\dispSFNumberedEquationmath{\frac{{d^3}p}{E({\bf p})}=\frac{1}{{E({\bf p})}}\Bigg|\det \big[\frac{\partial {\bf p}}{\partial {\bf u}}\big]\Bigg|{d^3}u,}\label{momentumElement2}
\end{equation}
where a transformation from orthonormal momentum components \inlineSFinmath${\big\{{p^{\overvar{i}{\RawWedge }}}\big\}}$ to some other set of coordinates
(e.g. momentum space spherical coordinates \inlineSFinmath${\big\{{u^{\overvar{i}{\RawWedge }}}\big\}=\{|{\bf p}|,\vartheta ,\varphi \}}$) has been performed.
Multiplying Eqs. (\ref{conservative1}) and (\ref{conservative2}) by Eq. (\ref{momentumElement2}) and integrating, the terms with momentum space derivatives are obviously transformed into vanishing
surface terms; the results are Eqs. (\ref{numberConservation}) and (\ref{energyConservation}) for total particle number and 4-momentum balance.

In
terms of differential forms, the procedure for obtaining the  conservative formulations
of kinetic theory is straightforward. First, express the volume element \inlineSFinmath${\Omega }$ in the phase space for particles of definite mass
in terms of the desired spacetime and 3-momentum coordinates. Next, by contraction with the Liouville vector, form the hypersurface element \inlineSFinmath${\omega ={L_m}\cdot \Omega }$. Then bring the exterior derivative \inlineSFinmath${d(f\multsp \omega )}$ into the form \inlineSFinmath${\mathbb{N}
[f]\Omega }$ by direct computation; Eq. (\ref{conservative1}) results on comparison with the Boltzmann equation. This result can then be used in conjunction with
an evaluation of \inlineSFinmath${d({v_{\mu }}{p^{\mu }}f\multsp \omega )}$ for arbitrary \inlineSFinmath${{v_{\mu }}}$ to obtain Eq. (\ref{conservative2}). The
reason the procedure is straightforward is that the ``heavy lifting'' of transforming the Boltzmann equation into  conservative forms is handled by two ``levers''
of considerable power: the generalized Stokes theorem, and the key relation \inlineSFinmath${\Mvariable{d\omega }=0}$, which is closely related to
the relativistic Liouville theorem.

A concrete example of our formalism is provided in the appendix, 
which contains \inlineSFinmath{$O
(v)$} equations for the specific particle number density, specific
particle energy density, and their angular moments---all in flat spacetime, but in coordinates sufficiently general to represent rectangular, spherical, and cylindrical coordinate systems.

We now comment on the connection of Eqs. (\ref{conservative1}) and (\ref{conservative2}) to moment formalisms. In the usual treatments, if one writes the distribution function
as a function of momentum variables as measured in a given frame (lab or comoving), it is natural to form moments by multiplying the distribution
function by, for example, energies and angles measured in that frame, and integrating. Lo and behold, it turns out that these moments are number
densities and fluxes, and energy/momentum densities and fluxes: components of a particle number flux vector and stress-energy tensor, measured
in the same frame chosen to measure angles and energies. Traditional, then, are treatments in which the components of conserved tensors as measured
in a given frame are expressed as functions of momentum variables as measured in that same frame.

But this traditional approach to moments may not be the most convenient, and Eqs. (\ref{conservative1}) and (\ref{conservative2}) provide an attractive alternative. For example, Liebend\"orfer
et al. \cite{liebendoerfer01,liebendoerfer02} form moments in the traditional way, resulting in components of conserved tensors as measured in an {\em orthonormal comoving frame.} 
But it is the tensor components in the {\em lab frame} that one would like
to check, either because it is the coordinate basis (at least in a 
spatially multidimensional simulation) and therefore natural to deal with, or
because it is the basis in which energy is conserved (in a simulation in
comoving coordinates in spherical symmetry).
The required transformation of the tensor components between these frames leads to the numerical complexity mentioned in Sec. \ref{sec:introduction}, and
discussed further in the latter parts of Subsecs. 
\ref{subsec:conservativeGeneral} and \ref{subsec:conservativeComoving}.
In contrast, when the coordinate basis is in the lab frame as expected
in spatially multidimensional simulations,
{\itshape integration of Eqs. (\ref{conservative1})
and (\ref{conservative2}) over momentum variables leads directly to the tensor components in the desired coordinate basis, even though the specific particle number
and stress-energy are functions of comoving frame momentum variables.} The insight here is that the frame in which the tensor components are measured
need not be the same as that employed to obtain the momentum variables used to parametrize the particle distributions.
The appendix provides an example of a moment formalism of this kind.

In simulations of systems like core-collapse supernovae---in which careful attention to energetics is critical---a number of possible approaches, based on
the conservative formulations of kinetic theory presented in this paper, 
might be suggested.

First, the general approach of, for example, Refs.
\cite{mezzacappa01,liebendoerfer01b,liebendoerfer02} could be followed, 
in which the conservative particle number distribution equation 
(Eq. (\ref{conservative1}))
is solved, and the conservative particle
energy equation (the time component of Eq. (\ref{conservative2})) 
is used as a consistency check. The quantitiy $\sqrt{-g}\,{\cal N}^0$, which
might be called the {\em specific particle number density}, would be the
primary neutrino distribution variable: It is the contribution of each {\em comoving frame}
momentum bin to the {\em lab frame} particle number density.
This approach makes number conservation a somewhat natural outcome, 
but energy conservation would require
finite-differenced representations of various quantities to be ``matched''
in order that Eq. (\ref{matching}) be satisified numerically. This might
be considered the most rigorous and self-consistent method.
(Note that if one solves the ``plain'', non-conservative 
Boltzmann equation---Eq. 
(\ref{fullBoltzmann})---for the scalar distribution function $f$ as
a function of comoving frame momentum variables, conservation of {\em neither}
lab frame particle number or energy is straightforward. The same is true
of methods (e.g., Ref. \cite{burrows00}) based on a non-conservative form of the transport equation
for the comoving frame specific intensity.)

In order to avoid the intricate finite differencing of numerous 
terms demanded by
this method, a second option would be the
use of \inlineSFinmath$\sqrt{-g}\, {{{\ScriptCapitalT }^{00}}}$, which might be called the {\itshape specific particle energy density, }as the primary neutrino distribution
variable. Designing a code around a differenced version of the \inlineSFinmath${\mu =0}$ component
of Eq. (\ref{conservative2}) would make accurate accounting of total neutrino energy (as represented by the \inlineSFinmath${\mu =0}$ component of Eq. (\ref{energyConservation}))
relatively straightfoward.
%
%
Of course, the
neutrino number balance equation expressed in terms of \inlineSFinmath$\sqrt{-g}\,{{{\ScriptCapitalT }^{00}}}$ would have numerical errors unless certain finite differencings
were carefully designed. But with respect to the crucial energetics of the physical system, it is worth noting that there are a couple of factors
mitigating the impact of errors in number conservation in comparison with errors in energy conservation. 
Errors in number conservation translate into errors in the electron fraction \inlineSFinmath${{Y_e}}$,
which affect energy conservation through the equation of state, but: 
(1) Only \inlineSFinmath${{{\nu }_e}}$ and \inlineSFinmath${{{\left( \overvar{\nu }{\_} \right)
}_e}}$ affect \inlineSFinmath${{Y_e}}$, while all species impact the energy budget. Better to have two species contributing to error rather than six!
(2) The effects of \inlineSFinmath${{{\nu }_e}}$ and \inlineSFinmath${{{\left( \overvar{\nu }{\_} \right) }_e}}$ on \inlineSFinmath${{Y_e}}$ are opposite
in sign (unlike their contributions to energy), so that to the extent that their distributions are similar, the impact of their errors on \inlineSFinmath${{Y_e}}$ may approximately cancel.

A third possibility would be to solve for {\itshape both} the specific particle energy density \inlineSFinmath$\sqrt{-g}\,{{{\ScriptCapitalT }^{00}}}$ and 
the specific
particle number density 
\inlineSFinmath$\sqrt{-g}\,{{{\ScriptCapitalN }^0}}$. (Rampp and Janka \cite{rampp02} 
solve for
both number and energy distributions, but in the comoving frame; this limits
the utility of their approach with respect to accurate tracking of
lab frame quantities.) 
Instead of pre-defining both the boundaries and center values of bins in energy
space, one could define the boundaries only and use the values of 
$\sqrt{-g}\,{{{\ScriptCapitalT }^{00}}}$ and 
$\sqrt{-g}\,{{\ScriptCapitalN }^0}$ (along with the transformation
\inlineSFinmath${{{{{\ScriptCapitalL }^{\overvar{\mu }{\RawWedge }}}\InvisibleSpace }_{\mu }}}$) to obtain center values of the energy bins in each
spatial zone and each time step. 
The consistency of the solutions 
would be 
arguably
reasonable
as long as the derived center values of the energy bins do not wander outside the pre-defined bin boundaries. 
\appendix*

\section{Flat spacetime, $O(v)$ equations in a general coordinate
  system}

In this appendix we present an application of the  conservative formulations of kinetic theory derived in this paper: \inlineSFinmath{$O
(v)$} equations for the specific particle number density and specific particle energy density, as well as angular integrals of these equations---all
in flat spacetime, but in coordinates sufficiently general to represent rectangular, spherical, and cylindrical coordinate systems. The angle-integrated
equations constitute ``monochromatic'' moment formalisms that provide an alternative to traditional variable Eddington factor methods of handling
radiation transport. We specialize to massless, electrically neutral particles.

We begin by describing our spacetime and momentum space coordinate systems. While we assume flat spacetime, in order to accomodate curvilinear coordinate
systems we employ a general spacetime coordinate labeling \inlineSFinmath{$({x^{\mu }})={{\big({x^1},{x^2},{x^3},t\big)}^T}$}. The line element is
\begin{equation}
\dispSFNumberedEquationmath{{{\Mvariable{ds}}^2}={g_{\Mvariable{\mu \nu }}}{{\Mvariable{dx}}^{\InvisibleComma \mu }}{{\Mvariable{dx}}^{\nu }},}
\end{equation}
with 
\begin{equation}
\dispSFNumberedEquationmath{({g_{\Mvariable{\mu \nu }}})=\pmatrix{
	1&0&0&0 \cr
	0&{a^2}({x^1})&0&0 \cr
	0&0&{b^2}({x^1}){c^2}({x^2})&0 \cr
	0&0&0&-1 
  }.}\label{flatMetric}
\end{equation}
In this matrix expression for the metric components, rows and columns are ordered \inlineSFinmath${1,2,3,0}$. In Cartesian coordinates, \inlineSFinmath${\big({x^1},{x^2},{x^3}\big)=(x,y,z)}$,
and \inlineSFinmath${(a,b,c)=(1,1,1)}$. In spherical coordinates,  \inlineSFinmath${\big({x^1},{x^2},{x^3}\big)=(r,\theta ,\phi )}$, and \inlineSFinmath${(a,b,c)=(r,r,\Mvariable{\sin}\theta
)}$. In cylindrical coordinates,  \inlineSFinmath${\big({x^1},{x^2},{x^3}\big)=(r,z,\phi )}$, and \inlineSFinmath${(a,b,c)=(1,r,1)}$. Our spacetime coordinate
systems are ``lab frames,'' so that the equations we derive are Eulerian. Orthonormal ``lab frame'' coordinates, indicated by barred indices, are
obtained by the transformation
\begin{equation}
\dispSFNumberedEquationmath{{{\Mvariable{dx}}^{\overvar{\mu }{\_}}}={{{e^{\overvar{\mu }{\_}}}\InvisibleSpace }_{\mu }}{{\Mvariable{dx}}^{\mu }},}
\end{equation}
with
\begin{equation}
\dispSFNumberedEquationmath{\big({{{e^{\overvar{\mu }{\_}}}\InvisibleSpace }_{\mu }}\big)=\pmatrix{
	1&0&0&0 \cr
	0&a({x^1})&0&0 \cr
	0&0&b({x^1})c({x^2})&0 \cr
	0&0&0&1 
  }.}
\end{equation}
The transformation to an orthonormal frame comoving with the fluid, indicated by indices adorned with a hat ($\hat{}$), is
\begin{equation}
\dispSFNumberedEquationmath{{{\Mvariable{dx}}^{\overvar{\mu }{\RawWedge }}}={{{{\Lambda }^{\overvar{\mu }{\RawWedge }}}\InvisibleSpace }_{\overvar{\mu
}{\_}}}{{\Mvariable{dx}}^{\overvar{\mu }{\_}}},}
\end{equation}
where to \inlineSFinmath${O (v)}$
\begin{equation}
\dispSFNumberedEquationmath{\big({{{{\Lambda }^{\overvar{\mu }{\RawWedge }}}\InvisibleSpace }_{\overvar{\mu }{\_}}}\big)=\pmatrix{
	1&0&0&-{v_{\overvar{1}{\_}}} \cr
	0&1&0&-{v_{\overvar{2}{\_}}} \cr
	0&0&1&-{v_{\overvar{3}{\_}}} \cr
	-{v_{\overvar{1}{\_}}}&-{v_{\overvar{2}{\_}}}&-{v_{\overvar{3}{\_}}}&1 
  }.}
\end{equation}
The bars on the velocity indices are a reminder that the fluid velocity is expressed in the orthonormal lab frame coordinate system. The combined
transformation from the lab coordinate frame to the orthonormal comoving frame is 
\begin{equation}
\dispSFNumberedEquationmath{{{\Mvariable{dx}}^{\overvar{\mu }{\RawWedge }}}={{{{\ScriptCapitalL }^{\overvar{\mu }{\RawWedge }}}\InvisibleSpace }_{\mu
}}{{\Mvariable{dx}}^{\mu }},}
\end{equation}
where
\begin{equation}
\dispSFNumberedEquationmath{{{{{\ScriptCapitalL }^{\overvar{\mu }{\RawWedge }}}\InvisibleSpace }_{\mu }}={{{{\Lambda }^{\overvar{\mu }{\RawWedge
}}}\InvisibleSpace }_{\overvar{\mu }{\_}}}{{{e^{\overvar{\mu }{\_}}}\InvisibleSpace }_{\mu }}.}
\end{equation}
The neutrino 4-momentum is described in terms of its comoving frame components, \inlineSFinmath${\Big({p^{\overvar{\mu }{\RawWedge }}}\Big)=\Big({p^{\overvar{1}{\RawWedge
}}},{p^{\overvar{2}{\RawWedge }}},{p^{\overvar{3}{\RawWedge }}},{p^{\overvar{0}{\RawWedge }}}\Big)^T}$. Only three momentum variables are independent;
we choose polar coordinates in momentum space, defined by
\begin{eqnarray}
{p^{\overvar{1}{\RawWedge }}}&=&\epsilon\, \multsp \Mvariable{\cos}\vartheta ,  \\
{p^{\overvar{2}{\RawWedge }}}&=&\epsilon\, \multsp \Mvariable{\sin}\vartheta\, \Mvariable{\cos}\varphi ,\IndentingNewLine  
\\
{p^{\overvar{3}{\RawWedge }}}&=&\epsilon\, \multsp \Mvariable{\sin}\vartheta\, \Mvariable{\sin}\varphi .
\end{eqnarray}
The radiation field is a function of the variables \inlineSFinmath${t,{x^1},{x^2},{x^3},\epsilon ,\vartheta ,\varphi }$. The invariant spacetime
volume element in Eq. (\ref{spacetimeElement}) is \inlineSFinmath${a\multsp\, b\multsp\, c\multsp\, {{\Mvariable{dx}}^1}{{\Mvariable{dx}}^2}{{\Mvariable{dx}}^3}} dt$, and the invariant volume
element on the mass shell in momentum space in Eq. (\ref{momentumElement})
is \inlineSFinmath${\epsilon\, \multsp \Mvariable{\sin}\vartheta\, \Mvariable{d\epsilon }\multsp\, \Mvariable{d\vartheta
}\multsp\, \Mvariable{d\varphi }}$.

We first present an equation for the specific particle density \inlineSFinmath${\ScriptCapitalN }$, defined by
\begin{equation}
\dispSFNumberedEquationmath{\ScriptCapitalN ={p^t}f={{{{\ScriptCapitalL }^t}\InvisibleSpace }_{\overvar{\mu }{\RawWedge }}}{p^{\overvar{\mu }{\RawWedge
}}}f,}\label{specificParticleDensity}
\end{equation}
where \inlineSFinmath${f}$ is the invariant particle distribution function defined in Sec. \ref{sec:distribution}.  The specific particle density is related to the lab frame particle density
\inlineSFinmath${\ScriptN }\equiv N^t$ by
\begin{equation}
\dispSFNumberedEquationmath{\ScriptN =\int \ScriptCapitalN \multsp \epsilon \multsp\, \Mvariable{\sin}\vartheta \,\Mvariable{d\epsilon }\,\multsp \Mvariable{d\vartheta\,
}\multsp \Mvariable{d\varphi }.}
\end{equation}
From Eq. (\ref{conservative1}), we find that the specific particle density satisfies
\begin{equation}
\dispSFNumberedEquationmath{\mathbb{D}_t[\ScriptCapitalN] +{\mathbb{D}_{{x^1}}}[\ScriptCapitalN] +{\mathbb{D}_{{x^2}}}[\ScriptCapitalN] +{\mathbb{D}_{{x^3}}}[\ScriptCapitalN] +{\mathbb{D}_{\vartheta
}}[\ScriptCapitalN] +{\mathbb{D}_{\varphi }}[\ScriptCapitalN] +{\mathbb{O}_{\epsilon }}[\ScriptCapitalN] +{\mathbb{O}_{\vartheta }}[\ScriptCapitalN] +{\mathbb{O}_{\varphi }}[\ScriptCapitalN]
=\mathbb{C},}\label{appendixNumber}
\end{equation}
where \inlineSFinmath$\mathbb{C}$ is the invariant collision integral appearing in the Boltzmann equation (Eq. (\ref{fullBoltzmann})), and
\begin{eqnarray}
\dispSFNumberedEquationmath{\mathbb{D}_t}[\ScriptCapitalN]& =&\frac{\partial \ScriptCapitalN }{\partial t},\label{dndt} \\
\dispSFNumberedEquationmath{\mathbb{D}_{{x^1}}}[\ScriptCapitalN]& =&\frac{1}{a\multsp b}\multsp \frac{\partial }{\partial {x^1}}(a\multsp b\multsp ((1-c_{\vartheta
}^{2})\multsp {v_{\overvar{1}{\_ }}}+{c_{\vartheta }}\multsp (1-{s_{\vartheta }}{c_{\varphi }}\multsp {v_{\overvar{2}{\_ }}}-{s_{\vartheta
}}\multsp {s_{\varphi }}\multsp {v_{\overvar{3}{\_ }}}))\ScriptCapitalN ),\\
{\mathbb{D}_{{x^2}}}[\ScriptCapitalN]& =&\frac{1}{a\multsp c}\frac{\partial }{\partial {x^2}}(c\multsp ((1-s_{\vartheta }^{2}c_{\varphi
}^{2})\multsp {v_{\overvar{2}{\_ }}}+{c_{\varphi }}\multsp {s_{\vartheta }}\multsp (1-{c_{\vartheta }}\multsp {v_{\overvar{1}{\_ }}}-{s_{\vartheta
}}\multsp {s_{\varphi }}\multsp {v_{\overvar{3}{\_ }}}))\ScriptCapitalN ),\\
{\mathbb{D}_{{x^3}}}[\ScriptCapitalN]& =&\frac{1}{b\multsp c}\multsp \frac{\partial }{\partial {x^3}}(((1-s_{\vartheta }^{2}\multsp
s_{\varphi }^{2})\multsp {v_{\overvar{3}{\_ }}}+{s_{\vartheta }}\multsp {s_{\varphi }}\multsp (1-{c_{\vartheta }}\multsp {v_{\overvar{1}{\_
}}}-\multsp {s_{\vartheta }}\multsp {c_{\varphi }}{v_{\overvar{2}{\_ }}}))\ScriptCapitalN ),\\
{\mathbb{D}_{\vartheta }}[\ScriptCapitalN]& =&-\frac{1}{{s_{\vartheta }}}\frac{\partial }{\partial \vartheta }\Big(\Big(\frac{1}{a}\multsp
\frac{\partial a}{\partial {x^1}}\multsp s_{\vartheta }^{2}\multsp c_{\varphi }^{2}+\frac{1}{b}\multsp \frac{\partial b}{\partial {x^1}}\multsp s_{\vartheta
}^{2}\multsp s_{\varphi }^{2}\Big)\ScriptCapitalN \Big), \\
{\mathbb{D}_{\varphi }}[\ScriptCapitalN]& =&-\frac{1}{{s_{\vartheta }}}\frac{\partial }{\partial \varphi }\Big(\Big(-\frac{1}{a}\multsp
\frac{\partial a}{\partial {x^1}}\multsp {c_{\vartheta }}{s_{\vartheta }}\multsp {c_{\varphi }}\multsp {s_{\varphi }}+\frac{1}{b}\multsp \frac{\partial
b}{\partial {x^1}}\multsp {c_{\vartheta }}{s_{\vartheta }}\multsp {c_{\varphi }}\multsp {s_{\varphi }}+\frac{1}{a\multsp c}\multsp \frac{\partial
c}{\partial {x^2}}\multsp s_{\vartheta }^{2}\multsp {s_{\varphi }}\Big)\ScriptCapitalN \Big),\multsp\\
{\mathbb{O}_{\epsilon }}[\ScriptCapitalN]& =&-\frac{1}{\epsilon }\frac{\partial }{\partial \epsilon }\Big(\multsp {{\epsilon }^2}\multsp \Big(\frac{\partial {v_{\overvar{1}{\_
}}}}{\partial {x^1}}\multsp c_{\vartheta }^{2}+\frac{\partial {v_{\overvar{2}{\_ }}}}{\partial {x^1}}\multsp {c_{\vartheta }}{s_{\vartheta
}}\multsp {c_{\varphi }}+ \frac{\partial {v_{\overvar{3}{\_ }}}}{\partial {x^1}}\multsp {c_{\vartheta }}\multsp {s_{\vartheta }}\multsp {s_{\varphi }}+ \nonumber\\
& & \frac{1}{a}\Big(\frac{\partial
{v_{\overvar{1}{\_ }}}}{\partial {x^2}}\multsp {c_{\vartheta }}\multsp {s_{\vartheta }}{c_{\varphi }}\multsp \multsp +\multsp \frac{\partial
{v_{\overvar{2}{\_ }}}}{\partial {x^2}}\multsp s_{\vartheta }^{2}c_{\varphi }^{2}\multsp +\multsp \frac{\partial {v_{\overvar{3}{\_
}}}}{\partial {x^2}}\multsp s_{\vartheta }^{2}{c_{\varphi }}\multsp {s_{\varphi }}\Big)+ \nonumber \\
& &  \frac{1}{b\multsp c}\multsp \Big(\frac{\partial {v_{\overvar{1}{\_ }}}}{\partial {x^3}}\multsp {c_{\vartheta }}\multsp {s_{\vartheta
}}\multsp {s_{\varphi }}+\frac{\partial {v_{\overvar{2}{\_ }}}}{\partial {x^3}}\multsp s_{\vartheta }^{2}{c_{\varphi }}\multsp {s_{\varphi
}}+\frac{\partial {v_{\overvar{3}{\_ }}}}{\partial {x^3}}\multsp s_{\vartheta }^{2}\multsp s_{\varphi }^{2}\Big)+  \nonumber \\
& & \frac{1}{a}\multsp \frac{\partial a}{\partial {x^1}}\multsp (s_{\vartheta }^{2}c_{\varphi }^{2}\multsp {v_{\overvar{1}{\_ }}}-{c_{\vartheta
}}\multsp {s_{\vartheta }}{c_{\varphi }}\multsp {v_{\overvar{2}{\_ }}})+\nonumber \\
& &\frac{1}{b}\multsp \frac{\partial b}{\partial {x^1}}\multsp (s_{\vartheta
}^{2}\multsp s_{\varphi }^{2}\multsp {v_{\overvar{1}{\_ }}}-{c_{\vartheta }}\multsp {s_{\vartheta }}\multsp {s_{\varphi }}\multsp {v_{\overvar{3}{\_
}}})+\nonumber\\
& &  \frac{1}{a\multsp c}\multsp \frac{\partial c}{\partial {x^2}}\multsp (s_{\vartheta }^{2}\multsp s_{\varphi }^{2}\multsp {v_{\overvar{2}{\_
}}}-\multsp s_{\vartheta }^{2}{c_{\varphi }}\multsp {s_{\varphi }}\multsp {v_{\overvar{3}{\_ }}})\Big)\ScriptCapitalN \Big),
\\
{\mathbb{O}_{\vartheta }}[\ScriptCapitalN] & =&-\frac{1}{{s_{\vartheta }}}\frac{\partial }{\partial \vartheta }\Big(\Big(-\frac{\partial {v_{\overvar{1}{\_
}}}}{\partial {x^1}}\multsp {c_{\vartheta }}\multsp s_{\vartheta }^{2}+\frac{\partial {v_{\overvar{2}{\_ }}}}{\partial {x^1}}\multsp c_{\vartheta
}^{2}{s_{\vartheta }}\multsp {c_{\varphi }}+\frac{\partial {v_{\overvar{3}{\_ }}}}{\partial {x^1}}\multsp c_{\vartheta }^{2}\multsp {s_{\vartheta
}}\multsp {s_{\varphi }}+  \nonumber\\
& & \frac{1}{a}\Big(-\frac{\partial {v_{\overvar{1}{\_ }}}}{\partial {x^2}}\multsp s_{\vartheta }^{3}{c_{\varphi }}\multsp +\multsp
\frac{\partial {v_{\overvar{2}{\_ }}}}{\partial {x^2}}\multsp {c_{\vartheta }}\multsp s_{\vartheta }^{2}c_{\varphi }^{2}\multsp +\multsp \frac{\partial
{v_{\overvar{3}{\_ }}}}{\partial {x^2}}\multsp {c_{\vartheta }}\multsp s_{\vartheta }^{2}\multsp {c_{\varphi }}\multsp {s_{\varphi }}\Big)+\nonumber \\
& &  \frac{1}{b\multsp c}\Big(\multsp - \frac{\partial {v_{\overvar{1}{\_ }}}}{\partial {x^3}}\multsp s_{\vartheta }^{3}\multsp {s_{\varphi
}}+\frac{\partial {v_{\overvar{2}{\_ }}}}{\partial {x^3}}\multsp {c_{\vartheta }}s_{\vartheta }^{2}\multsp {c_{\varphi }}\multsp {s_{\varphi
}}+\frac{\partial {v_{\overvar{3}{\_ }}}}{\partial {x^3}}\multsp {c_{\vartheta }}\multsp s_{\vartheta }^{2}\multsp s_{\varphi }^{2}\Big)+
 \nonumber \\
& & \frac{1}{a}\multsp \frac{\partial a}{\partial {x^1}}\multsp ({c_{\varphi }}\multsp {s_{\vartheta }}\multsp {v_{\overvar{2}{\_
}}}+{c_{\varphi }}\multsp s_{\vartheta }^{3}\multsp s_{\varphi }^{2}\multsp {v_{\overvar{2}{\_ }}}-c_{\varphi }^{2}\multsp s_{\vartheta }^{3}\multsp
{s_{\varphi }}\multsp {v_{\overvar{3}{\_ }}})+\nonumber\\
& & \frac{1}{b}\multsp \frac{\partial b}{\partial {x^1}}\multsp (-{c_{\varphi }}\multsp s_{\vartheta }^{3}\multsp s_{\varphi }^{2}\multsp
{v_{\overvar{2}{\_ }}}+{s_{\vartheta }}\multsp {s_{\varphi }}\multsp {v_{\overvar{3}{\_ }}}+c_{\varphi }^{2}\multsp s_{\vartheta }^{3}\multsp
{s_{\varphi }}\multsp {v_{\overvar{3}{\_ }}})+\nonumber\\
& & \frac{1}{a\multsp c}\multsp \frac{\partial c}{\partial {x^2}}\multsp ({c_{\vartheta }}\multsp s_{\vartheta }^{2}\multsp s_{\varphi
}^{2}\multsp {v_{\overvar{2}{\_ }}}-{c_{\vartheta }}\multsp s_{\vartheta }^{2}{c_{\varphi }}\multsp {s_{\varphi }}\multsp {v_{\overvar{3}{\_
}}})
\Big)\ScriptCapitalN \multsp \Big),
\\
{\mathbb{O}_{\varphi }}[\ScriptCapitalN] &=&  -\frac{1}{{s_{\vartheta }}}\frac{\partial }{\partial \varphi }\Big(\Big(-\frac{\partial {v_{\overvar{2}{\_ }}}}{\partial {x^1}}\multsp
{c_{\vartheta }}\multsp {s_{\varphi }}+\frac{\partial {v_{\overvar{3}{\_ }}}}{\partial {x^1}}\multsp {c_{\vartheta }}\multsp {c_{\varphi }}+
\nonumber \\
& &\frac{1}{a}\Big(-\frac{\partial
{v_{\overvar{2}{\_ }}}}{\partial {x^2}}{s_{\vartheta }}\multsp {c_{\varphi }}\multsp {s_{\varphi }}+\multsp \frac{\partial {v_{\overvar{3}{\_
}}}}{\partial {x^2}}\multsp {s_{\vartheta }}c_{\varphi }^{2}\multsp \Big)+
\nonumber \\
& &\frac{1}{b\multsp c}  \Big(-\multsp \frac{\partial {v_{\overvar{2}{\_ }}}}{\partial {x^3}}\multsp {s_{\vartheta }}\multsp s_{\varphi }^{2}+\multsp
\frac{\partial {v_{\overvar{3}{\_ }}}}{\partial {x^3}}\multsp \multsp {s_{\vartheta }}{c_{\varphi }}\multsp {s_{\varphi }}\Big)+ \nonumber \\
& & \frac{1}{b}\multsp \frac{\partial b}{\partial {x^1}}\multsp (s_{\vartheta }^{3}{c_{\varphi }}\multsp {s_{\varphi }}\multsp {v_{\overvar{1}{\_
}}}-{c_{\vartheta }}s_{\vartheta }^{2}\multsp c_{\varphi }^{2}\multsp {s_{\varphi }}\multsp {v_{\overvar{2}{\_ }}}+c_{\vartheta }^{3}\multsp
{c_{\varphi }}\multsp {v_{\overvar{3}{\_ }}}+{c_{\vartheta }}\multsp s_{\vartheta }^{2}c_{\varphi }^{3}\multsp \multsp {v_{\overvar{3}{\_
}}})+  \nonumber \\
& & \frac{1}{a\multsp c}\multsp \frac{\partial c}{\partial {x^2}}\multsp (-{c_{\vartheta }}\multsp s_{\vartheta }^{2}\multsp {s_{\varphi
}}\multsp {v_{\overvar{1}{\_ }}}+c_{\vartheta }^{2}\multsp {s_{\vartheta }}{c_{\varphi }}\multsp {s_{\varphi }}\multsp {v_{\overvar{2}{\_
}}}+{s_{\vartheta }}\multsp {v_{\overvar{3}{\_ }}}+c_{\vartheta }^{2}\multsp {s_{\vartheta }}\multsp s_{\varphi }^{2}\multsp {v_{\overvar{3}{\_
}}})+  \nonumber\\
& & \frac{1}{a}\multsp \frac{\partial a}{\partial {x^1}}\multsp (-s_{\vartheta }^{3}{c_{\varphi }}\multsp {s_{\varphi }}\multsp {v_{\overvar{1}{\_
}}}-c_{\vartheta }^{3}\multsp {s_{\varphi }}\multsp {v_{\overvar{2}{\_ }}}-{c_{\vartheta }}s_{\vartheta }^{2}\multsp s_{\varphi }^{3}\multsp
{v_{\overvar{2}{\_ }}}+{c_{\vartheta }}\multsp s_{\vartheta }^{2}{c_{\varphi }}\multsp s_{\varphi }^{2}\multsp {v_{\overvar{3}{\_ }}})\Big)\ScriptCapitalN
\multsp \Big).\label{observerNphi}
\end{eqnarray}
In these expressions we have employed the notation \inlineSFinmath${{s_{\vartheta }}=\Mvariable{\sin}\vartheta }$, \inlineSFinmath${{c_{\vartheta }}=\Mvariable{\cos}\vartheta
}$. In accordance with the \inlineSFinmath${O(v)}$ limit, we have ignored acceleration terms and terms nonlinear in velocities and/or velocity spatial
derivatives.

Next we consider an equation for the specific particle energy density ${\cal E}$, defined by
\begin{equation}
\dispSFNumberedEquationmath{\ScriptCapitalE ={p^t}{p^t}f={{{{\ScriptCapitalL }^t}\InvisibleSpace }_{\overvar{\mu }{\RawWedge }}}{{{{\ScriptCapitalL
}^t}\InvisibleSpace }_{\overvar{\nu }{\RawWedge }}}{p^{\overvar{\mu }{\RawWedge }}}{p^{\overvar{\nu }{\RawWedge }}}f.}\label{specificEnergyDensity}
\end{equation}
The specific particle energy density is related to the lab frame particle energy density \inlineSFinmath${\ScriptE }\equiv T^{tt}$ by
\begin{equation}
\dispSFNumberedEquationmath{\ScriptE =\int \ScriptCapitalE \multsp\, \epsilon \multsp\, \Mvariable{\sin}\vartheta \,\Mvariable{d\epsilon }\multsp\, \Mvariable{d\vartheta
}\multsp\, \Mvariable{d\varphi }.}
\end{equation}
From Eq. (\ref{conservative2}), we find that the specific particle density satisfies
\begin{equation}
{\mathbb{D}_t}[\ScriptCapitalE] +{\mathbb{D}_{{x^1}}}[\ScriptCapitalE] +{\mathbb{D}_{{x^2}}}[\ScriptCapitalE] +{\mathbb{D}_{{x^3}}}[\ScriptCapitalE] +{\mathbb{D}_{\vartheta }}[\ScriptCapitalE] +{\mathbb{D}_{\varphi
}}[\ScriptCapitalE] +{\mathbb{O}_{\epsilon }}[\ScriptCapitalE] +{\mathbb{O}_{\vartheta }}[\ScriptCapitalE] +{\mathbb{O}_{\varphi }}[\ScriptCapitalE] = \epsilon \multsp (1+\multsp {v_{\overvar{1}{\_}}}{c_{\vartheta }}+{v_{\overvar{2}{\_}}}{s_{\vartheta }}{c_{\varphi }}+\multsp {v_{\overvar{3}{\_}}}{s_{\vartheta
}}{s_{\varphi }})\mathbb{C}.\label{appendixEnergy}
\end{equation}
The terms on the left-hand side are given by Eqs. (\ref{dndt})-(\ref{observerNphi}) with \inlineSFinmath${\ScriptCapitalN }$ replaced by \inlineSFinmath${\ScriptCapitalE }$.\footnote{Because the relations between $\ScriptCapitalN$, 
$\ScriptCapitalE$ and $f$ have different dependencies on velocity, expressions
of Eqs. (\ref{appendixNumber}) and (\ref{appendixEnergy}) in terms of $f$
would differ by terms of $O(v^2)$ and higher. While in principle these
discrepancies vanish to $O(v)$, a simulation solving for both 
$\ScriptCapitalN$ and $\ScriptCapitalE$ in which $v^2$ turns out to
be non-negligible compared to unity can be expected to exhibit some 
inconsistency between values of $f$ derived from Eq. 
(\ref{specificParticleDensity}) and Eq. (\ref{specificEnergyDensity}). }
The \inlineSFinmath${t}$ component of the second term of Eq. (\ref{conservative2}), which we refer to as the ``acceleration term'', vanishes for the flat spacetime
metric of Eq. (\ref{flatMetric}).

We now consider angle-integrated versions of Eqs. (\ref{appendixNumber}) and (\ref{appendixEnergy}), which constitute equations for ``monochromatic moments'' of the radiation field.
These moments are functions of the variables \inlineSFinmath${t,{x^1},{x^2},{x^3},\epsilon}$. In deriving and presenting these expressions,
it is convenient to define
\begin{eqnarray}
{n_{\overvar{1}{\RawWedge }}}&=&\Mvariable{\cos}\vartheta ,  \\
{n_{\overvar{2}{\RawWedge }}}&=&\Mvariable{\sin}\vartheta \,\Mvariable{\cos}\varphi ,\IndentingNewLine   \\
{n_{\overvar{3}{\RawWedge }}}&=&\Mvariable{\sin}\vartheta \,\Mvariable{\sin}\varphi ,
\end{eqnarray}
the orthonormal comoving frame components of the unit 3-vector  specifying the particle direction.  In accordance with the \inlineSFinmath${O(v)}$
limit, we ignore acceleration terms and terms nonlinear in velocities and/or velocity spatial derivatives.

Our first equation of this type dictates the evolution of the lab frame monochromatic particle density \inlineSFinmath${\mathfrak{N}}$, defined by
\begin{equation}
\dispSFNumberedEquationmath{\mathfrak{N}=\int \ScriptCapitalN \multsp \Mvariable{\sin}\vartheta \,\Mvariable{d\vartheta }\multsp\, \Mvariable{d\varphi }.}
\end{equation}
It is related to the lab frame particle density \inlineSFinmath${\ScriptN }\equiv N^t$ by
\begin{equation}
\dispSFNumberedEquationmath{\ScriptN =\int \mathfrak{N}\multsp\, \epsilon\, \multsp \Mvariable{d\epsilon }\multsp ,}
\end{equation}
and its evolution is governed by
\begin{equation}
\dispSFNumberedEquationmath{{\mathbb{D}_t}[\mathfrak{N}]+{\mathbb{D}_{{x^1}}}[\mathfrak{N}]+{\mathbb{D}_{{x^2}}}[\mathfrak{N}]+{\mathbb{D}_{{x^3}}}[\mathfrak{N}]+{\mathbb{O}_{\epsilon }}[\mathfrak{N}]=\int \mathbb{C}\multsp\, \Mvariable{\sin}\vartheta \,\multsp \Mvariable{d\vartheta
}\multsp\, \Mvariable{d\varphi },}
\end{equation}
where
\begin{eqnarray}
{\mathbb{D}_t}[\mathfrak{N}]&=&\frac{\partial \mathfrak{N}}{\partial t},
\\
{\mathbb{D}_{{x^1}}}[\mathfrak{N}]&=&\frac{1}{a\multsp b}\frac{\partial }{\partial {x^1}}(a\multsp b\multsp ({f_{\overvar{1}{\RawWedge }}}+{v_{\overvar{1}{\_}}}-{f_{\overvar{1}{\RawWedge
}\multsp \overvar{1}{\RawWedge }}}\multsp {v_{\overvar{1}{\_}}}-{f_{\overvar{1}{\RawWedge }\multsp \overvar{2}{\RawWedge }}}\multsp {v_{\overvar{2}{\_}}}-{f_{\overvar{1}{\RawWedge
}\multsp \overvar{3}{\RawWedge }}}\multsp {v_{\overvar{3}{\_}}})\mathfrak{N}),
\\
{\mathbb{D}_{{x^2}}}[\mathfrak{N}]&=&\frac{1}{a\multsp c}\multsp \frac{\partial }{\partial {x^2}}(c\multsp ({f_{\overvar{2}{\RawWedge }}}+{v_{\overvar{2}{\_}}}-{f_{\overvar{1}{\RawWedge
}\multsp \overvar{2}{\RawWedge }}}\multsp {v_{\overvar{1}{\_}}}-{f_{\overvar{2}{\RawWedge }\multsp \overvar{2}{\RawWedge }}}\multsp {v_{\overvar{2}{\_}}}-{f_{\overvar{2}{\RawWedge
}\multsp \overvar{3}{\RawWedge }}}\multsp {v_{\overvar{3}{\_}}})\mathfrak{N}),
\\
{\mathbb{D}_{{x^3}}}[\mathfrak{N}]&=&\frac{1}{b\multsp c}\multsp \frac{\partial }{\partial {x^3}}(({f_{\overvar{3}{\RawWedge }}}+{v_{\overvar{3}{\_}}}-{f_{\overvar{1}{\RawWedge
}\multsp \overvar{3}{\RawWedge }}}\multsp {v_{\overvar{1}{\_}}}-{f_{\overvar{2}{\RawWedge }\multsp \overvar{3}{\RawWedge }}}\multsp {v_{\overvar{2}{\_}}}-{f_{\overvar{3}{\RawWedge
}\multsp \overvar{3}{\RawWedge }}}\multsp {v_{\overvar{3}{\_}}})\mathfrak{N}),
\\
{\mathbb{O}_{\epsilon }}[\mathfrak{N}]&=&-\frac{1}{\epsilon }\frac{\partial }{\partial \epsilon } \Big({{\epsilon }^2}\multsp \Big(\frac{\partial {v_{\overvar{1}{\_}}}}{\partial {x^1}}\multsp {f_{\overvar{1}{\RawWedge }\multsp \overvar{1}{\RawWedge
}}}+\frac{\partial {v_{\overvar{2}{\_}}}}{\partial {x^1}}\multsp {f_{\overvar{1}{\RawWedge }\multsp \overvar{2}{\RawWedge }}}+\frac{\partial {v_{\overvar{3}{\_}}}}{\partial
{x^1}}\multsp {f_{\overvar{1}{\RawWedge }\multsp \overvar{3}{\RawWedge }}}+
\nonumber \\
& & \frac{1}{a}\multsp \Big(\frac{\partial {v_{\overvar{1}{\_}}}}{\partial
{x^2}}\multsp {f_{\overvar{1}{\RawWedge }\multsp \overvar{2}{\RawWedge }}}+\frac{\partial {v_{\overvar{2}{\_}}}}{\partial {x^2}}\multsp {f_{\overvar{2}{\RawWedge
}\multsp \overvar{2}{\RawWedge }}}+\frac{\partial {v_{\overvar{3}{\_}}}}{\partial {x^2}}\multsp {f_{\overvar{2}{\RawWedge }\multsp \overvar{3}{\RawWedge
}}}\Big)+  \nonumber \\
& & \frac{1}{b\multsp c}\multsp \Big(\frac{\partial {v_{\overvar{1}{\_}}}}{\partial {x^3}}\multsp {f_{\overvar{1}{\RawWedge }\multsp \overvar{3}{\RawWedge
}}}+\frac{\partial {v_{\overvar{2}{\_}}}}{\partial {x^3}}\multsp {f_{\overvar{2}{\RawWedge }\multsp \overvar{3}{\RawWedge }}}+\frac{\partial {v_{\overvar{3}{\_}}}}{\partial
{x^3}}\multsp {f_{\overvar{3}{\RawWedge }\multsp \overvar{3}{\RawWedge }}}\Big)+\nonumber \\
& &\frac{1}{a}\frac{\partial a}{\partial {x^1}}\multsp ({f_{\overvar{2}{\RawWedge
}\multsp \overvar{2}{\RawWedge }}}\multsp {v_{\overvar{1}{\_}}}-{f_{\overvar{1}{\RawWedge }\multsp \overvar{2}{\RawWedge }}}\multsp {v_{\overvar{2}{\_}}})+
\nonumber \\
& & \frac{1}{b}\multsp \frac{\partial b}{\partial {x^1}}\multsp ({f_{\overvar{3}{\RawWedge }\multsp \overvar{3}{\RawWedge }}}\multsp {v_{\overvar{1}{\_}}}-{f_{\overvar{1}{\RawWedge
}\multsp \overvar{3}{\RawWedge }}}\multsp {v_{\overvar{3}{\_}}})+
\nonumber \\
& &\frac{1}{a\multsp c}\multsp \frac{\partial c}{\partial {x^2}}\multsp ({f_{\overvar{3}{\RawWedge
}\multsp \overvar{3}{\RawWedge }}}\multsp {v_{\overvar{2}{\_}}}-{f_{\overvar{2}{\RawWedge }\multsp \overvar{3}{\RawWedge }}}\multsp {v_{\overvar{3}{\_}}})\Big)\mathfrak{N}\Big).
\end{eqnarray}
In these expressions, we have defined
\begin{eqnarray}
{f_{\overvar{i}{\RawWedge }}}&\equiv& 
\frac{1}{\mathfrak{N}}\int \ScriptCapitalN\, \multsp {n_{\overvar{i}{\RawWedge }}}\,\Mvariable{\sin}\vartheta
\Mvariable\,{d\vartheta }\multsp\, \Mvariable{d\varphi },\nonumber \\
{f_{\overvar{i}{\RawWedge }\overvar{j}{\RawWedge }}}&\equiv& 
\frac{1}{\mathfrak{N}}\int \ScriptCapitalN\, \multsp {n_{\overvar{i}{\RawWedge
}}}{n_{\overvar{j}{\RawWedge }}}\,\Mvariable{\sin}\vartheta \,\Mvariable{d\vartheta }\multsp\, \Mvariable{d\varphi },
\end{eqnarray}
which are akin to Eddington factors in traditional moment approaches to radiation transport.

Finally, we present a set of coupled equations for the lab frame monochromatic particle energy density \inlineSFinmath${\mathfrak{E}}$ and orthonormal lab frame monochromatic particle momentum
density \inlineSFinmath${{\mathfrak{P}^{\overvar{i}{\_}}}}$, defined by
\begin{eqnarray}
\mathfrak{E}&=&\int \ScriptCapitalE \multsp\, \Mvariable{\sin}\vartheta \,\Mvariable{d\vartheta }\multsp\, \Mvariable{d\varphi }, \\
{\mathfrak{P}^{\overvar{i}{\_}}}&=&{{{e^{\overvar{i}{\_}}}\InvisibleSpace }_{\mu }}\int f\multsp\, {p^t}{p^{\mu }}\,\Mvariable{\sin}\vartheta
\Mvariable{d\vartheta }\,\multsp \Mvariable{d\varphi }.
\end{eqnarray}
They are related to the lab frame particle energy density \inlineSFinmath${\ScriptE }\equiv T^{tt}$ and orthonormal lab frame particle momentum density \inlineSFinmath${{{\ScriptP
}^{\overvar{i}{\_}}}}\equiv {{{e^{\overvar{i}{\_}}}\InvisibleSpace }_{\mu }} 
T^{t\mu}$ by 
\begin{eqnarray}
\ScriptE& =&\int \mathfrak{E}\,\multsp \epsilon\, \multsp \Mvariable{d\epsilon }\multsp ,\\
{{\ScriptP }^{\overvar{i}{\_}}}&=&\int {\mathfrak{P}^{\overvar{i}{\_}}}\,\epsilon\, \multsp \Mvariable{d\epsilon }\multsp .
\end{eqnarray}
The evolution of the lab frame monochromatic particle energy density is governed by
\begin{equation}
\dispSFNumberedEquationmath{{\mathbb{D}_t}[\mathfrak{E}]+{\mathbb{D}_{{x^1}}}[\mathfrak{E}]+{\mathbb{D}_{{x^2}}}[\mathfrak{E}]+{\mathbb{D}_{{x^3}}}[\mathfrak{E}]+{\mathbb{O}_{\epsilon }}[\mathfrak{E}]=\epsilon \int \mathbb{C}\multsp (1+{v_{\overvar{1}{\_}}}{n_{\overvar{1}{\RawWedge
}}}+{v_{\overvar{2}{\_}}}{n_{\overvar{2}{\RawWedge }}}+{v_{\overvar{3}{\_}}}{n_{\overvar{3}{\RawWedge }}})\Mvariable{\sin}\vartheta \multsp\, \Mvariable{d\vartheta\,
}\multsp \Mvariable{d\varphi },}
\end{equation}
where
\begin{eqnarray}
{\mathbb{D}_t}[\mathfrak{E}]&=&\frac{\partial \mathfrak{E}}{\partial t},\\
{\mathbb{D}_{{x^1}}}[\mathfrak{E}]&=&\frac{1}{a\multsp b}\multsp \frac{\partial }{\partial {x^1}}\big(a\multsp b\multsp {\mathfrak{P}^{\overvar{1}{\_}}}\big),\\
{\mathbb{D}_{{x^2}}}[\mathfrak{E}]&=&\frac{1}{a\multsp c}\multsp \frac{\partial }{\partial {x^2}}\big(c\multsp {\mathfrak{P}^{\overvar{2}{\_}}}\big),\\
{\mathbb{D}_{{x^3}}}[\mathfrak{E}]&=&\frac{1}{b\multsp c}\multsp \frac{\partial {\mathfrak{P}^{\overvar{3}{\_}}}}{\partial {x^3}},\\
{\mathbb{O}_{\epsilon }}[\mathfrak{E}]&=&-\frac{1}{\epsilon }\frac{\partial }{\partial \epsilon } \Big(\multsp {{\epsilon }^2}\multsp \Big(\frac{\partial {v_{\overvar{1}{\_}}}}{\partial {x^1}}\multsp {h_{\overvar{1}{\RawWedge }\multsp
\overvar{1}{\RawWedge }}}+\frac{\partial {v_{\overvar{2}{\_}}}}{\partial {x^1}}\multsp {h_{\overvar{1}{\RawWedge }\multsp \overvar{2}{\RawWedge }}}+\frac{\partial
{v_{\overvar{3}{\_}}}}{\partial {x^1}}\multsp {h_{\overvar{1}{\RawWedge }\multsp \overvar{3}{\RawWedge }}}+\nonumber 
\\
& &\frac{1}{a}\multsp \Big(\frac{\partial
{v_{\overvar{1}{\_}}}}{\partial {x^2}}\multsp {h_{\overvar{1}{\RawWedge }\multsp \overvar{2}{\RawWedge }}}+\frac{\partial {v_{\overvar{2}{\_}}}}{\partial
{x^2}}\multsp {h_{\overvar{2}{\RawWedge }\multsp \overvar{2}{\RawWedge }}}+\frac{\partial {v_{\overvar{3}{\_}}}}{\partial {x^2}}\multsp {h_{\overvar{2}{\RawWedge
}\multsp \overvar{3}{\RawWedge }}}\Big)+  \nonumber\\
& & \frac{1}{b\multsp c}\multsp \Big(\frac{\partial {v_{\overvar{1}{\_}}}}{\partial {x^3}}\multsp {h_{\overvar{1}{\RawWedge }\multsp \overvar{3}{\RawWedge
}}}+\frac{\partial {v_{\overvar{2}{\_}}}}{\partial {x^3}}\multsp {h_{\overvar{2}{\RawWedge }\multsp \overvar{3}{\RawWedge }}}+\frac{\partial {v_{\overvar{3}{\_}}}}{\partial
{x^3}}\multsp {h_{\overvar{3}{\RawWedge }\multsp \overvar{3}{\RawWedge }}}\Big)+\nonumber
\\
& &\frac{1}{a}\frac{\partial a}{\partial {x^1}}\multsp ({h_{\overvar{2}{\RawWedge
}\multsp \overvar{2}{\RawWedge }}}\multsp {v_{\overvar{1}{\_}}}-{h_{\overvar{1}{\RawWedge }\multsp \overvar{2}{\RawWedge }}}\multsp {v_{\overvar{2}{\_}}})+
\nonumber \\
& & \frac{1}{b}\multsp \frac{\partial b}{\partial {x^1}}\multsp ({h_{\overvar{3}{\RawWedge }\multsp \overvar{3}{\RawWedge }}}\multsp {v_{\overvar{1}{\_}}}-{h_{\overvar{1}{\RawWedge
}\multsp \overvar{3}{\RawWedge }}}\multsp {v_{\overvar{3}{\_}}})+\nonumber
\\
& &\frac{1}{a\multsp c}\multsp \frac{\partial c}{\partial {x^2}}\multsp ({h_{\overvar{3}{\RawWedge
}\multsp \overvar{3}{\RawWedge }}}\multsp {v_{\overvar{2}{\_}}}-{h_{\overvar{2}{\RawWedge }\multsp \overvar{3}{\RawWedge }}}\multsp {v_{\overvar{3}{\_}}})\Big)\mathfrak{E}\Big).
\end{eqnarray}
The evolution of the first component of the orthonormal lab frame monochromatic particle momentum density is governed by
\begin{equation}
{\mathbb{D}_t}[{\mathfrak{P}^{\overvar{1}{\_}}}]+{\mathbb{D}_{{x^1}}}[{\mathfrak{P}^{\overvar{1}{\_}}}]+{\mathbb{D}_{{x^2}}}[{\mathfrak{P}^{\overvar{1}{\_}}}]+{\mathbb{D}_{{x^3}}}[{\mathfrak{P}^{\overvar{1}{\_}}}]+{\mathbb{F}_{22}}[{\mathfrak{P}^{\overvar{1}{\_}}}]+{\mathbb{F}_{33}}[{\mathfrak{P}^{\overvar{1}{\_}}}]+{\mathbb{O}_{\epsilon
}}[{\mathfrak{P}^{\overvar{1}{\_}}}]= 
 \epsilon \int \mathbb{C}\multsp\, ({n_{\overvar{1}{\RawWedge }}}+{v_{\overvar{1}{\_}}})\,\Mvariable{\sin}\vartheta \,\multsp \Mvariable{d\vartheta
}\,\multsp \Mvariable{d\varphi },
\end{equation}
where
\begin{eqnarray}
{\mathbb{D}_t}[{\mathfrak{P}^{\overvar{1}{\_}}}]&=&\frac{\partial {\mathfrak{P}^{\overvar{1}{\_}}}}{\partial t},
\\
{\mathbb{D}_{{x^1}}}[{\mathfrak{P}^{\overvar{1}{\_}}}]&=&\frac{1}{a\multsp b}\multsp \frac{\partial }{\partial {x^1}}\big(a\multsp b\multsp \big(2\multsp
\multsp {v_{\overvar{1}{\_}}}{\mathfrak{P}^{\overvar{1}{\_}}}+\multsp ({h_{\overvar{1}{\RawWedge }\multsp \overvar{1}{\RawWedge }}}-2\multsp ({h_{\overvar{1}{\RawWedge
}\multsp \overvar{1}{\RawWedge }\multsp \overvar{1}{\RawWedge }}}\multsp {v_{\overvar{1}{\_}}}+{h_{\overvar{1}{\RawWedge }\multsp \overvar{1}{\RawWedge
}\multsp \overvar{2}{\RawWedge }}}\multsp {v_{\overvar{2}{\_}}}+{h_{\overvar{1}{\RawWedge }\multsp \overvar{1}{\RawWedge }\multsp \overvar{3}{\RawWedge
}}}\multsp {v_{\overvar{3}{\_}}}))\mathfrak{E}\big)\big),
\\
{\mathbb{D}_{{x^2}}}[{\mathfrak{P}^{\overvar{1}{\_}}}]&=& 
 \frac{1}{a\multsp c}\multsp \frac{\partial }{\partial {x^2}}\big(c\multsp \big({\mathfrak{P}^{\overvar{2}{\_}}}\multsp {v_{\overvar{1}{\_}}}+{\mathfrak{P}^{\overvar{1}{\_}}}\multsp
{v_{\overvar{2}{\_}}}+({h_{\overvar{1}{\RawWedge }\multsp \overvar{2}{\RawWedge }}}-2\multsp ({h_{\overvar{1}{\RawWedge }\multsp \overvar{1}{\RawWedge
}\multsp \overvar{2}{\RawWedge }}}\multsp {v_{\overvar{1}{\_}}}+{h_{\overvar{1}{\RawWedge }\multsp \overvar{2}{\RawWedge }\multsp \overvar{2}{\RawWedge
}}}\multsp {v_{\overvar{2}{\_}}}+{h_{\overvar{1}{\RawWedge }\multsp \overvar{2}{\RawWedge }\multsp \overvar{3}{\RawWedge }}}\multsp {v_{\overvar{3}{\_}}}))\mathfrak{E}\big)\big),\\
{\mathbb{D}_{{x^3}}}[{\mathfrak{P}^{\overvar{1}{\_}}}]&=&\frac{1}{b\multsp c}\multsp \frac{\partial }{\partial {x^3}}\big({\mathfrak{P}^{\overvar{3}{\_}}}\multsp
{v_{\overvar{1}{\_}}}+{\mathfrak{P}^{\overvar{1}{\_}}}\multsp {v_{\overvar{3}{\_}}}+({h_{\overvar{1}{\RawWedge }\multsp \overvar{3}{\RawWedge }}}-2\multsp ({h_{\overvar{1}{\RawWedge
}\multsp \overvar{1}{\RawWedge }\multsp \overvar{3}{\RawWedge }}}\multsp {v_{\overvar{1}{\_}}}+{h_{\overvar{1}{\RawWedge }\multsp \overvar{2}{\RawWedge
}\multsp \overvar{3}{\RawWedge }}}\multsp {v_{\overvar{2}{\_}}}+{h_{\overvar{1}{\RawWedge }\multsp \overvar{3}{\RawWedge }\multsp \overvar{3}{\RawWedge
}}}\multsp {v_{\overvar{3}{\_}}}))\mathfrak{E}\big),
\\
{\mathbb{F}_{22}}[{\mathfrak{P}^{\overvar{1}{\_}}}]&=&-\frac{1}{a}\multsp \frac{\partial a}{\partial {x^1}}\multsp \big(2\multsp {\mathfrak{P}^{\overvar{2}{\_}}}\multsp
{v_{\overvar{2}{\_}}}+({h_{\overvar{2}{\RawWedge }\multsp \overvar{2}{\RawWedge }}}-2\multsp ({h_{\overvar{1}{\RawWedge }\multsp \overvar{2}{\RawWedge
}\multsp \overvar{2}{\RawWedge }}}\multsp {v_{\overvar{1}{\_}}}+{h_{\overvar{2}{\RawWedge }\multsp \overvar{2}{\RawWedge }\multsp \overvar{2}{\RawWedge
}}}\multsp {v_{\overvar{2}{\_}}}+{h_{\overvar{2}{\RawWedge }\multsp \overvar{2}{\RawWedge }\multsp \overvar{3}{\RawWedge }}}\multsp {v_{\overvar{3}{\_}}}))\mathfrak{E}\big),
\\
{\mathbb{F}_{33}}[{\mathfrak{P}^{\overvar{1}{\_}}}]&=&-\frac{1}{b}\multsp \frac{\partial b}{\partial {x^1}}\multsp \big(2\multsp {\mathfrak{P}^{\overvar{3}{\_}}}\multsp
{v_{\overvar{3}{\_}}}+({h_{\overvar{3}{\RawWedge }\multsp \overvar{3}{\RawWedge }}}-2\multsp ({h_{\overvar{1}{\RawWedge }\multsp \overvar{3}{\RawWedge
}\multsp \overvar{3}{\RawWedge }}}\multsp {v_{\overvar{1}{\_}}}+{h_{\overvar{2}{\RawWedge }\multsp \overvar{3}{\RawWedge }\multsp \overvar{3}{\RawWedge
}}}\multsp {v_{\overvar{2}{\_}}}+{h_{\overvar{3}{\RawWedge }\multsp \overvar{3}{\RawWedge }\multsp \overvar{3}{\RawWedge }}}\multsp {v_{\overvar{3}{\_}}}))\mathfrak{E}\big),
\\
{\mathbb{O}_{\epsilon }}[{\mathfrak{P}^{\overvar{1}{\_}}}]&=&  -\frac{1}{\epsilon }\frac{\partial }{\partial \epsilon }\Big({{\epsilon }^2}\multsp \Big(\frac{\partial {v_{\overvar{1}{\_}}}}{\partial
{x^1}}\multsp {h_{\overvar{1}{\RawWedge }\multsp \overvar{1}{\RawWedge }\multsp \overvar{1}{\RawWedge }}}+\frac{\partial {v_{\overvar{2}{\_}}}}{\partial
{x^1}}\multsp {h_{\overvar{1}{\RawWedge }\multsp \overvar{1}{\RawWedge }\multsp \overvar{2}{\RawWedge }}}+\frac{\partial {v_{\overvar{3}{\_}}}}{\partial
{x^1}}\multsp {h_{\overvar{1}{\RawWedge }\multsp \overvar{1}{\RawWedge }\multsp \overvar{3}{\RawWedge }}}+\nonumber
\\
& &\frac{1}{a}\multsp \Big(\frac{\partial
{v_{\overvar{1}{\_}}}}{\partial {x^2}}\multsp {h_{\overvar{1}{\RawWedge }\multsp \overvar{1}{\RawWedge }\multsp \overvar{2}{\RawWedge }}}+\frac{\partial
{v_{\overvar{2}{\_}}}}{\partial {x^2}}\multsp {h_{\overvar{1}{\RawWedge }\multsp \overvar{2}{\RawWedge }\multsp \overvar{2}{\RawWedge }}}+  \frac{\partial {v_{\overvar{3}{\_}}}}{\partial {x^2}}\multsp {h_{\overvar{1}{\RawWedge }\multsp \overvar{2}{\RawWedge }\multsp \overvar{3}{\RawWedge
}}}\Big)+\nonumber
\\
& &\frac{1}{b\multsp c}\multsp \Big(\frac{\partial {v_{\overvar{1}{\_}}}}{\partial {x^3}}\multsp {h_{\overvar{1}{\RawWedge }\multsp \overvar{1}{\RawWedge
}\multsp \overvar{3}{\RawWedge }}}+\frac{\partial {v_{\overvar{2}{\_}}}}{\partial {x^3}}\multsp {h_{\overvar{1}{\RawWedge }\multsp \overvar{2}{\RawWedge
}\multsp \overvar{3}{\RawWedge }}}+\frac{\partial {v_{\overvar{3}{\_}}}}{\partial {x^3}}\multsp {h_{\overvar{1}{\RawWedge }\multsp \overvar{3}{\RawWedge
}\multsp \overvar{3}{\RawWedge }}}\Big)+ \nonumber \\
& & \frac{1}{a}\frac{\partial a}{\partial {x^1}}\multsp ({h_{\overvar{1}{\RawWedge }\multsp \overvar{2}{\RawWedge }\multsp \overvar{2}{\RawWedge
}}}\multsp {v_{\overvar{1}{\_}}}-{h_{\overvar{1}{\RawWedge }\multsp \overvar{1}{\RawWedge }\multsp \overvar{2}{\RawWedge }}}\multsp {v_{\overvar{2}{\_}}})+
\nonumber
\\
& & \frac{1}{b}\multsp
\frac{\partial b}{\partial {x^1}}\multsp ({h_{\overvar{1}{\RawWedge }\multsp \overvar{3}{\RawWedge }\multsp \overvar{3}{\RawWedge }}}\multsp {v_{\overvar{1}{\_}}}-{h_{\overvar{1}{\RawWedge
}\multsp \overvar{1}{\RawWedge }\multsp \overvar{3}{\RawWedge }}}\multsp {v_{\overvar{3}{\_}}})+  \nonumber \\
& & \frac{1}{a\multsp c}\multsp \frac{\partial c}{\partial {x^2}}\multsp ({h_{\overvar{1}{\RawWedge }\multsp \overvar{3}{\RawWedge }\multsp
\overvar{3}{\RawWedge }}}\multsp {v_{\overvar{2}{\_}}}-{h_{\overvar{1}{\RawWedge }\multsp \overvar{2}{\RawWedge }\multsp \overvar{3}{\RawWedge }}}\multsp
{v_{\overvar{3}{\_}}})\Big)\mathfrak{E}\Big).
\end{eqnarray}
The evolution of the second component of the orthonormal lab frame monochromatic particle momentum density is governed by
\begin{equation}
{\mathbb{D}_t}[{\mathfrak{P}^{\overvar{2}{\_}}}]+{\mathbb{D}_{{x^1}}}[{\mathfrak{P}^{\overvar{2}{\_}}}]+{\mathbb{D}_{{x^2}}}[{\mathfrak{P}^{\overvar{2}{\_}}}]+{\mathbb{D}_{{x^3}}}[{\mathfrak{P}^{\overvar{2}{\_}}}]+{\mathbb{F}_{12}}[{\mathfrak{P}^{\overvar{2}{\_}}}]+{\mathbb{F}_{33}}[{\mathfrak{P}^{\overvar{2}{\_}}}]+{\mathbb{O}_{\epsilon
}}[{\mathfrak{P}^{\overvar{2}{\_}}}]=  
 \epsilon \int \mathbb{C}\multsp\, ({n_{\overvar{2}{\RawWedge }}}+{v_{\overvar{2}{\_}}})\,\Mvariable{\sin}\vartheta \,\multsp \Mvariable{d\vartheta
}\multsp\, \Mvariable{d\varphi },
\end{equation}
where
\begin{eqnarray}
{\mathbb{D}_t}[{\mathfrak{P}^{\overvar{2}{\_}}}]&=&\frac{\partial {\mathfrak{P}^{\overvar{2}{\_}}}}{\partial t},\\
{\mathbb{D}_{{x^1}}}[{\mathfrak{P}^{\overvar{2}{\_}}}]&=&  \frac{1}{a\multsp b}\multsp \frac{\partial }{\partial {x^1}}\big(a\multsp b\multsp \big({\mathfrak{P}^{\overvar{2}{\_}}}\multsp {v_{\overvar{1}{\_}}}+{\mathfrak{P}^{\overvar{1}{\_}}}\multsp
{v_{\overvar{2}{\_}}}+({h_{\overvar{1}{\RawWedge }\multsp \overvar{2}{\RawWedge }}}-2\multsp ({h_{\overvar{1}{\RawWedge }\multsp \overvar{1}{\RawWedge
}\multsp \overvar{2}{\RawWedge }}}\multsp {v_{\overvar{1}{\_}}}+{h_{\overvar{1}{\RawWedge }\multsp \overvar{2}{\RawWedge }\multsp \overvar{2}{\RawWedge
}}}\multsp {v_{\overvar{2}{\_}}}+{h_{\overvar{1}{\RawWedge }\multsp \overvar{2}{\RawWedge }\multsp \overvar{3}{\RawWedge }}}\multsp {v_{\overvar{3}{\_}}}))\mathfrak{E}\big)\big),\\
{\mathbb{D}_{{x^2}}}[{\mathfrak{P}^{\overvar{2}{\_}}}]&=&\frac{1}{a\multsp c}\multsp \frac{\partial }{\partial {x^2}}\big(c\multsp \big(2\multsp
{\mathfrak{P}^{\overvar{2}{\_}}}\multsp {v_{\overvar{2}{\_}}}+({h_{\overvar{2}{\RawWedge }\multsp \overvar{2}{\RawWedge }}}-2\multsp ({h_{\overvar{1}{\RawWedge
}\multsp \overvar{2}{\RawWedge }\multsp \overvar{2}{\RawWedge }}}\multsp {v_{\overvar{1}{\_}}}+{h_{\overvar{2}{\RawWedge }\multsp \overvar{2}{\RawWedge
}\multsp \overvar{2}{\RawWedge }}}\multsp {v_{\overvar{2}{\_}}}+{h_{\overvar{2}{\RawWedge }\multsp \overvar{2}{\RawWedge }\multsp \overvar{3}{\RawWedge
}}}\multsp {v_{\overvar{3}{\_}}}))\mathfrak{E}\big)\big),\\
{\mathbb{D}_{{x^3}}}[{\mathfrak{P}^{\overvar{2}{\_}}}]&=&\frac{1}{b\multsp c}\multsp \frac{\partial }{\partial {x^3}}\big({\mathfrak{P}^{\overvar{3}{\_}}}\multsp
{v_{\overvar{2}{\_}}}+{\mathfrak{P}^{\overvar{2}{\_}}}\multsp {v_{\overvar{3}{\_}}}+({h_{\overvar{2}{\RawWedge }\multsp \overvar{3}{\RawWedge }}}-2\multsp ({h_{\overvar{1}{\RawWedge
}\multsp \overvar{2}{\RawWedge }\multsp \overvar{3}{\RawWedge }}}\multsp {v_{\overvar{1}{\_}}}+{h_{\overvar{2}{\RawWedge }\multsp \overvar{2}{\RawWedge
}\multsp \overvar{3}{\RawWedge }}}\multsp {v_{\overvar{2}{\_}}}+{h_{\overvar{2}{\RawWedge }\multsp \overvar{3}{\RawWedge }\multsp \overvar{3}{\RawWedge
}}}\multsp {v_{\overvar{3}{\_}}}))\mathfrak{E}\big),
\\
{\mathbb{F}_{12}}[{\mathfrak{P}^{\overvar{2}{\_}}}]&=&\frac{1}{a}\multsp \frac{\partial a}{\partial {x^1}}\multsp \big({\mathfrak{P}^{\overvar{2}{\_}}}\multsp
{v_{\overvar{1}{\_}}}+{\mathfrak{P}^{\overvar{1}{\_}}}\multsp {v_{\overvar{2}{\_}}}+({h_{\overvar{1}{\RawWedge }\multsp \overvar{2}{\RawWedge }}}-2\multsp ({h_{\overvar{1}{\RawWedge
}\multsp \overvar{1}{\RawWedge }\multsp \overvar{2}{\RawWedge }}}\multsp {v_{\overvar{1}{\_}}}+{h_{\overvar{1}{\RawWedge }\multsp \overvar{2}{\RawWedge
}\multsp \overvar{2}{\RawWedge }}}\multsp {v_{\overvar{2}{\_}}}+{h_{\overvar{1}{\RawWedge }\multsp \overvar{2}{\RawWedge }\multsp \overvar{3}{\RawWedge
}}}\multsp {v_{\overvar{3}{\_}}}))\mathfrak{E}\big),
\\
{\mathbb{F}_{33}}[{\mathfrak{P}^{\overvar{2}{\_}}}]&=&-\frac{1}{a\multsp c}\multsp \frac{\partial c}{\partial {x^2}}\multsp \big(2\multsp {\mathfrak{P}^{\overvar{3}{\_}}}\multsp
{v_{\overvar{3}{\_}}}+({h_{\overvar{3}{\RawWedge }\multsp \overvar{3}{\RawWedge }}}-2\multsp ({h_{\overvar{1}{\RawWedge }\multsp \overvar{3}{\RawWedge
}\multsp \overvar{3}{\RawWedge }}}\multsp {v_{\overvar{1}{\_}}}+{h_{\overvar{2}{\RawWedge }\multsp \overvar{3}{\RawWedge }\multsp \overvar{3}{\RawWedge
}}}\multsp {v_{\overvar{2}{\_}}}+{h_{\overvar{3}{\RawWedge }\multsp \overvar{3}{\RawWedge }\multsp \overvar{3}{\RawWedge }}}\multsp {v_{\overvar{3}{\_}}}))\mathfrak{E}\big),
\\
{\mathbb{O}_{\epsilon }}[{\mathfrak{P}^{\overvar{2}{\_}}}]&=&-\frac{1}{\epsilon }\frac{\partial }{\partial \epsilon }\Big({{\epsilon }^2}\multsp \multsp \Big(\frac{\partial
{v_{\overvar{1}{\_}}}}{\partial {x^1}}\multsp {h_{\overvar{1}{\RawWedge }\multsp \overvar{1}{\RawWedge }\multsp \overvar{2}{\RawWedge }}}+\frac{\partial
{v_{\overvar{2}{\_}}}}{\partial {x^1}}\multsp {h_{\overvar{1}{\RawWedge }\multsp \overvar{2}{\RawWedge }\multsp \overvar{2}{\RawWedge }}}+  \frac{\partial {v_{\overvar{3}{\_}}}}{\partial {x^1}}\multsp {h_{\overvar{1}{\RawWedge }\multsp \overvar{2}{\RawWedge }\multsp \overvar{3}{\RawWedge
}}}+ \nonumber 
\\
& &\frac{1}{b\multsp c}\multsp \Big(\frac{\partial {v_{\overvar{1}{\_}}}}{\partial {x^3}}\multsp {h_{\overvar{1}{\RawWedge }\multsp \overvar{2}{\RawWedge
}\multsp \overvar{3}{\RawWedge }}}+\frac{\partial {v_{\overvar{2}{\_}}}}{\partial {x^3}}\multsp {h_{\overvar{2}{\RawWedge }\multsp \overvar{2}{\RawWedge
}\multsp \overvar{3}{\RawWedge }}}+\frac{\partial {v_{\overvar{3}{\_}}}}{\partial {x^3}}\multsp {h_{\overvar{2}{\RawWedge }\multsp \overvar{3}{\RawWedge
}\multsp \overvar{3}{\RawWedge }}}\Big)+ \nonumber \\
& & \frac{1}{a}\multsp \Big(\frac{\partial {v_{\overvar{1}{\_}}}}{\partial {x^2}}\multsp {h_{\overvar{1}{\RawWedge }\multsp \overvar{2}{\RawWedge
}\multsp \overvar{2}{\RawWedge }}}+\frac{\partial {v_{\overvar{2}{\_}}}}{\partial {x^2}}\multsp {h_{\overvar{2}{\RawWedge }\multsp \overvar{2}{\RawWedge
}\multsp \overvar{2}{\RawWedge }}}+\frac{\partial {v_{\overvar{3}{\_}}}}{\partial {x^2}}\multsp {h_{\overvar{2}{\RawWedge }\multsp \overvar{2}{\RawWedge
}\multsp \overvar{3}{\RawWedge }}}\Big)+\nonumber
\\
& & \frac{1}{a}\frac{\partial a}{\partial {x^1}}\multsp ({h_{\overvar{2}{\RawWedge }\multsp \overvar{2}{\RawWedge
}\multsp \overvar{2}{\RawWedge }}}\multsp {v_{\overvar{1}{\_}}}-{h_{\overvar{1}{\RawWedge }\multsp \overvar{2}{\RawWedge }\multsp \overvar{2}{\RawWedge
}}}\multsp {v_{\overvar{2}{\_}}})+ \nonumber \\
& & \frac{1}{b}\multsp \frac{\partial b}{\partial {x^1}}\multsp ({h_{\overvar{2}{\RawWedge }\multsp \overvar{3}{\RawWedge }\multsp \overvar{3}{\RawWedge
}}}\multsp {v_{\overvar{1}{\_}}}-{h_{\overvar{1}{\RawWedge }\multsp \overvar{2}{\RawWedge }\multsp \overvar{3}{\RawWedge }}}\multsp {v_{\overvar{3}{\_}}})+
\nonumber
\\
& &\frac{1}{a\multsp
c}\multsp \frac{\partial c}{\partial {x^2}}\multsp ({h_{\overvar{2}{\RawWedge }\multsp \overvar{3}{\RawWedge }\multsp \overvar{3}{\RawWedge }}}\multsp
{v_{\overvar{2}{\_}}}-{h_{\overvar{2}{\RawWedge }\multsp \overvar{2}{\RawWedge }\multsp \overvar{3}{\RawWedge }}}\multsp {v_{\overvar{3}{\_}}})\Big)\mathfrak{E}\Big).
\end{eqnarray}
The evolution of the third component of the orthonormal lab frame monochromatic particle momentum density is governed by
\begin{equation}
{\mathbb{D}_t}{\mathfrak{P}^{\overvar{3}{\_}}}+{\mathbb{D}_{{x^1}}}[{\mathfrak{P}^{\overvar{3}{\_}}}]+{\mathbb{D}_{{x^2}}}[{\mathfrak{P}^{\overvar{3}{\_}}}]+{\mathbb{D}_{{x^3}}}[{\mathfrak{P}^{\overvar{3}{\_}}}]+{\mathbb{F}_{13}}[{\mathfrak{P}^{\overvar{3}{\_}}}]+{\mathbb{F}_{23}}[{\mathfrak{P}^{\overvar{3}{\_}}}]+{\mathbb{O}_{\epsilon
}}[{\mathfrak{P}^{\overvar{3}{\_}}}]= 
 \epsilon \int \mathbb{C}\multsp\, ({n_{\overvar{3}{\RawWedge }}}+{v_{\overvar{3}{\_}}})\,\Mvariable{\sin}\vartheta \,\multsp \Mvariable{d\vartheta
}\multsp\, \Mvariable{d\varphi },
\end{equation}
where
\begin{eqnarray}
{\mathbb{D}_t}[{\mathfrak{P}^{\overvar{3}{\_}}}]&=&\frac{\partial {\mathfrak{P}^{\overvar{3}{\_}}}}{\partial t},\\
{\mathbb{D}_{{x^1}}}[{\mathfrak{P}^{\overvar{3}{\_}}}]&=&  \frac{1}{a\multsp b}\multsp \frac{\partial }{\partial {x^1}}\big(a\multsp b\multsp \big({\mathfrak{P}^{\overvar{3}{\_}}}\multsp {v_{\overvar{1}{\_}}}+{\mathfrak{P}^{\overvar{1}{\_}}}\multsp
{v_{\overvar{3}{\_}}}+({h_{\overvar{1}{\RawWedge }\multsp \overvar{3}{\RawWedge }}}-2\multsp ({h_{\overvar{1}{\RawWedge }\multsp \overvar{1}{\RawWedge
}\multsp \overvar{3}{\RawWedge }}}\multsp {v_{\overvar{1}{\_}}}+{h_{\overvar{1}{\RawWedge }\multsp \overvar{2}{\RawWedge }\multsp \overvar{3}{\RawWedge
}}}\multsp {v_{\overvar{2}{\_}}}+{h_{\overvar{1}{\RawWedge }\multsp \overvar{3}{\RawWedge }\multsp \overvar{3}{\RawWedge }}}\multsp {v_{\overvar{3}{\_}}}))\mathfrak{E}\big)\big),\\
{\mathbb{D}_{{x^2}}}[{\mathfrak{P}^{\overvar{3}{\_}}}]&=&  \frac{1}{a\multsp c}\multsp \frac{\partial }{\partial {x^2}}\big(c\multsp \big({\mathfrak{P}^{\overvar{3}{\_}}}\multsp {v_{\overvar{2}{\_}}}+{\mathfrak{P}^{\overvar{2}{\_}}}\multsp
{v_{\overvar{3}{\_}}}+({h_{\overvar{2}{\RawWedge }\multsp \overvar{3}{\RawWedge }}}-2\multsp ({h_{\overvar{1}{\RawWedge }\multsp \overvar{2}{\RawWedge
}\multsp \overvar{3}{\RawWedge }}}\multsp {v_{\overvar{1}{\_}}}+{h_{\overvar{2}{\RawWedge }\multsp \overvar{2}{\RawWedge }\multsp \overvar{3}{\RawWedge
}}}\multsp {v_{\overvar{2}{\_}}}+{h_{\overvar{2}{\RawWedge }\multsp \overvar{3}{\RawWedge }\multsp \overvar{3}{\RawWedge }}}\multsp {v_{\overvar{3}{\_}}}))\mathfrak{E}\big)\big),\\
{\mathbb{D}_{{x^3}}}[{\mathfrak{P}^{\overvar{3}{\_}}}]&=&\frac{1}{b\multsp c}\multsp \frac{\partial }{\partial {x^3}}\big(2\multsp {\mathfrak{P}^{\overvar{3}{\_}}}\multsp
{v_{\overvar{3}{\_}}}+({h_{\overvar{3}{\RawWedge }\multsp \overvar{3}{\RawWedge }}}-2\multsp ({h_{\overvar{1}{\RawWedge }\multsp \overvar{3}{\RawWedge
}\multsp \overvar{3}{\RawWedge }}}\multsp {v_{\overvar{1}{\_}}}+{h_{\overvar{2}{\RawWedge }\multsp \overvar{3}{\RawWedge }\multsp \overvar{3}{\RawWedge
}}}\multsp {v_{\overvar{2}{\_}}}+{h_{\overvar{3}{\RawWedge }\multsp \overvar{3}{\RawWedge }\multsp \overvar{3}{\RawWedge }}}\multsp {v_{\overvar{3}{\_}}}))\mathfrak{E}\big),\\
{\mathbb{F}_{13}}[{\mathfrak{P}^{\overvar{3}{\_}}}]&=&\frac{1}{b}\multsp \frac{\partial b}{\partial {x^1}}\multsp \big({\mathfrak{P}^{\overvar{3}{\_}}}\multsp
{v_{\overvar{1}{\_}}}+{\mathfrak{P}^{\overvar{1}{\_}}}\multsp {v_{\overvar{3}{\_}}}+({h_{\overvar{1}{\RawWedge }\multsp \overvar{3}{\RawWedge }}}-2\multsp ({h_{\overvar{1}{\RawWedge
}\multsp \overvar{1}{\RawWedge }\multsp \overvar{3}{\RawWedge }}}\multsp {v_{\overvar{1}{\_}}}+{h_{\overvar{1}{\RawWedge }\multsp \overvar{2}{\RawWedge
}\multsp \overvar{3}{\RawWedge }}}\multsp {v_{\overvar{2}{\_}}}+{h_{\overvar{1}{\RawWedge }\multsp \overvar{3}{\RawWedge }\multsp \overvar{3}{\RawWedge
}}}\multsp {v_{\overvar{3}{\_}}}))\mathfrak{E}\big),
\\
{\mathbb{F}_{23}}[{\mathfrak{P}^{\overvar{3}{\_}}}]&=&\frac{1}{a\multsp c}\multsp \frac{\partial c}{\partial {x^2}}\multsp \big({\mathfrak{P}^{\overvar{3}{\_}}}\multsp
{v_{\overvar{2}{\_}}}+{\mathfrak{P}^{\overvar{2}{\_}}}\multsp {v_{\overvar{3}{\_}}}+({h_{\overvar{2}{\RawWedge }\multsp \overvar{3}{\RawWedge }}}-2\multsp ({h_{\overvar{1}{\RawWedge
}\multsp \overvar{2}{\RawWedge }\multsp \overvar{3}{\RawWedge }}}\multsp {v_{\overvar{1}{\_}}}+{h_{\overvar{2}{\RawWedge }\multsp \overvar{2}{\RawWedge
}\multsp \overvar{3}{\RawWedge }}}\multsp {v_{\overvar{2}{\_}}}+{h_{\overvar{2}{\RawWedge }\multsp \overvar{3}{\RawWedge }\multsp \overvar{3}{\RawWedge
}}}\multsp {v_{\overvar{3}{\_}}}))\mathfrak{E}\big),
\\
{\mathbb{O}_{\epsilon }}[{\mathfrak{P}^{\overvar{3}{\_}}}]&=&-\frac{1}{\epsilon }\frac{\partial }{\partial \epsilon }   \Big({{\epsilon }^2}\multsp \multsp \Big(\frac{\partial {v_{\overvar{1}{\_}}}}{\partial {x^1}}\multsp {h_{\overvar{1}{\RawWedge }\multsp
\overvar{1}{\RawWedge }\multsp \overvar{3}{\RawWedge }}}+\frac{\partial {v_{\overvar{2}{\_}}}}{\partial {x^1}}\multsp {h_{\overvar{1}{\RawWedge }\multsp
\overvar{2}{\RawWedge }\multsp \overvar{3}{\RawWedge }}}+\frac{\partial {v_{\overvar{3}{\_}}}}{\partial {x^1}}\multsp {h_{\overvar{1}{\RawWedge }\multsp
\overvar{3}{\RawWedge }\multsp \overvar{3}{\RawWedge }}}+\multsp\nonumber
\\
& & \frac{1}{a}\Big(\frac{\partial {v_{\overvar{1}{\_}}}}{\partial {x^2}}\multsp {h_{\overvar{1}{\RawWedge
}\multsp \overvar{2}{\RawWedge }\multsp \overvar{3}{\RawWedge }}}+\frac{\partial {v_{\overvar{2}{\_}}}}{\partial {x^2}}\multsp {h_{\overvar{2}{\RawWedge
}\multsp \overvar{2}{\RawWedge }\multsp \overvar{3}{\RawWedge }}}+  \frac{\partial {v_{\overvar{3}{\_}}}}{\partial {x^2}}\multsp {h_{\overvar{2}{\RawWedge }\multsp \overvar{3}{\RawWedge }\multsp \overvar{3}{\RawWedge
}}}\Big)+\nonumber
\\
& &\frac{1}{b\multsp c}\multsp \Big(\frac{\partial {v_{\overvar{1}{\_}}}}{\partial {x^3}}\multsp {h_{\overvar{1}{\RawWedge }\multsp \overvar{3}{\RawWedge
}\multsp \overvar{3}{\RawWedge }}}+\frac{\partial {v_{\overvar{2}{\_}}}}{\partial {x^3}}\multsp {h_{\overvar{2}{\RawWedge }\multsp \overvar{3}{\RawWedge
}\multsp \overvar{3}{\RawWedge }}}+\frac{\partial {v_{\overvar{3}{\_}}}}{\partial {x^3}}\multsp {h_{\overvar{3}{\RawWedge }\multsp \overvar{3}{\RawWedge
}\multsp \overvar{3}{\RawWedge }}}\Big)+\nonumber  \\
& & \frac{1}{a}\frac{\partial a}{\partial {x^1}}\multsp ({h_{\overvar{2}{\RawWedge }\multsp \overvar{2}{\RawWedge }\multsp \overvar{3}{\RawWedge
}}}\multsp {v_{\overvar{1}{\_}}}-{h_{\overvar{1}{\RawWedge }\multsp \overvar{2}{\RawWedge }\multsp \overvar{3}{\RawWedge }}}\multsp {v_{\overvar{2}{\_}}})+
\nonumber \\
& &\frac{1}{b}\multsp
\frac{\partial b}{\partial {x^1}}\multsp ({h_{\overvar{3}{\RawWedge }\multsp \overvar{3}{\RawWedge }\multsp \overvar{3}{\RawWedge }}}\multsp {v_{\overvar{1}{\_}}}-{h_{\overvar{1}{\RawWedge
}\multsp \overvar{3}{\RawWedge }\multsp \overvar{3}{\RawWedge }}}\multsp {v_{\overvar{3}{\_}}})+ \nonumber \\
& & \frac{1}{a\multsp c}\multsp \frac{\partial c}{\partial {x^2}}\multsp ({h_{\overvar{3}{\RawWedge }\multsp \overvar{3}{\RawWedge }\multsp
\overvar{3}{\RawWedge }}}\multsp {v_{\overvar{2}{\_}}}-{h_{\overvar{2}{\RawWedge }\multsp \overvar{3}{\RawWedge }\multsp \overvar{3}{\RawWedge }}}\multsp
{v_{\overvar{3}{\_}}})\Big)\mathfrak{E}\Big).
\end{eqnarray}
In these energy and momentum equations, we have defined
\begin{eqnarray}
{h_{\overvar{i}{\RawWedge }\overvar{j}{\RawWedge }}}&\equiv& \frac{1}{\mathfrak{E}}\int \ScriptCapitalE\, \multsp {n_{\overvar{i}{\RawWedge
}}}{n_{\overvar{j}{\RawWedge }}}\,\Mvariable{\sin}\vartheta \Mvariable{d\vartheta }\,\multsp \Mvariable{d\varphi },\\
{h_{\overvar{i}{\RawWedge }\overvar{j}{\RawWedge }\overvar{k}{\RawWedge }}}&\equiv& \frac{1}{\mathfrak{E}}\int \ScriptCapitalE \multsp\,
{n_{\overvar{i}{\RawWedge }}}{n_{\overvar{j}{\RawWedge }}}{n_{\overvar{k}{\RawWedge }}}\,\Mvariable{\sin}\vartheta \,\Mvariable{d\vartheta }\multsp\,
\Mvariable{d\varphi },
\end{eqnarray}
which are akin to Eddington factors in traditional moment approaches to radiation transport. We point out that $f_{\overvar{i}{\RawWedge}}$ is not equal to $\mathfrak{P}^{\overvar{i}{\_}}/\mathfrak{E}$, and that the second angular moment factors
\inlineSFinmath${{f_{\overvar{i}{\RawWedge }\overvar{j}{\RawWedge }}}}$ and \inlineSFinmath${{h_{\overvar{i}{\RawWedge }\overvar{j}{\RawWedge }}}}$ are
not equal to each other. This is because \inlineSFinmath${\ScriptCapitalN }$ and \inlineSFinmath${\ScriptCapitalE }$ have different angular dependencies,
as can be seen from Eqs. (\ref{specificParticleDensity}) and (\ref{specificEnergyDensity}). This complication is a result of taking moments of {\em lab frame} number and energy distributions with respect to {\em comoving frame} angular variables.

\begin{acknowledgments}
We are grateful to 
Matthias Liebend\"orfer, Bronson Messer, and Raph Hix for insightful
discussions about various aspects of this work, and
helpful suggestions on the manuscript. 
We thank Dimitri Mihalas and Steve Bruenn 
for discussions about momentum space
coordinate systems. This work was supported 
by the U.S. Department of Energy (DoE) HENP Scientific Discovery Through
Advanced Computing Program; Oak Ridge National Laboratory, managed by UT-Battelle, LLC, for the DoE under contract DE-AC05-00OR22725; the Joint Institute
for Heavy Ion Research; and a DoE PECASE grant.
\end{acknowledgments}

\bibliography{bibliography}

\begin{thebibliography}{27}
\expandafter\ifx\csname natexlab\endcsname\relax\def\natexlab#1{#1}\fi
\expandafter\ifx\csname bibnamefont\endcsname\relax
  \def\bibnamefont#1{#1}\fi
\expandafter\ifx\csname bibfnamefont\endcsname\relax
  \def\bibfnamefont#1{#1}\fi
\expandafter\ifx\csname citenamefont\endcsname\relax
  \def\citenamefont#1{#1}\fi
\expandafter\ifx\csname url\endcsname\relax
  \def\url#1{\texttt{#1}}\fi
\expandafter\ifx\csname urlprefix\endcsname\relax\def\urlprefix{URL }\fi
\providecommand{\bibinfo}[2]{#2}
\providecommand{\eprint}[2][]{\url{#2}}

\bibitem[{\citenamefont{{Rampp} and {Janka}}(2000)}]{rampp00}
\bibinfo{author}{\bibfnamefont{M.}~\bibnamefont{{Rampp}}} \bibnamefont{and}
  \bibinfo{author}{\bibfnamefont{H.-T.} \bibnamefont{{Janka}}},
  \bibinfo{journal}{Astrophys. J. Lett.} \textbf{\bibinfo{volume}{539}},
  \bibinfo{pages}{33} (\bibinfo{year}{2000}).

\bibitem[{\citenamefont{{Mezzacappa} et~al.}(2001)\citenamefont{{Mezzacappa},
  {Liebend\"orfer}, {Messer}, {Hix}, {Thielemann}, and
  {Bruenn}}}]{mezzacappa01}
\bibinfo{author}{\bibfnamefont{A.}~\bibnamefont{{Mezzacappa}}},
  \bibinfo{author}{\bibfnamefont{M.}~\bibnamefont{{Liebend\"orfer}}},
  \bibinfo{author}{\bibfnamefont{O.~E.~B.} \bibnamefont{{Messer}}},
  \bibinfo{author}{\bibfnamefont{W.~R.} \bibnamefont{{Hix}}},
  \bibinfo{author}{\bibfnamefont{F.-K.} \bibnamefont{{Thielemann}}},
  \bibnamefont{and} \bibinfo{author}{\bibfnamefont{S.~W.}
  \bibnamefont{{Bruenn}}}, \bibinfo{journal}{\prl}
  \textbf{\bibinfo{volume}{86}}, \bibinfo{pages}{1935} (\bibinfo{year}{2001}).

\bibitem[{\citenamefont{{Liebend{\" o}rfer}
  et~al.}(2001)\citenamefont{{Liebend{\" o}rfer}, {Mezzacappa}, {Thielemann},
  {Messer}, {Hix}, and {Bruenn}}}]{liebendoerfer01b}
\bibinfo{author}{\bibfnamefont{M.}~\bibnamefont{{Liebend{\" o}rfer}}},
  \bibinfo{author}{\bibfnamefont{A.}~\bibnamefont{{Mezzacappa}}},
  \bibinfo{author}{\bibfnamefont{F.}~\bibnamefont{{Thielemann}}},
  \bibinfo{author}{\bibfnamefont{O.~E.} \bibnamefont{{Messer}}},
  \bibinfo{author}{\bibfnamefont{W.~R.} \bibnamefont{{Hix}}}, \bibnamefont{and}
  \bibinfo{author}{\bibfnamefont{S.~W.} \bibnamefont{{Bruenn}}},
  \bibinfo{journal}{\prd} \textbf{\bibinfo{volume}{63}},
  \bibinfo{pages}{103004} (\bibinfo{year}{2001}).

\bibitem[{\citenamefont{{Rampp} and {Janka}}(2002)}]{rampp02}
\bibinfo{author}{\bibfnamefont{M.}~\bibnamefont{{Rampp}}} \bibnamefont{and}
  \bibinfo{author}{\bibfnamefont{H.-T.} \bibnamefont{{Janka}}},
  \bibinfo{journal}{Astron. Astrophys.} \textbf{\bibinfo{volume}{396}},
  \bibinfo{pages}{361} (\bibinfo{year}{2002}).

\bibitem[{\citenamefont{{Liebend\"orfer}
  et~al.}(2002)\citenamefont{{Liebend\"orfer}, {Messer}, {Mezzacappa},
  {Bruenn}, {Cardall}, and {Thielemann}}}]{liebendoerfer02}
\bibinfo{author}{\bibfnamefont{M.}~\bibnamefont{{Liebend\"orfer}}},
  \bibinfo{author}{\bibfnamefont{O.~E.~B.} \bibnamefont{{Messer}}},
  \bibinfo{author}{\bibfnamefont{A.}~\bibnamefont{{Mezzacappa}}},
  \bibinfo{author}{\bibfnamefont{S.~W.} \bibnamefont{{Bruenn}}},
  \bibinfo{author}{\bibfnamefont{C.~Y.} \bibnamefont{{Cardall}}},
  \bibnamefont{and} \bibinfo{author}{\bibfnamefont{F.-K.}
  \bibnamefont{{Thielemann}}} (\bibinfo{year}{2002}),
  \eprint[http://arXiv.org/abs]{astro-ph/0207036}.

\bibitem[{\citenamefont{Thompson et~al.}(2002)\citenamefont{Thompson, Burrows,
  and Pinto}}]{thompson02}
\bibinfo{author}{\bibfnamefont{T.~A.} \bibnamefont{Thompson}},
  \bibinfo{author}{\bibfnamefont{A.}~\bibnamefont{Burrows}}, \bibnamefont{and}
  \bibinfo{author}{\bibfnamefont{P.~A.} \bibnamefont{Pinto}}
  (\bibinfo{year}{2002}), \eprint{astro-ph/0211194}.

\bibitem[{\citenamefont{{Janka}
  et~al.}(2002{\natexlab{a}})\citenamefont{{Janka}, {Buras}, and
  {Rampp}}}]{janka02}
\bibinfo{author}{\bibfnamefont{H.-T.} \bibnamefont{{Janka}}},
  \bibinfo{author}{\bibfnamefont{R.}~\bibnamefont{{Buras}}}, \bibnamefont{and}
  \bibinfo{author}{\bibfnamefont{M.}~\bibnamefont{{Rampp}}}
  (\bibinfo{year}{2002}{\natexlab{a}}), \eprint{astro-ph/0212317}.

\bibitem[{\citenamefont{{Janka}
  et~al.}(2002{\natexlab{b}})\citenamefont{{Janka}, {Buras}, {Kifonidis},
  {Plewa}, and {Rampp}}}]{janka02b}
\bibinfo{author}{\bibfnamefont{H.-T.} \bibnamefont{{Janka}}},
  \bibinfo{author}{\bibfnamefont{R.}~\bibnamefont{{Buras}}},
  \bibinfo{author}{\bibfnamefont{K.}~\bibnamefont{{Kifonidis}}},
  \bibinfo{author}{\bibfnamefont{T.}~\bibnamefont{{Plewa}}}, \bibnamefont{and}
  \bibinfo{author}{\bibfnamefont{M.}~\bibnamefont{{Rampp}}}
  (\bibinfo{year}{2002}{\natexlab{b}}), \eprint{astro-ph/0212316}.

\bibitem[{\citenamefont{{Mezzacappa} and
  {Bruenn}}(1993{\natexlab{a}})}]{mezzacappa93b}
\bibinfo{author}{\bibfnamefont{A.}~\bibnamefont{{Mezzacappa}}}
  \bibnamefont{and} \bibinfo{author}{\bibfnamefont{S.~W.}
  \bibnamefont{{Bruenn}}}, \bibinfo{journal}{\apj}
  \textbf{\bibinfo{volume}{405}}, \bibinfo{pages}{669}
  (\bibinfo{year}{1993}{\natexlab{a}}).

\bibitem[{\citenamefont{{Colgate} and {White}}(1966)}]{colgate66}
\bibinfo{author}{\bibfnamefont{S.~A.} \bibnamefont{{Colgate}}}
  \bibnamefont{and} \bibinfo{author}{\bibfnamefont{R.~H.}
  \bibnamefont{{White}}}, \bibinfo{journal}{\apj}
  \textbf{\bibinfo{volume}{143}}, \bibinfo{pages}{626} (\bibinfo{year}{1966}).

\bibitem[{\citenamefont{{Bethe} and {Wilson}}(1985)}]{bethe85}
\bibinfo{author}{\bibfnamefont{H.~A.} \bibnamefont{{Bethe}}} \bibnamefont{and}
  \bibinfo{author}{\bibfnamefont{J.~R.} \bibnamefont{{Wilson}}},
  \bibinfo{journal}{\apj} \textbf{\bibinfo{volume}{295}}, \bibinfo{pages}{14}
  (\bibinfo{year}{1985}).

\bibitem[{\citenamefont{{Janka} and {Mueller}}(1996)}]{janka96}
\bibinfo{author}{\bibfnamefont{H.-T.} \bibnamefont{{Janka}}} \bibnamefont{and}
  \bibinfo{author}{\bibfnamefont{E.}~\bibnamefont{{Mueller}}},
  \bibinfo{journal}{Astron. Astrophys.} \textbf{\bibinfo{volume}{306}},
  \bibinfo{pages}{167} (\bibinfo{year}{1996}).

\bibitem[{\citenamefont{{Janka}}(2001)}]{janka01}
\bibinfo{author}{\bibfnamefont{H.-T.} \bibnamefont{{Janka}}},
  \bibinfo{journal}{Astron. Astrophys.} \textbf{\bibinfo{volume}{368}},
  \bibinfo{pages}{527} (\bibinfo{year}{2001}).

\bibitem[{\citenamefont{{Herant} et~al.}(1994)\citenamefont{{Herant}, {Benz},
  {Hix}, {Fryer}, and {Colgate}}}]{herant94}
\bibinfo{author}{\bibfnamefont{M.}~\bibnamefont{{Herant}}},
  \bibinfo{author}{\bibfnamefont{W.}~\bibnamefont{{Benz}}},
  \bibinfo{author}{\bibfnamefont{W.~R.} \bibnamefont{{Hix}}},
  \bibinfo{author}{\bibfnamefont{C.~L.} \bibnamefont{{Fryer}}},
  \bibnamefont{and} \bibinfo{author}{\bibfnamefont{S.~A.}
  \bibnamefont{{Colgate}}}, \bibinfo{journal}{\apj}
  \textbf{\bibinfo{volume}{435}}, \bibinfo{pages}{339} (\bibinfo{year}{1994}).

\bibitem[{\citenamefont{{Burrows} et~al.}(1995)\citenamefont{{Burrows},
  {Hayes}, and {Fryxell}}}]{burrows95}
\bibinfo{author}{\bibfnamefont{A.}~\bibnamefont{{Burrows}}},
  \bibinfo{author}{\bibfnamefont{J.}~\bibnamefont{{Hayes}}}, \bibnamefont{and}
  \bibinfo{author}{\bibfnamefont{B.~A.} \bibnamefont{{Fryxell}}},
  \bibinfo{journal}{\apj} \textbf{\bibinfo{volume}{450}}, \bibinfo{pages}{830}
  (\bibinfo{year}{1995}).

\bibitem[{\citenamefont{{Fryer} and {Warren}}(2002)}]{fryer02}
\bibinfo{author}{\bibfnamefont{C.~L.} \bibnamefont{{Fryer}}} \bibnamefont{and}
  \bibinfo{author}{\bibfnamefont{M.~S.} \bibnamefont{{Warren}}},
  \bibinfo{journal}{Astrophys. J. Lett.} \textbf{\bibinfo{volume}{574}},
  \bibinfo{pages}{L65} (\bibinfo{year}{2002}).

\bibitem[{\citenamefont{{Mezzacappa}
  et~al.}(1998{\natexlab{a}})\citenamefont{{Mezzacappa}, {Calder}, {Bruenn},
  {Blondin}, {Guidry}, {Strayer}, and {Umar}}}]{mezzacappa98}
\bibinfo{author}{\bibfnamefont{A.}~\bibnamefont{{Mezzacappa}}},
  \bibinfo{author}{\bibfnamefont{A.~C.} \bibnamefont{{Calder}}},
  \bibinfo{author}{\bibfnamefont{S.~W.} \bibnamefont{{Bruenn}}},
  \bibinfo{author}{\bibfnamefont{J.~M.} \bibnamefont{{Blondin}}},
  \bibinfo{author}{\bibfnamefont{M.~W.} \bibnamefont{{Guidry}}},
  \bibinfo{author}{\bibfnamefont{M.~R.} \bibnamefont{{Strayer}}},
  \bibnamefont{and} \bibinfo{author}{\bibfnamefont{A.~S.}
  \bibnamefont{{Umar}}}, \bibinfo{journal}{\apj}
  \textbf{\bibinfo{volume}{493}}, \bibinfo{pages}{848}
  (\bibinfo{year}{1998}{\natexlab{a}}).

\bibitem[{\citenamefont{{Mezzacappa}
  et~al.}(1998{\natexlab{b}})\citenamefont{{Mezzacappa}, {Calder}, {Bruenn},
  {Blondin}, {Guidry}, {Strayer}, and {Umar}}}]{mezzacappa98b}
\bibinfo{author}{\bibfnamefont{A.}~\bibnamefont{{Mezzacappa}}},
  \bibinfo{author}{\bibfnamefont{A.~C.} \bibnamefont{{Calder}}},
  \bibinfo{author}{\bibfnamefont{S.~W.} \bibnamefont{{Bruenn}}},
  \bibinfo{author}{\bibfnamefont{J.~M.} \bibnamefont{{Blondin}}},
  \bibinfo{author}{\bibfnamefont{M.~W.} \bibnamefont{{Guidry}}},
  \bibinfo{author}{\bibfnamefont{M.~R.} \bibnamefont{{Strayer}}},
  \bibnamefont{and} \bibinfo{author}{\bibfnamefont{A.~S.}
  \bibnamefont{{Umar}}}, \bibinfo{journal}{\apj}
  \textbf{\bibinfo{volume}{495}}, \bibinfo{pages}{911}
  (\bibinfo{year}{1998}{\natexlab{b}}).

\bibitem[{\citenamefont{{Liebend\"orfer}
  et~al.}(2001)\citenamefont{{Liebend\"orfer}, {Mezzacappa}, and
  {Thielemann}}}]{liebendoerfer01}
\bibinfo{author}{\bibfnamefont{M.}~\bibnamefont{{Liebend\"orfer}}},
  \bibinfo{author}{\bibfnamefont{A.}~\bibnamefont{{Mezzacappa}}},
  \bibnamefont{and} \bibinfo{author}{\bibfnamefont{F.-K.}
  \bibnamefont{{Thielemann}}}, \bibinfo{journal}{\prd}
  \textbf{\bibinfo{volume}{63}}, \bibinfo{pages}{104003}
  (\bibinfo{year}{2001}).

\bibitem[{\citenamefont{{Lindquist}}(1966)}]{lindquist66}
\bibinfo{author}{\bibfnamefont{R.~W.} \bibnamefont{{Lindquist}}},
  \bibinfo{journal}{Ann. Phys. (NY)} \textbf{\bibinfo{volume}{37}},
  \bibinfo{pages}{487} (\bibinfo{year}{1966}).

\bibitem[{\citenamefont{{Mezzacappa} and {Matzner}}(1989)}]{mezzacappa89}
\bibinfo{author}{\bibfnamefont{A.}~\bibnamefont{{Mezzacappa}}}
  \bibnamefont{and} \bibinfo{author}{\bibfnamefont{R.~A.}
  \bibnamefont{{Matzner}}}, \bibinfo{journal}{\apj}
  \textbf{\bibinfo{volume}{343}}, \bibinfo{pages}{853} (\bibinfo{year}{1989}).

\bibitem[{\citenamefont{{Ehlers}}(1971)}]{ehlers71}
\bibinfo{author}{\bibfnamefont{J.}~\bibnamefont{{Ehlers}}},
  \emph{\bibinfo{title}{Proceedings of the International School of Physics
  ``Enrico Fermi'' Course XLVII: General Relativity and Cosmology}}
  (\bibinfo{publisher}{Academic Press}, \bibinfo{address}{New York},
  \bibinfo{year}{1971}), pp. \bibinfo{pages}{1--70}.

\bibitem[{\citenamefont{{Israel}}(1972)}]{israel72}
\bibinfo{author}{\bibfnamefont{W.}~\bibnamefont{{Israel}}},
  \emph{\bibinfo{title}{General Relativity: Papers in Honour of J. L. Synge}}
  (\bibinfo{publisher}{Clarendon Press}, \bibinfo{address}{Oxford},
  \bibinfo{year}{1972}), pp. \bibinfo{pages}{201--241}.

\bibitem[{\citenamefont{{Mezzacappa} and
  {Bruenn}}(1993{\natexlab{b}})}]{mezzacappa93}
\bibinfo{author}{\bibfnamefont{A.}~\bibnamefont{{Mezzacappa}}}
  \bibnamefont{and} \bibinfo{author}{\bibfnamefont{S.~W.}
  \bibnamefont{{Bruenn}}}, \bibinfo{journal}{\apj}
  \textbf{\bibinfo{volume}{405}}, \bibinfo{pages}{637}
  (\bibinfo{year}{1993}{\natexlab{b}}).

\bibitem[{\citenamefont{{Misner} et~al.}(1973)\citenamefont{{Misner}, {Thorne},
  and {Wheeler}}}]{mtw}
\bibinfo{author}{\bibfnamefont{C.~W.} \bibnamefont{{Misner}}},
  \bibinfo{author}{\bibfnamefont{K.~S.} \bibnamefont{{Thorne}}},
  \bibnamefont{and} \bibinfo{author}{\bibfnamefont{J.~A.}
  \bibnamefont{{Wheeler}}}, \emph{\bibinfo{title}{Gravitation}}
  (\bibinfo{publisher}{W. H. Freeman and Company}, \bibinfo{address}{New York},
  \bibinfo{year}{1973}).

\bibitem[{\citenamefont{{Castor}}(1972)}]{castor72}
\bibinfo{author}{\bibfnamefont{J.~I.} \bibnamefont{{Castor}}},
  \bibinfo{journal}{\apj} \textbf{\bibinfo{volume}{178}}, \bibinfo{pages}{779}
  (\bibinfo{year}{1972}).

\bibitem[{\citenamefont{{Burrows} et~al.}(2000)\citenamefont{{Burrows},
  {Young}, {Pinto}, {Eastman}, and {Thompson}}}]{burrows00}
\bibinfo{author}{\bibfnamefont{A.}~\bibnamefont{{Burrows}}},
  \bibinfo{author}{\bibfnamefont{T.}~\bibnamefont{{Young}}},
  \bibinfo{author}{\bibfnamefont{P.}~\bibnamefont{{Pinto}}},
  \bibinfo{author}{\bibfnamefont{R.}~\bibnamefont{{Eastman}}},
  \bibnamefont{and} \bibinfo{author}{\bibfnamefont{T.~A.}
  \bibnamefont{{Thompson}}}, \bibinfo{journal}{\apj}
  \textbf{\bibinfo{volume}{539}}, \bibinfo{pages}{865} (\bibinfo{year}{2000}).

\end{thebibliography}

\end{document}